\documentclass{appolb}
\usepackage{epsfig}
\usepackage{wrapfig}

\begin{document}
\pagestyle{plain}
\newcount\eLiNe\eLiNe=\inputlineno\advance\eLiNe by -1
\title{TESTING NEW STRONGLY INTENSIVE MEASURES \\ OF TRANSVERSE MOMENTUM 
FLUCTUATIONS %
\thanks{Send any remarks to {\tt kperl@if.pw.edu.pl}}%
}

\author{Katarzyna GREBIESZKOW 
\address{Faculty of Physics, Warsaw University of Technology,
Koszykowa 75, 00-662~Warsaw, Poland}}
\maketitle

\begin{abstract}

Recently, the new strongly intensive measures of fluctuations $\Delta$ and $\Sigma$ have been proposed. In this publication their properties are tested using an example of event-by-event transverse momentum fluctuations. The obtained values are compared to the long used $\Phi$ measure of $p_T$ fluctuations. Several tests are preformed within data produced by fast generators, as well as by the UrQMD model. The UrQMD calculations are presented for the systems and energies which are planned to be studied in the NA61/SHINE experiment at the CERN Super Proton Synchrotron.   
\end{abstract}

\section{Introduction}

The main motivation of colliding relativistic heavy ions is to create and study the properties of the system composed by deconfined quarks and gluons (QGP). The data suggest that the energy threshold for deconfinement (onset of deconfinement) is located at low SPS energies \cite{mg_model, na49_kpi}. The phase diagram of strongly interacting matter can be presented in terms of temperature $T$ and baryochemical potential $\mu_B$. The bulk of theoretical calculations suggest that the phase boundary between hadrons and QGP is of first order at large values of $\mu_B$, ending in a critical point of second order and then turning into a continuous rapid transition at low $\mu_B$. Lattice QCD calculations indicate that the critical point (CP) can be located in the SPS energy range, i.e. $T^{CP} = 162 \pm 2$ MeV, $\mu_B^{CP} = 360 \pm 40 $ MeV \cite{fodor_latt_2004} or $(T^{CP}, \mu_{B}^{CP})
= (0.927(5)T_c, 2.60(8)T_c) = (\sim 157,\sim 441)$ MeV \cite{lat2011}, where $T_c$ is the critical temperature of hadron gas $\leftrightarrow$ QGP transition at vanishing baryochemical potential.

The analysis of dynamical fluctuations can be an important tool for localizing the phase boundary and the critical point. In particular, significant transverse momentum and multiplicity fluctuations are expected to appear for systems freezing-out close to CP \cite{SRS}. The position of the freeze-out point in the phase diagram can be moved by varying the collision energy and the size of the colliding nuclei \cite{beccatini}. A non-monotonic evolution of fluctuations with such parameters 
can serve as a signature for the phase transition and the critical point. 

\begin{wrapfigure}{r}{7.cm} 
\vspace{-0.5cm}
\includegraphics[scale=0.35]{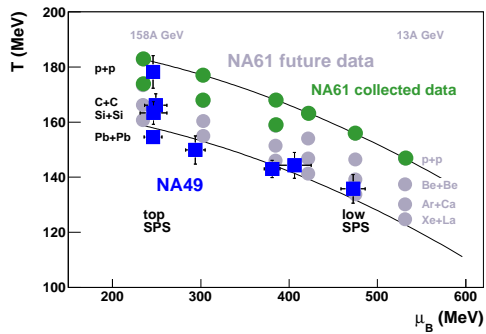}
\vspace{-0.6cm}
\caption[]{Positions of chemical freeze-out points obtained within hadron gas model by fitting NA49 data \cite{beccatini} (blue squares). Circles are those expected in NA61. Taken from \cite{KG_CPOD2011}.}
\label{phas_full}
\end{wrapfigure}

In fact, these considerations motivated an extensive program of 
fluctuation studies at the SPS and RHIC accelerators. The NA49 
\cite{na49_nim} experiment reported non-monotonic behavior of average 
$p_T$ and multiplicity fluctuations at the top SPS energy 
\cite{kg_qm09}. This intriguing result might be a first hint of the 
critical point. Therefore the efforts to look for CP will be continued 
within the NA61/SHINE \cite{shine} project where a 2D (energy and 
system size) scan of the phase diagram will be performed. Figure 
\ref{phas_full} shows the hypothetical chemical freeze-out points in 
the NA61 experiment.

Dynamical fluctuations can be measured by use of event-by-event
methods. However, while measuring event-by-event fluctuations in nucleus+nucleus ($A+A$) collisions one should remember about a trivial source of fluctuations
caused by event-by-event changes of the collision geometry. 
Therefore, a suitable choice of statistical tools for the study of 
event-by-event fluctuations is really important. In Ref.~\cite{Phi_measure} a 
strongly intensive measure $\Phi$ was introduced. In a 
superposition model $\Phi$ does not depend on the number of 
"sources" (e.g. wounded nucleons in the Wounded Nucleon Model \cite{wnm_model}) 
composing $A+A$ collision (intensive measure) and on the 
fluctuations of this number of sources (strongly intensive measure). 
In addition, in thermodynamical models $\Phi$ does not depend on volume 
and volume fluctuations provided that temperature and chemical potential are constant.
The $\Phi$ measure was already used by the NA49 experiment to calculate
transverse momentum fluctuations ($\Phi_{p_T}$) \cite{fluct_size, 
fluct_energy}. Recently, two new classes of strongly intensive 
measures have been proposed: $\Delta$ and $\Sigma$ \cite{strongly}. In fact, 
previously proposed $\Phi$ belongs to $\Sigma$-family measures.
In heavy ion experiments, the use of strongly intensive measures of fluctuations, such as $\Phi$, $\Delta$ and $\Sigma$, can be a remedy for an imperfect centrality selection of $A+A$ collisions. Therefore, the NA49 experiment was able to use relatively wide centrality bins while studying $\Phi_{p_T}$ measure (up to 0-15\% centrality), whereas the analysis of multiplicity fluctuations had to be limited to 1\% most central interactions only \cite{kg_qm09}.   
  
In this paper first basic tests of the newly proposed $\Delta$ and 
$\Sigma$ measures will be presented for transverse momentum 
fluctuations. The obtained values will be compared to the long used $\Phi$ 
measure of $p_T$ fluctuations. Several effects have been studied for events generated by use of so-called fast generators. Moreover, the analysis within a much more complex UrQMD3.3 model \cite{UrQMD} will be shown. The UrQMD calculations have been done for the systems and energies which are planned to be studied in the CERN NA61/SHINE experiment (see Fig. \ref{phas_full}).

\section{Strongly intensive measures}

We call intensive quantities those ones which do {\it not} depend on the volume of the system. In contrary, extensive quantities (for example mean multiplicity or variance of multiplicity distribution) are proportional to the system volume. 
Note, that it is useful to extend the notion of intensive and extensive quantities to the Wounded Nucleon Model. Namely, the intensive quantities can be called those ones which are independent of the number of wounded nucleons, and extensive ones those which are proportional to the number of wounded nucleons. The ratio of two extensive quantities is an intensive quantity  \cite{strongly}. Therefore the ratio of mean multiplicities, as well as the scaled variance of multiplicity distribution $\omega = (\langle N^2 \rangle - \langle N 
\rangle ^2)/ \langle N \rangle$, are intensive measures. In fact, due to 
its intensity property $\omega$ measure is quite commonly used to 
determine multiplicity fluctuations in heavy ion experiments.       

There is one more important problem which one should not forget about. 
In high energy heavy ion collisions the volume 
of the produced matter cannot be fixed. In fact the system volume 
changes significantly from event to event. Therefore it is very 
important to be able to measure the properties of the created matter 
independently of its volume fluctuations. The quantities which allow this 
are called strongly intensive measures. They do not depend on the volume 
and on volume fluctuations. 

Mean multiplicities ratios are both intensive and strongly intensive 
measures. The situation is however more difficult for fluctuation 
analysis. The scaled variance of multiplicity distribution is an 
intensive measure but not strongly intensive. Quite long ago a strongly 
intensive measure $\Phi$ was first introduced \cite{Phi_measure}. 
In the recent paper \cite{strongly} is was shown that there are at 
least two families of strongly intensive measures: $\Delta$ and 
$\Sigma$. The previously known $\Phi$ measure belongs to 
$\Sigma$-type family. They can be calculated for any two extensive quantities.
In this paper $\Delta$, $\Sigma$, and $\Phi$ measures calculated for particle multiplicity, $N$, and sum of their transverse momenta modules will be tested.

\subsection{$\Phi_{p_T}$ measure}

The $\Phi$ measure \cite{Phi_measure} was already successfully used by NA49 
to determine transverse momentum fluctuations ($\Phi_{p_T}$) 
\cite{fluct_size, fluct_energy}. Following the authors of 
\cite{Phi_measure} one defines the single-particle variable 
$z_{p_{T}}=p_{T}-\overline{p_{T}}$ with the bar denoting averaging over 
the single-particle inclusive distribution. As seen 
$\overline{z_{p_{T}}} = 0$. Further, one introduces the event variable 
$Z_{p_{T}}$, which is a multi-particle analog of $z_{p_{T}}$, defined as
$Z_{p_{T}}=\sum_{i=1}^{N}(p_{Ti}-\overline{p_{T}})$, where the summation 
runs over particles in a given event. Note, that   
$\langle Z_{p_{T}} \rangle = 0$, where $\langle ... \rangle$ represents
averaging over events. Finally, the $\Phi_{p_{T}}$ measure is defined as
\begin{equation}
\label{Phi}
\Phi_{p_{T}}=\sqrt{\frac{\langle
Z_{p_{T}}^{2} \rangle }{\langle N
\rangle }}-\sqrt{\overline{z_{p_{T}}^{2}}}.
\label{eq_phi}
\end{equation}

$\Phi_{p_T}$ is a strongly intensive measure and therefore if $A+A$ 
collision is represented by an incoherent superposition of independent nucleon+nucleon ($N+N$) interactions (superposition model), then $\Phi_{p_{T}}$ has a 
constant value, the same for $A+A$ and $N+N$ interactions. This implies 
that, in particular, $\Phi_{p_{T}}$ does not depend on the impact 
parameter (centrality), if the $A+A$ collision is a simple superposition 
of $N+N$ interactions.  

Another property of this measure is that $\Phi_{p_{T}}$ vanishes when  
the system consists of particles that are emitted independently (no 
inter-particle correlations) and the single particle momentum spectrum 
is independent of multiplicity. In contrary, $\Delta$ and $\Sigma$ 
measures, which will be shown below, do not assume zero values for 
independent particle production.

\subsection{$\Delta^{XN}$ and $\Sigma^{XN}$ measures}

Let $A$ and $B$ be two fluctuating extensive quantities. Then 
$\Delta^{AB}$ and $\Sigma^{AB}$ can be defined \cite{strongly}:

\begin{equation}
\Delta^{AB}= {\langle C \rangle}^{-1} [\langle B \rangle \omega_{A} - 
\langle A \rangle \omega_{B}],
\end{equation}

\begin{equation}
\Sigma^{AB} = {\langle C \rangle}^{-1} [ \langle B \rangle \omega_{A} +
\langle A \rangle \omega_{B} - 2 (\langle AB \rangle - \langle A 
\rangle \langle B \rangle ) ],
\end{equation}

where: 
\begin{equation}
\omega_{A} = \frac {\langle A^2 \rangle - {\langle A \rangle}^2 }  
{\langle A \rangle}  
\end{equation}
and  
\begin{equation}
\omega_{B} = \frac {\langle B^2 \rangle - {\langle B \rangle}^2 }
{\langle B \rangle}
\end{equation}

\noindent
are the scaled variances of two fluctuating extensive quantities $A$ and 
$B$. $\langle C \rangle$ is the average of any extensive quantity e.g., 
$\langle A \rangle$ or $\langle B \rangle$.  

There is an important difference between $\Delta^{AB}$ and $\Sigma^{AB}$. 
Only the first two moments: $\langle A \rangle$, $\langle B \rangle$, 
and $\langle A^2 \rangle$, $\langle B^2 \rangle$ are required to 
calculate  $\Delta^{AB}$, whereas $\Sigma^{AB}$ includes the 
correlation term $\langle AB \rangle - \langle A \rangle \langle B 
\rangle$. Thus $\Delta^{AB}$ and $\Sigma^{AB}$ measures can be 
sensitive to several physics effects in different ways. In publication 
\cite{strongly} all strongly intensive quantities including correlation 
term are named the $\Sigma$ family, and those including only mean values and 
variances the $\Delta$ family. The already used $\Phi$ measure belongs to 
$\Sigma$-type family. The definitions of new quantities $\Delta^{AB}$ and $\Sigma^{AB}$ are, however, more general because one can use here {\it any} two extensive measures $A$ and $B$, whereas in the original definition of $\Phi$ one of them was fixed to be particle multiplicity. Nevertheless, in this paper one of the extensive measures will be again multiplicity, in order to compare the results with known measure $\Phi$. Finally, one should mention that $\Delta^{AB}$ and $\Sigma^{AB}$ have also different properties with respect to exchange $A$ and $B$:
$\Sigma^{AB} = \Sigma^{BA}$ and  $\Delta^{AB} = - \Delta^{BA}$

For the analysis of transverse momentum fluctuations one can use 
\cite{strongly}: $A \equiv X = \sum_{i=1}^{N} x_i$ (where $x_i \equiv 
p_{T, i}$ and the summation runs over all {\it accepted} particles in a 
given event), $B \equiv N$, $C \equiv N $, and $X_2 = \sum_{i=1}^{N} 
x^2_i$. Then we obtain:

\begin{eqnarray}
\Delta^{XN}= \frac {1}{\langle N \rangle} [ \langle N \rangle \omega_X - 
\langle X \rangle \omega_N ] =   \cr 
= \frac {1}{\langle N \rangle} 
\left[ 
\langle N \rangle \left( 
\frac{\langle X^2 \rangle - {\langle X \rangle}^2}{ \langle X \rangle} 
\right) -
\langle X \rangle \left(
\frac{\langle N^2 \rangle - {\langle N \rangle}^2}{ \langle N \rangle} 
\right)
\right]
\end{eqnarray}

\noindent
and

\begin{equation}
\Sigma^{XN} = \frac {1}{\langle N \rangle} [\langle N \rangle \omega_X 
+ \langle X \rangle \omega_N - 2 ( \langle XN \rangle - \langle X 
\rangle \langle N \rangle ) ].
\end{equation}

\noindent
The $\Sigma^{XN}$ measure can be also expressed \cite{strongly} using known 
$\Phi_{p_T}$ quantity:
\begin{equation}
\Sigma^{XN}= \frac{\Phi^2_{p_T} \langle N \rangle}{ \langle X \rangle} +
\frac{\langle X_2 \rangle}{ \langle X \rangle} - \frac{ \langle X 
\rangle}{ \langle N \rangle}.
\label{sfromphi}
\end{equation}

In this paper $\Sigma^{XN}$ is calculated using Eqs. 
(\ref{sfromphi}). The statistical errors on $\Phi_{p_{T}}$, 
$\Delta^{XN}$ and $\Sigma^{XN}$ were estimated as follows. 
The whole sample of events was divided into 30 independent sub-samples. 
The values of $\Phi_{p_{T}}$, $\Delta^{XN}$, and $\Sigma^{XN}$ were 
evaluated for each sub-sample and the dispersions ($D_{\Phi}$, 
$D_{\Delta}$, and $D_{\Sigma}$) of the results were then calculated. 
The statistical error of $\Phi_{p_{T}}$ ($\Delta^{XN}$ or 
$\Sigma^{XN}$) is taken to be equal to $D_{\Phi}/\sqrt{30}$ 
($D_{\Delta}/\sqrt{30}$ or $D_{\Sigma}/\sqrt{30}$).


\section{Results of fast generators}

\subsection{Independent particle production}

\begin{figure}[ht]
\includegraphics[width=0.325\textwidth]{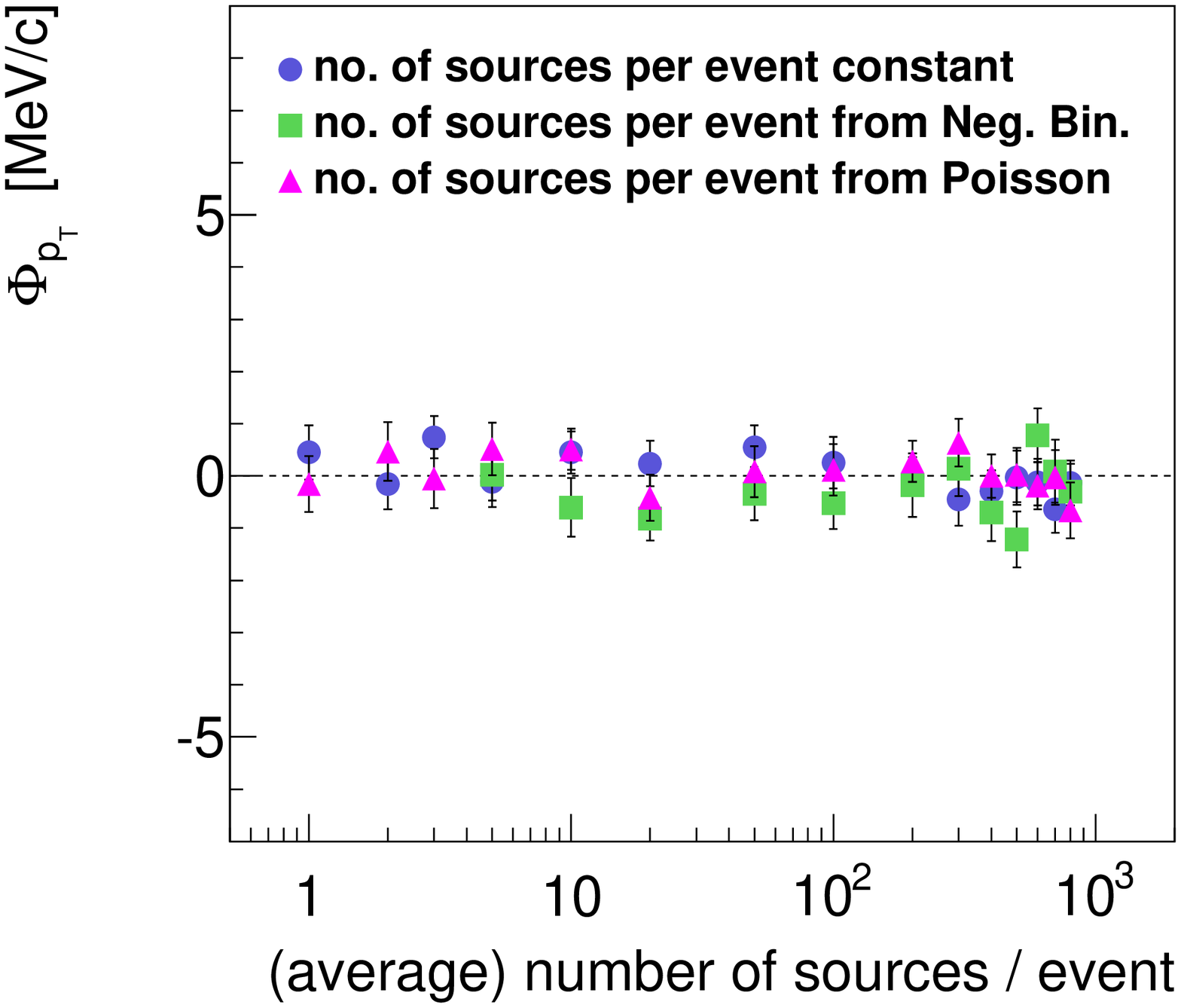}
\includegraphics[width=0.325\textwidth]{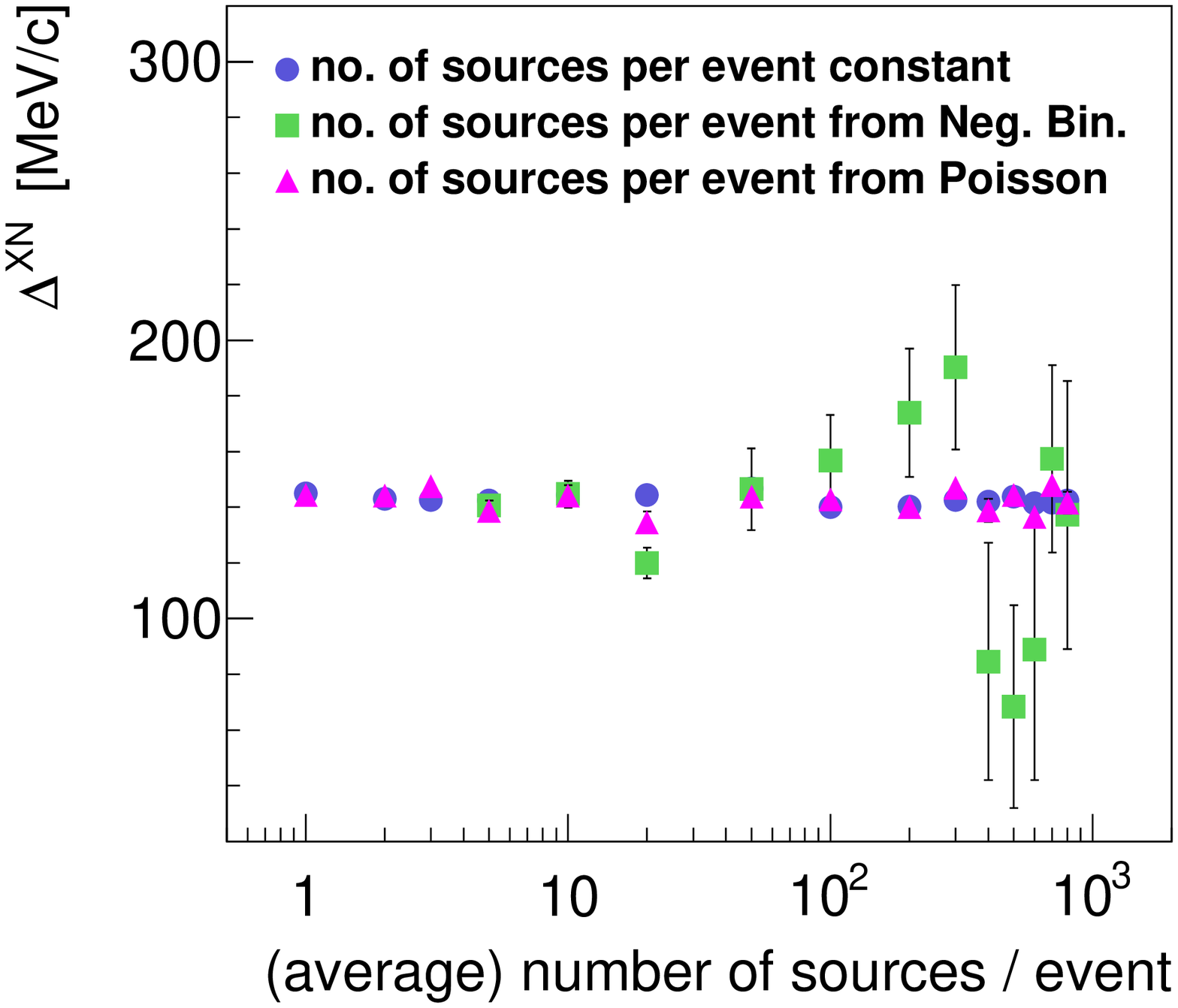}
\includegraphics[width=0.325\textwidth]{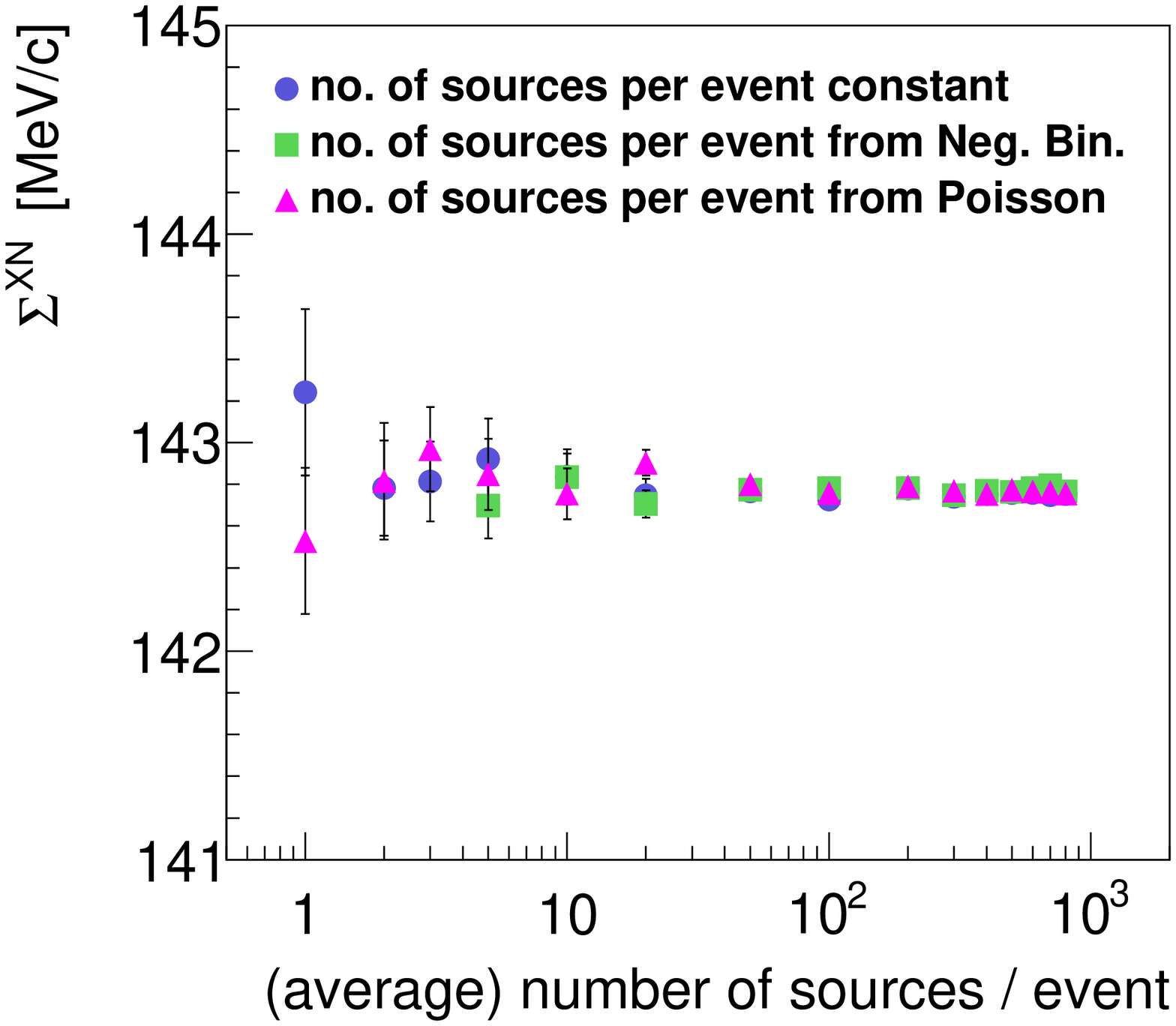}
\vspace{-0.6cm}
\caption[]{$\Phi_{p_{T}}$, $\Delta^{XN}$ and $\Sigma^{XN}$ versus number or mean number of sources composing one event. Example of independent particle production.}
\label{inten}
\end{figure}

The basic properties of $\Phi_{p_{T}}$, $\Delta^{XN}$ and $\Sigma^{XN}$ were tested by use of Monte Carlo models (so-called fast generators). Each interaction (event) was composed by a given number of identical single sources. For each source the number of particles was generated from the Poisson distribution with a mean value of 5. The particle transverse momentum was generated from exponential transverse mass spectrum with inverse slope parameter ("temperature") $T=150$ MeV. The number of sources composing an event was either constant (circles in Fig. \ref{inten}) or selected from Poisson (triangles) or from Negative Binomial distribution (squares). For Negative Binomial distribution its dispersion $\sqrt{Var(N_{S})}$ was large and taken to be equal $\langle N_{S} \rangle / 2$. 

Figure \ref{inten} presents $\Phi_{p_{T}}$, $\Delta^{XN}$ and $\Sigma^{XN}$ 
measures versus the number or the mean number of sources composing one event. As the simulation performed here represent independent particle production $\Phi_{p_{T}}$ measure is consistent with zero. In contrary, $\Delta^{XN}$ and $\Sigma^{XN}$ have non-zero values because their definitions do not assume zero for independent particle production. The circles nicely confirm the intensity property of all three measures, whereas the triangles and the squares show that these quantities are also strongly intensive (do not depend on the system size and the system size fluctuations). One can also mention here that for a constant number of sources per event (circles in Fig. \ref{inten}) the scaled variance of multiplicity distribution $\omega = (\langle N^2 \rangle - \langle N 
\rangle ^2)/ \langle N \rangle$ is close to 1 in the whole range of the horizontal axis. For Poisson number of sources distribution $\omega$ is approximately 6 also in the whole range of the mean number of sources per event. For Negative Binomial distribution of the number of sources $\omega$ increases from about 7 (for on average 5 sources per event), through 126 (for on average 100 sources per event) up to approximately 1000 (for on average 800 sources per event). Figure \ref{inten} shows that within statistical errors (which can be sometimes large) all three measures are strongly intensive even if the multiplicity distribution is extremely and abnormally wide. 

\subsection{"Temperature" fluctuations}

\begin{figure}[ht]
\includegraphics[width=0.325\textwidth]{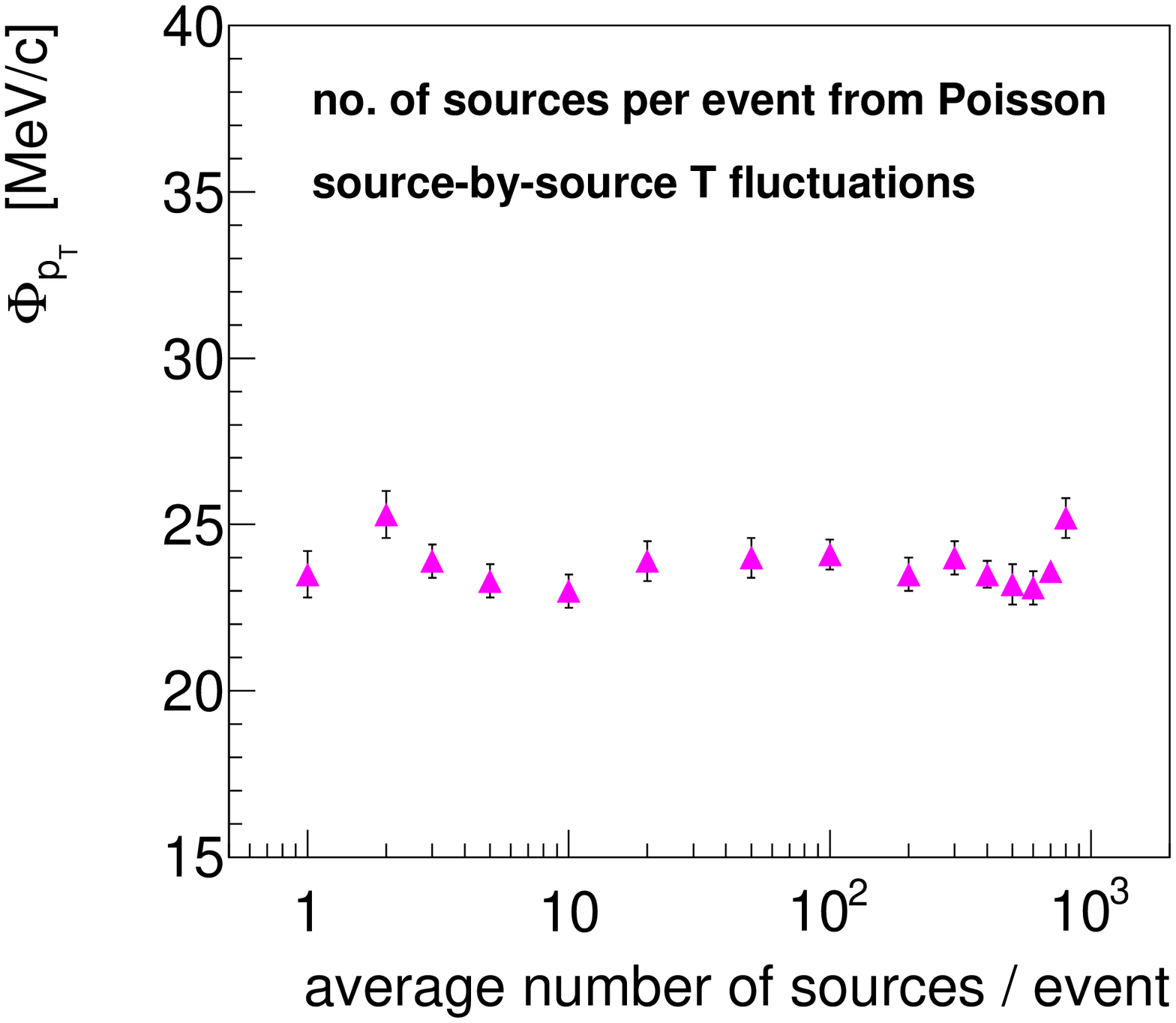}
\includegraphics[width=0.325\textwidth]{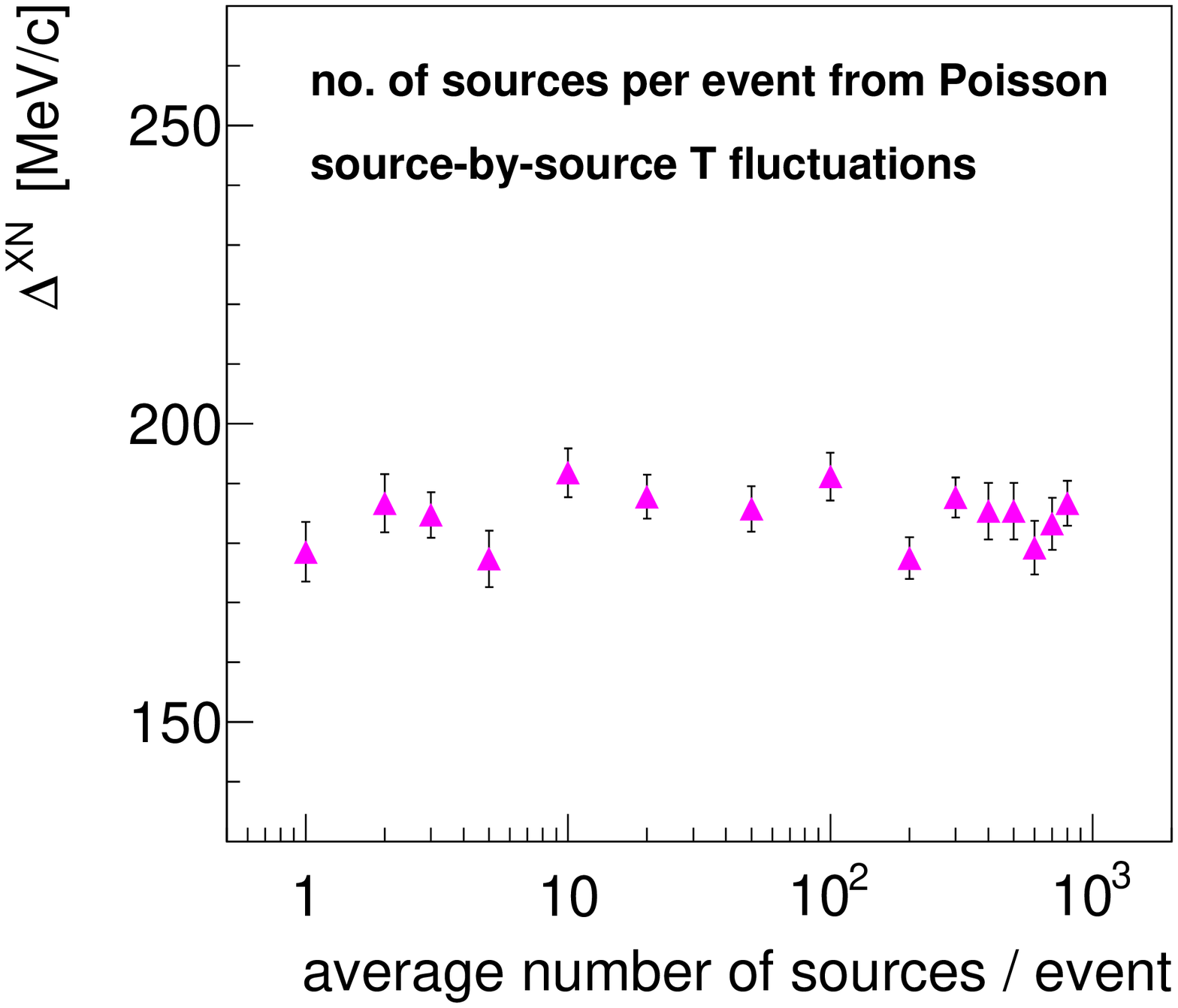}
\includegraphics[width=0.325\textwidth]{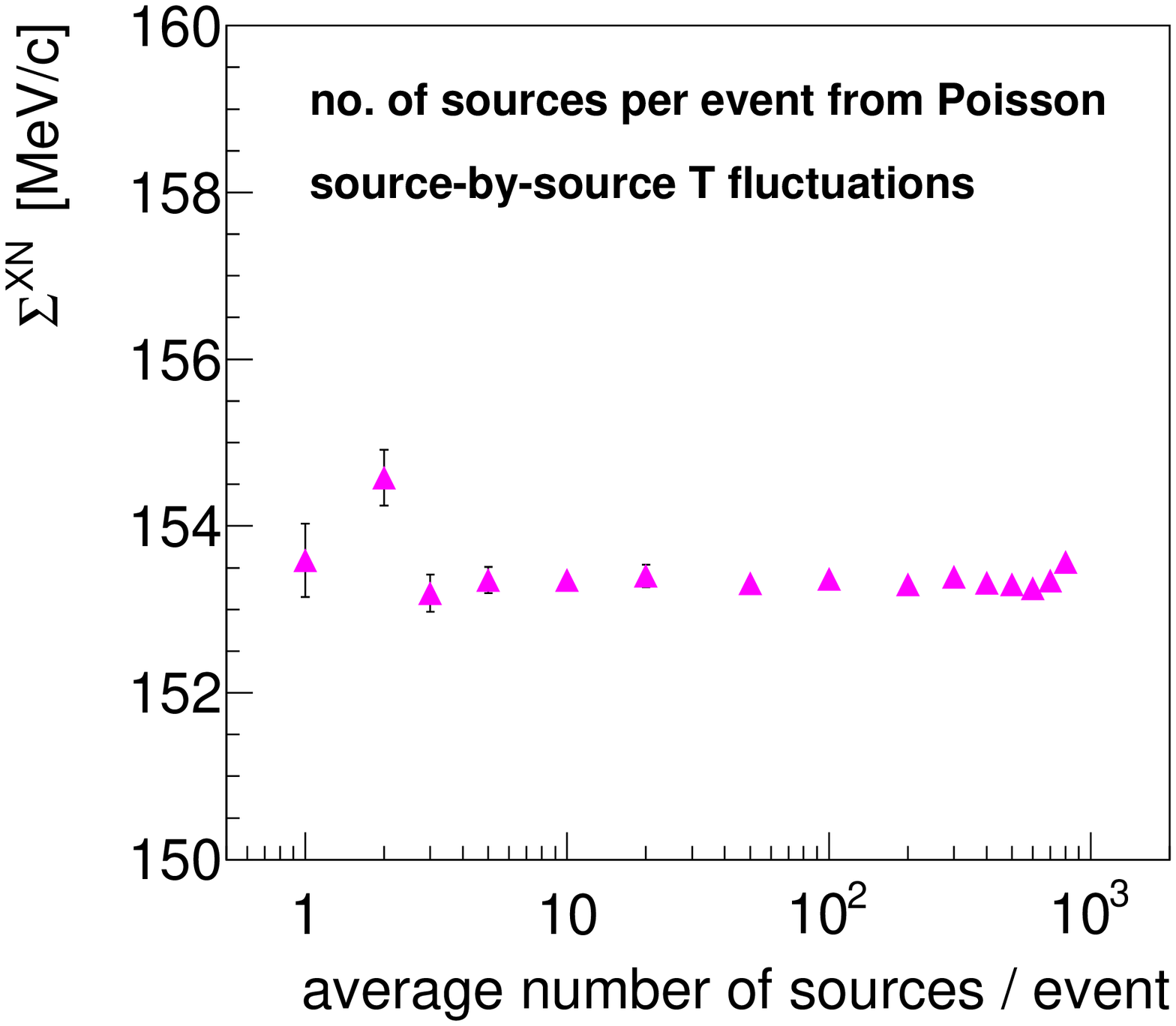}
\vspace{-0.6cm}
\caption[]{$\Phi_{p_{T}}$, $\Delta^{XN}$ and $\Sigma^{XN}$ versus mean number of sources composing one event. Simulation of the effect of source-by-source inverse slope parameter ($T$) fluctuations.}
\label{sbys}
\end{figure}

In the next simulation for each single source the number of particles was again selected from the Poisson distribution with a mean value of 5. The particle transverse momentum was generated from exponential transverse mass spectrum with average inverse slope parameter $\langle T \rangle =150$ MeV. The $T$ parameter was generated separately for each single source (source-by-source $T$ fluctuations) from Gaussian shape with dispersion $\sigma_{T}=25$ MeV. Finally, the number of sources composing an event was generated from the Poisson distribution. The results are presented in Fig. \ref{sbys}. The effect of "temperature" fluctuations results in positive $\Phi_{p_{T}}$ values and in higher (then in Fig. \ref{inten}) $\Delta^{XN}$ and $\Sigma^{XN}$ values. As the sources are identical (superposition model) $\Phi_{p_{T}}$, $\Delta^{XN}$ and $\Sigma^{XN}$ measures do not depend on the mean number of sources composing an event. Therefore Fig. \ref{sbys} also confirms that all three measures are strongly intensive.

\begin{figure}[ht]
\includegraphics[width=0.325\textwidth]{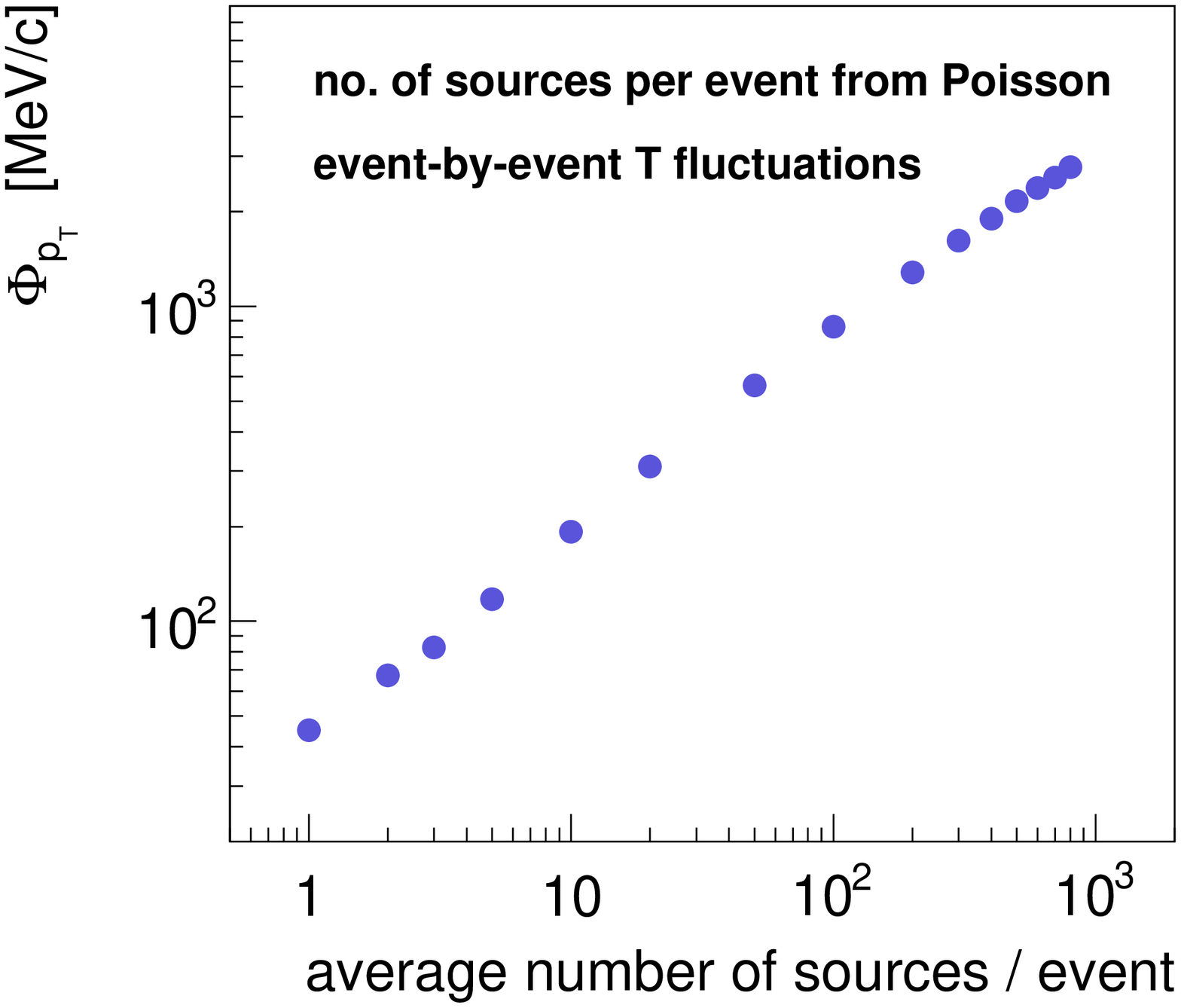}
\includegraphics[width=0.325\textwidth]{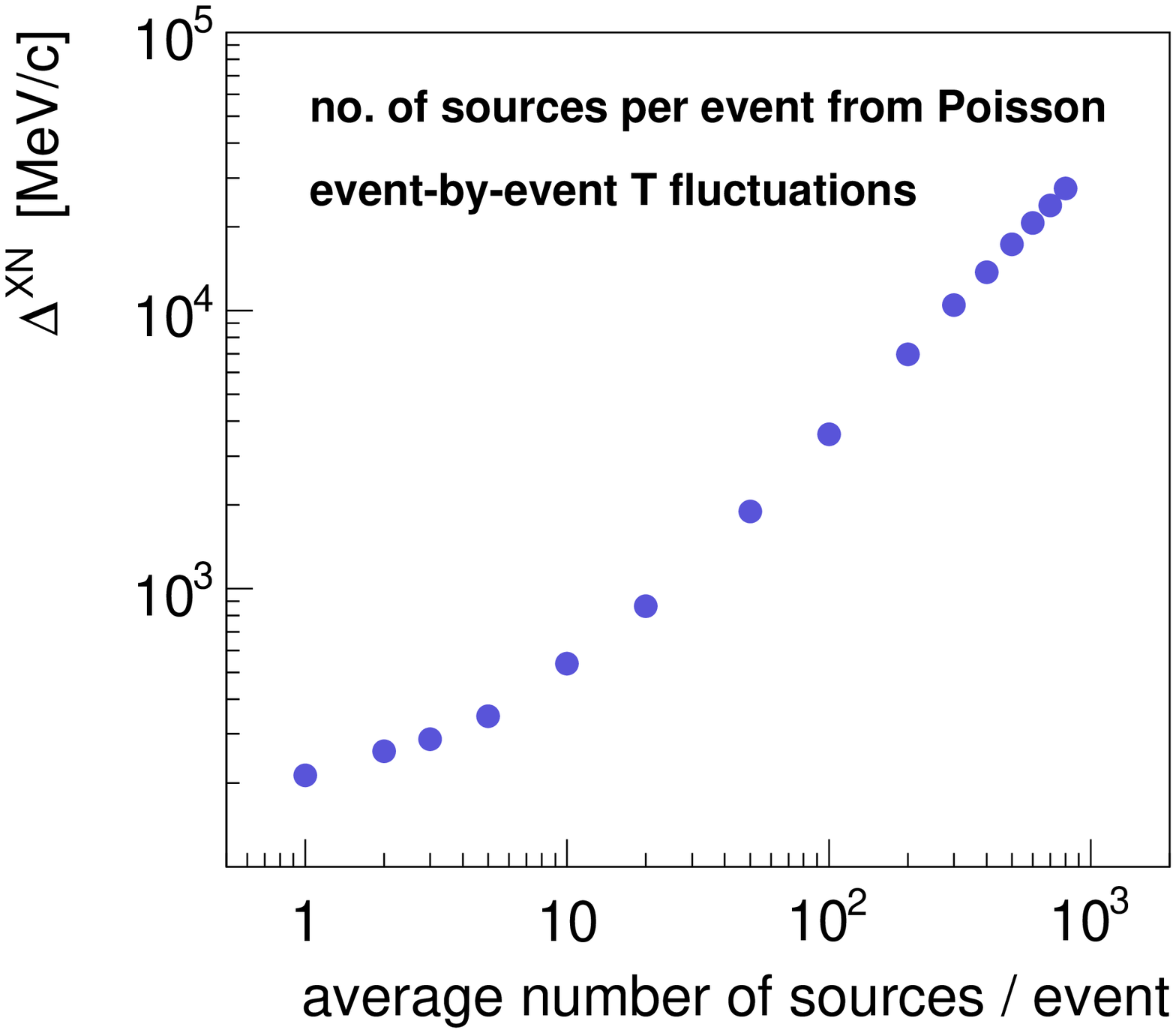}
\includegraphics[width=0.325\textwidth]{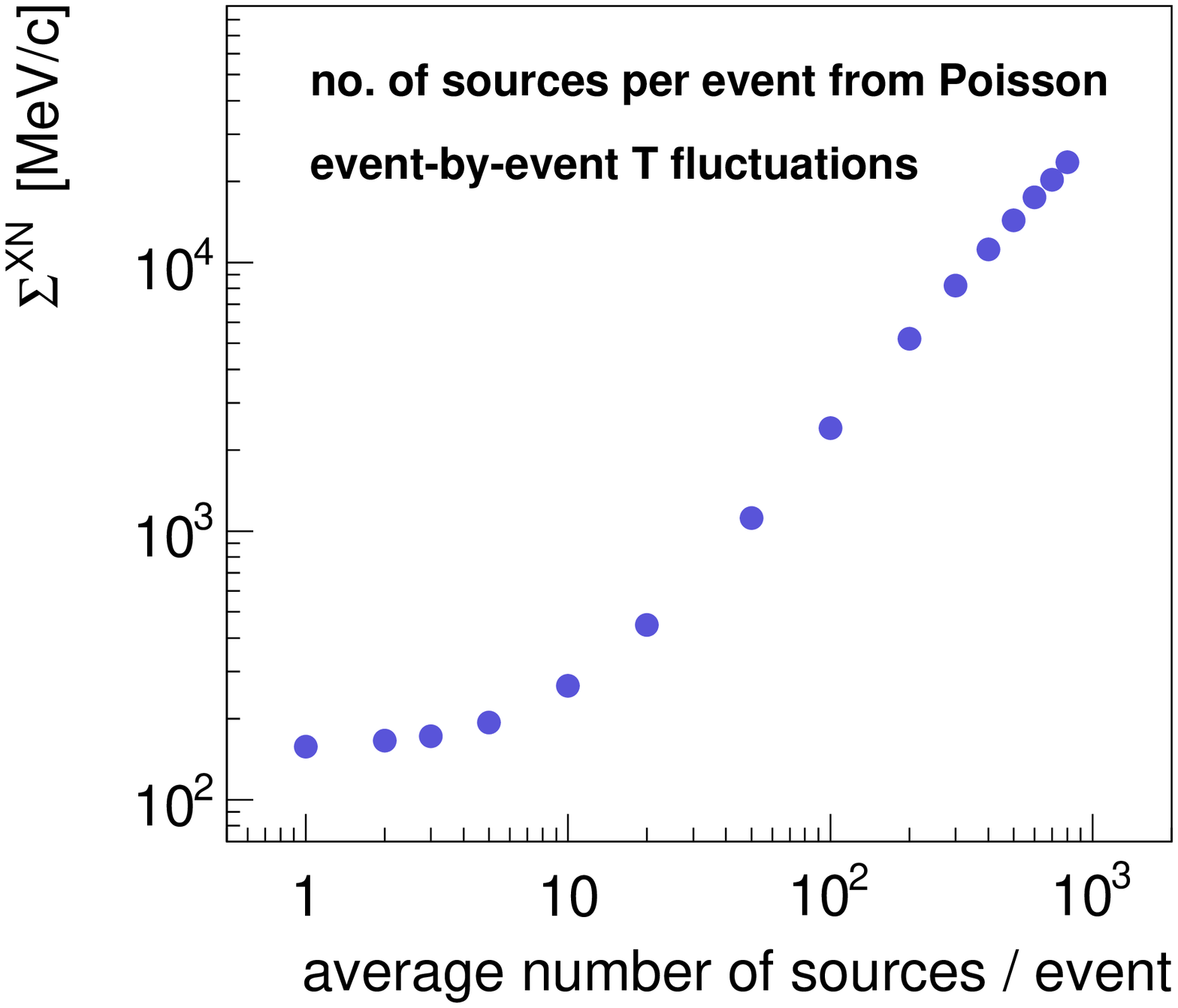}
\vspace{-0.6cm}
\caption[]{$\Phi_{p_{T}}$, $\Delta^{XN}$ and $\Sigma^{XN}$ versus mean number of sources composing one event. Simulation of the effect of event-by-event inverse slope parameter ($T$) fluctuations.}
\label{ebye}
\end{figure}

In the next step source-by-source $T$ fluctuations from the previous simulation were replaced by event-by-event $T$ fluctuations ($T$ parameter was the same for all sources composing a given event but was varied between events). Here the $T$ parameter was generated separately for each event from Gaussian shape with $\sigma_{T}=25$ MeV. Again, the number of sources composing an event was generated from the Poisson distribution. The results are shown in Fig. \ref{ebye}. In this simulation the values of $\Phi_{p_{T}}$, $\Delta^{XN}$ and $\Sigma^{XN}$ increase with increasing the number of sources composing an event. It suggests that in real experimental data the effect of event-by-event "temperature" fluctuations should be better detectable for more central (or for heavier $A$) collisions. One should also mention here that the relationship between temperature and multiplicity (or volume) fluctuations was studied in \cite{multip_T_fluct, volume_T_fluct, Tfluct_2012} .

\begin{figure}[ht]
\includegraphics[width=0.325\textwidth]{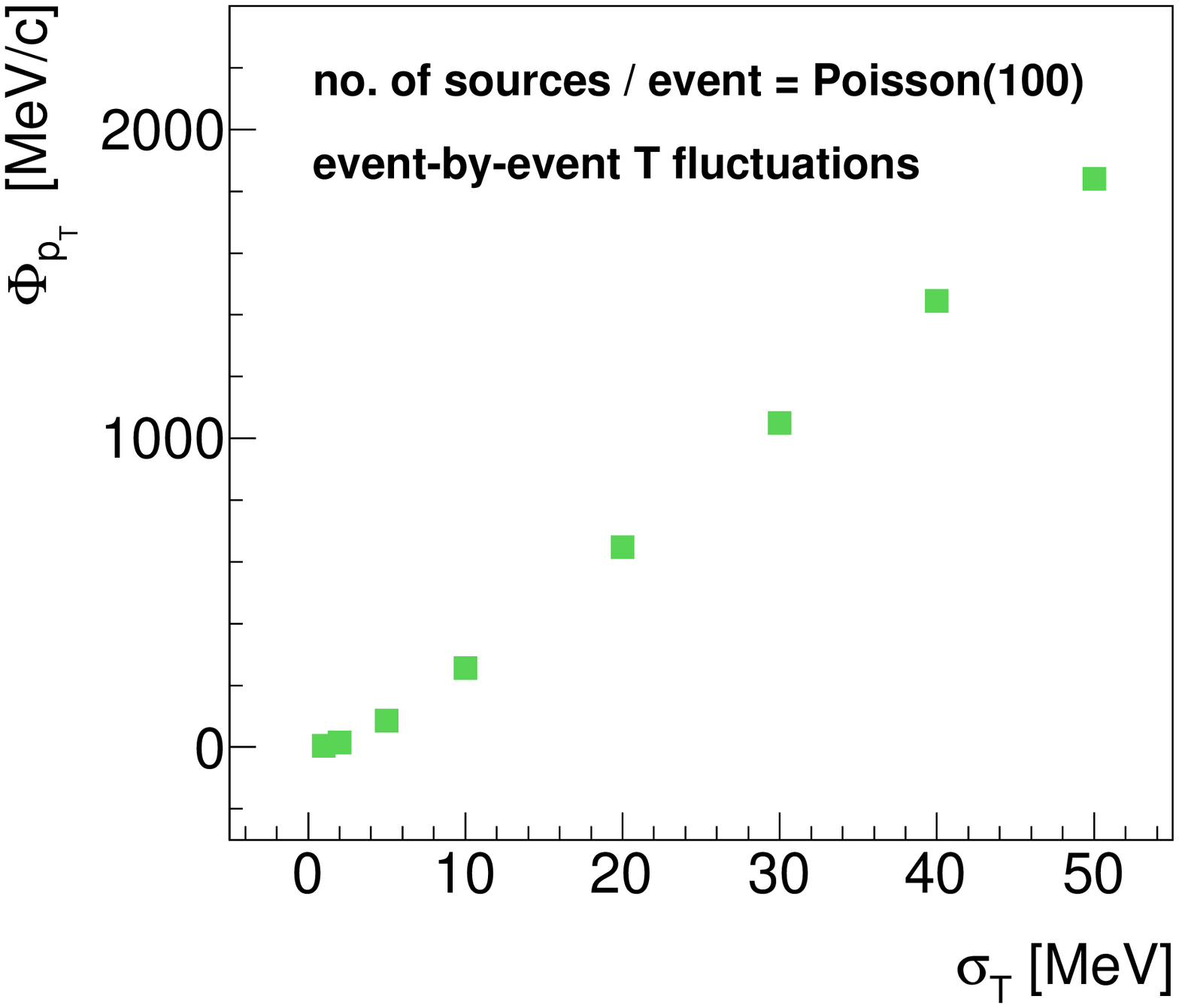}
\includegraphics[width=0.325\textwidth]{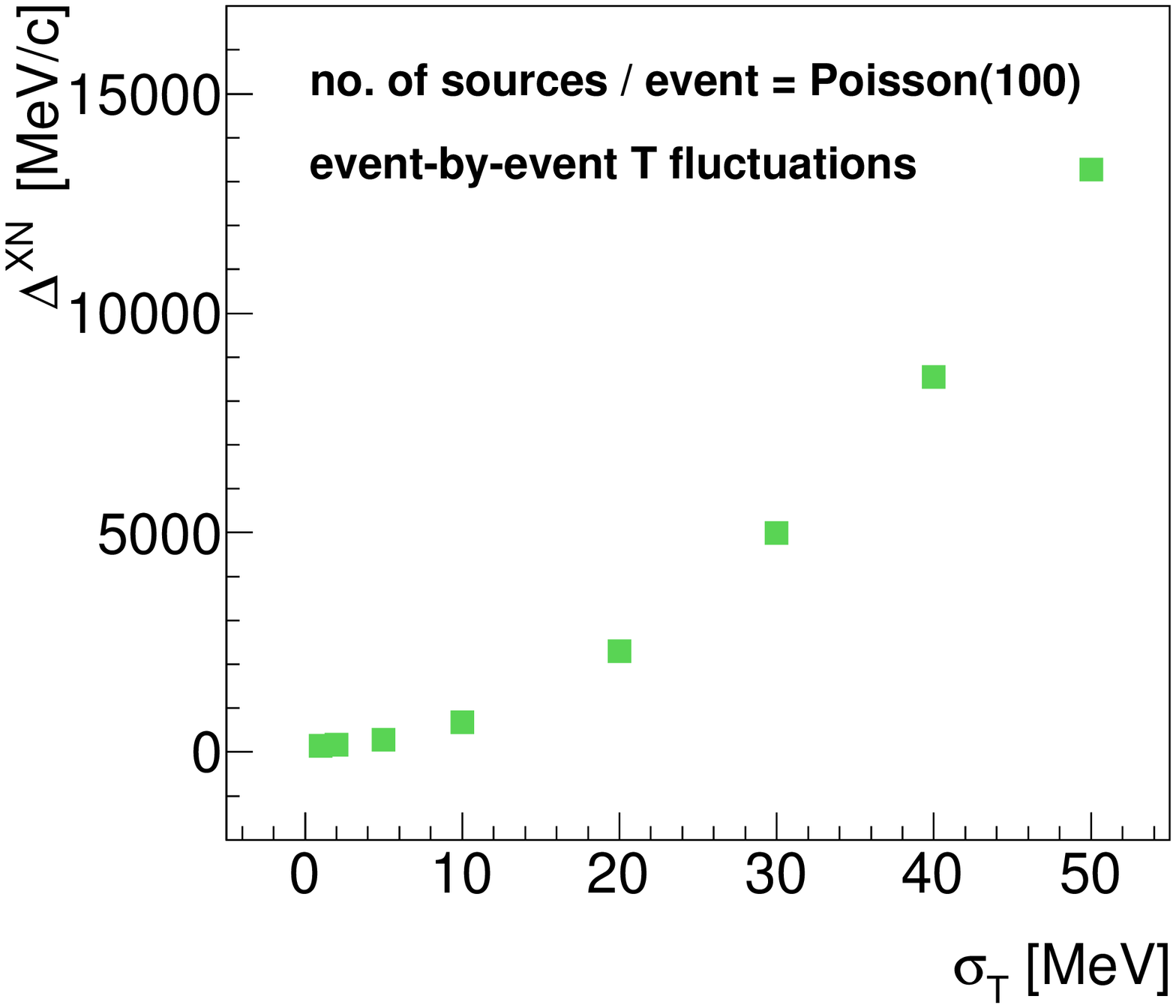}
\includegraphics[width=0.325\textwidth]{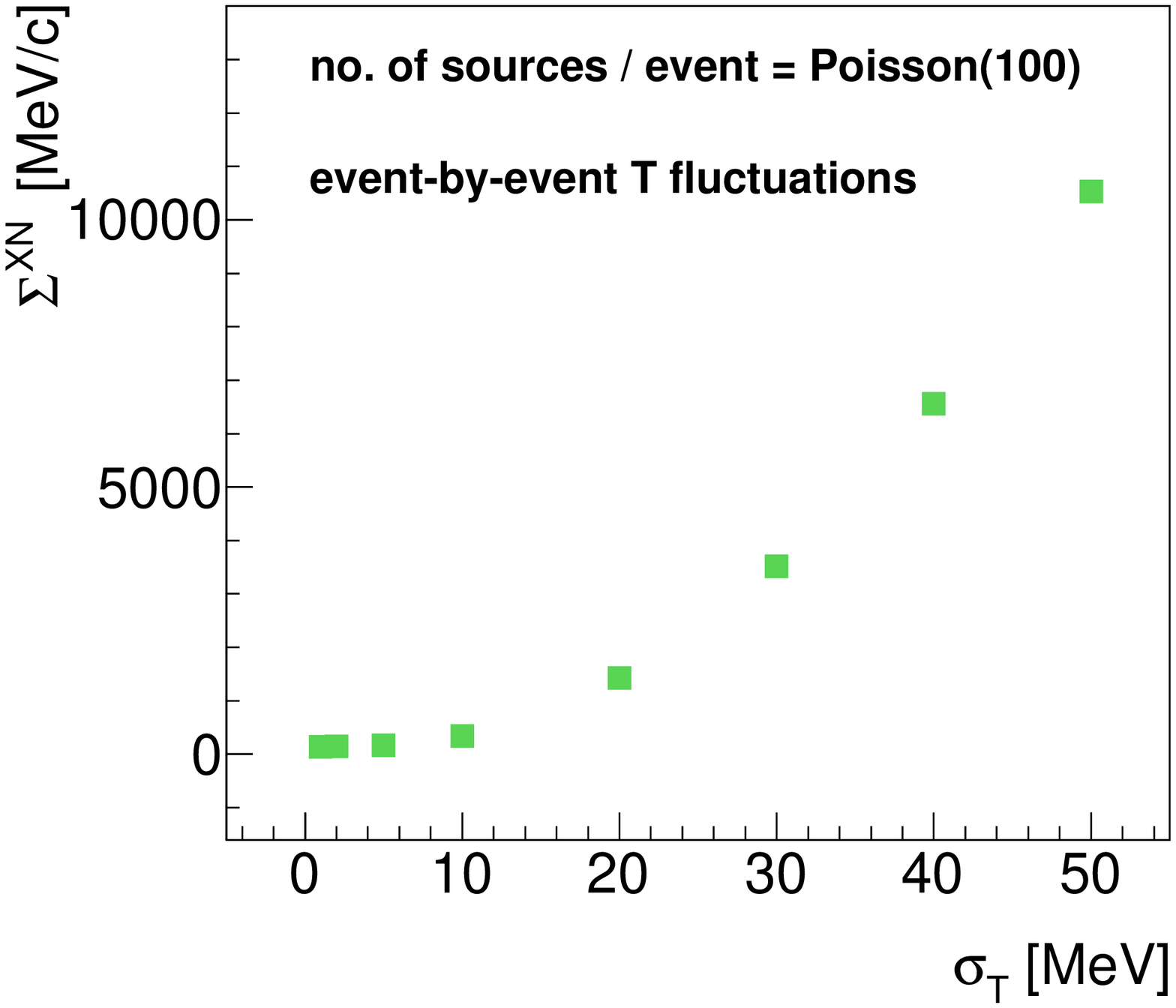}
\vspace{-0.6cm}
\caption[]{$\Phi_{p_{T}}$, $\Delta^{XN}$ and $\Sigma^{XN}$ versus width of $T$ parameter distribution.}
\label{sigmaT}
\end{figure}

In the last simulation different widths of $T$ parameter distribution were considered. As usually, for each single source the number of particles was selected from Poisson with a mean value of 5. The particle transverse momentum was generated from exponential transverse mass shape with average inverse slope parameter $\langle T \rangle =150$ MeV. The $T$ parameter was varied from event to event following Gaussian distribution with dispersion $\sigma_{T}$. In order to avoid negative $T$ values only events within $T=150 \pm 3 \sigma_{T}$ MeV were accepted. Finally, the number of sources composing an event was generated from the Poisson distribution with a mean value of 100. The results are shown in Fig. \ref{sigmaT}. As expected, the values of all three fluctuation measures increase when event-by-event "temperature" fluctuations are stronger (higher $\sigma_{T}$).  

\clearpage

\begin{wrapfigure}{r}{4.cm}
\vspace{-0.6cm}
\includegraphics[width=0.325\textwidth]{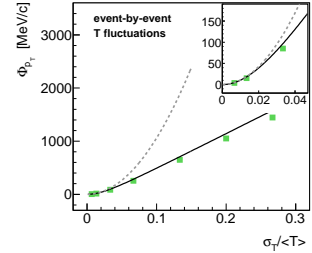}
\vspace{-0.6cm}
\caption[]{$\Phi_{p_{T}}$ versus relative width of $T$ parameter distribution. See text for details.}\label{sigmaTSM}
\end{wrapfigure}

The influence of "temperature" fluctuations on transverse momentum fluctuations was already studied in \cite{korus_staszek}. Following the authors of \cite{korus_staszek} one can easily derive that event-by-event "temperature" fluctuations result in $\Phi_{p_{T}}$ equal $\sqrt{\sigma_T^2 (2+ 4\langle N^2 \rangle / \langle N \rangle ) +2 \langle T \rangle ^2} - \sqrt{6 \sigma_T^2 +2 \langle T \rangle ^2}$. For the scaled variance of multiplicity distribution $\omega \approx 6$ (taken from Monte Carlo simulations in Fig. \ref{sigmaT}) the formula can be rewritten as: $\Phi_{p_{T}} = \sqrt{2 \sigma_T^2 (13 + 2 \langle N \rangle) + 2 \langle T \rangle ^2} - \sqrt{6 \sigma_T^2 + 2 \langle T \rangle ^2}$.
This function, for $\langle T \rangle =150$ MeV and $\langle N \rangle = 500$, is drawn as a solid line in Fig. \ref{sigmaTSM}, together with results from Monte Carlo simulations (taken from Fig. \ref{sigmaT} left). The Monte Carlo results are in a very good agreement with the analytical formula. In publication \cite{korus_staszek} it is also suggested that for sufficiently small $T$ parameter fluctuations, it is when $\langle N \rangle \langle T \rangle ^{2} \gg \langle N^2 \rangle (\langle T^2 \rangle - \langle T \rangle ^2)$, and for Poisson ($\omega = 1$) multiplicity distribution $\Phi_{p_{T}}$ simplifies to the formula $\sqrt{2} \langle N \rangle \frac{\langle T^2 \rangle - \langle T \rangle ^2} {\langle T \rangle}$. This simplified function, for $\langle T \rangle =150$ MeV and $\langle N \rangle = 500$, is drawn as a dashed line in Fig. \ref{sigmaTSM}. As seen, the simplified function is indeed valid only for small $T$ parameter fluctuations and starts to deviate from Monte Carlo simulations for $\sigma_{T}/\langle T \rangle $ higher than approximately 3 percent.


\section{Results of the UrQMD model}

In order to calculate $\Delta^{XN}$, $\Sigma^{XN}$ and $\Phi_{p_T}$ the UrQMD3.3 model \cite{UrQMD} was used. The model was already used to study event-by-event fluctuations of average transverse momentum \cite{urqmd_1, kg_imp}, charged particle ratio \cite{urqmd_2}, and particle number \cite{urqmd_3}.The UrQMD generator is a microscopic transport model producing hadrons via formation, decay, and rescattering of resonances and strings. The UrQMD approach simulates multiple interactions of both target/beam nucleons and newly produced particles, excitation, and fragmentation of color strings and the formation and decay of hadronic resonances. 
The samples of minimum bias events were generated for the systems which will be analyzed in the NA61/SHINE experiment, namely $Be+Be$, $Ar+Ca$, $Xe+La$ at beam 
energies 13$A$, 20$A$, 30$A$, 40$A$, 80$A$, and 158$A$ GeV (Fig. 
\ref{phas_full}). These beam energies correspond to center-of-mass energies for nucleon+nucleon pair $\sqrt{s_{NN}}$ equal 5.12, 6.27, 7.62, 8.76, 12.3, and 17.3 GeV, respectively. Only inelastic $A+A$ collisions were considered, it is the generated events with the number of collisions equal zero and events with the number of inelastic collisions equal zero were removed from the sample. The remaining events which had the final state multiplicities equal to the sum of the nucleons in both colliding nuclei were also rejected. 

From such prepared minimum bias events only 20\% of the most central 
collisions were selected (in agreement to NA61 data taking plans). This 
centrality selection corresponds to impact parameter cuts in UrQMD: $b<1.96$ fm 
for $Be+Be$, $b<3.81$ fm for $Ar+Ca$, and $b<5.71$ fm for $Xe+La$.

\subsection{Influence of high-$p_T$ particles (at SPS energies)}

\begin{figure}[ht]
\includegraphics[width=0.325\textwidth]{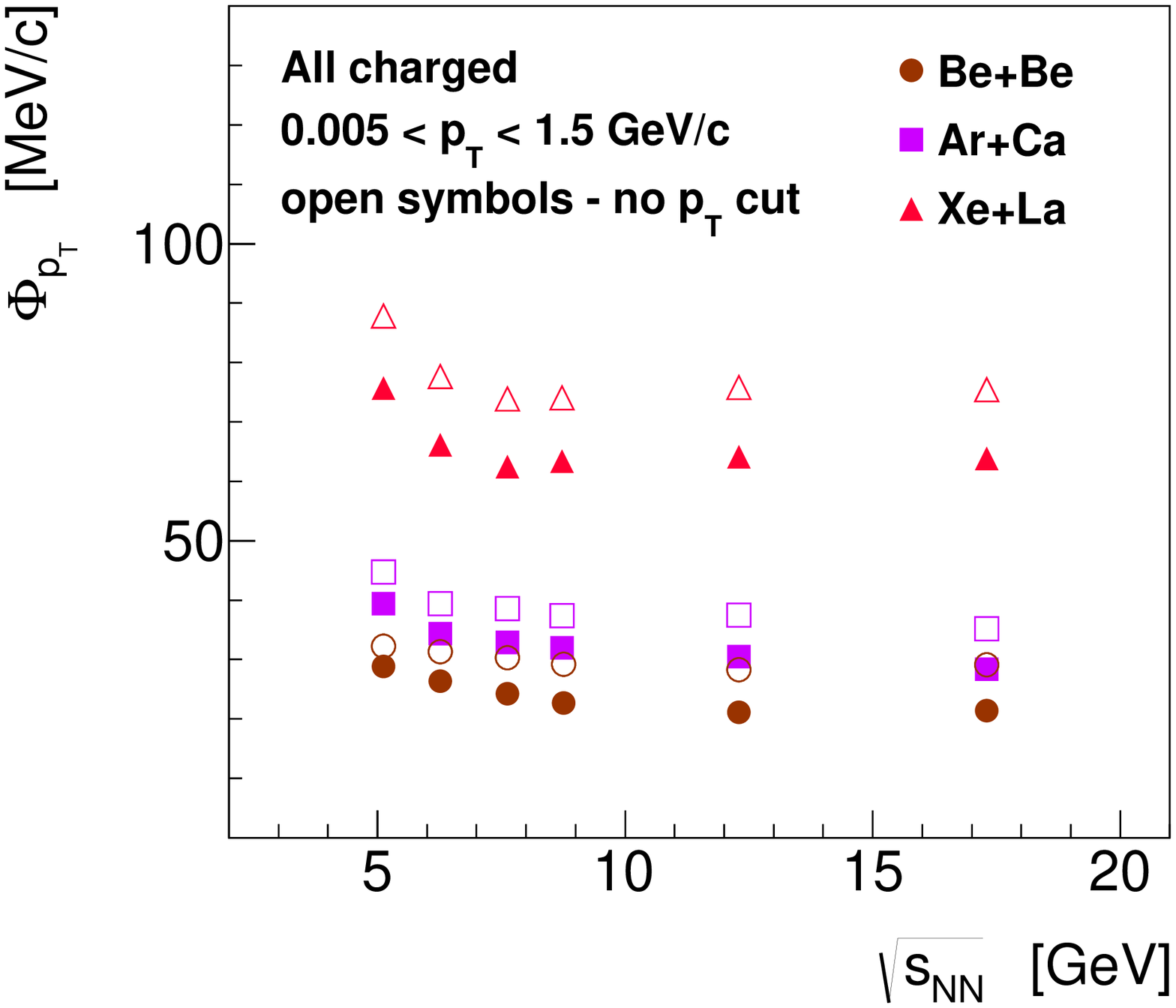}
\includegraphics[width=0.325\textwidth]{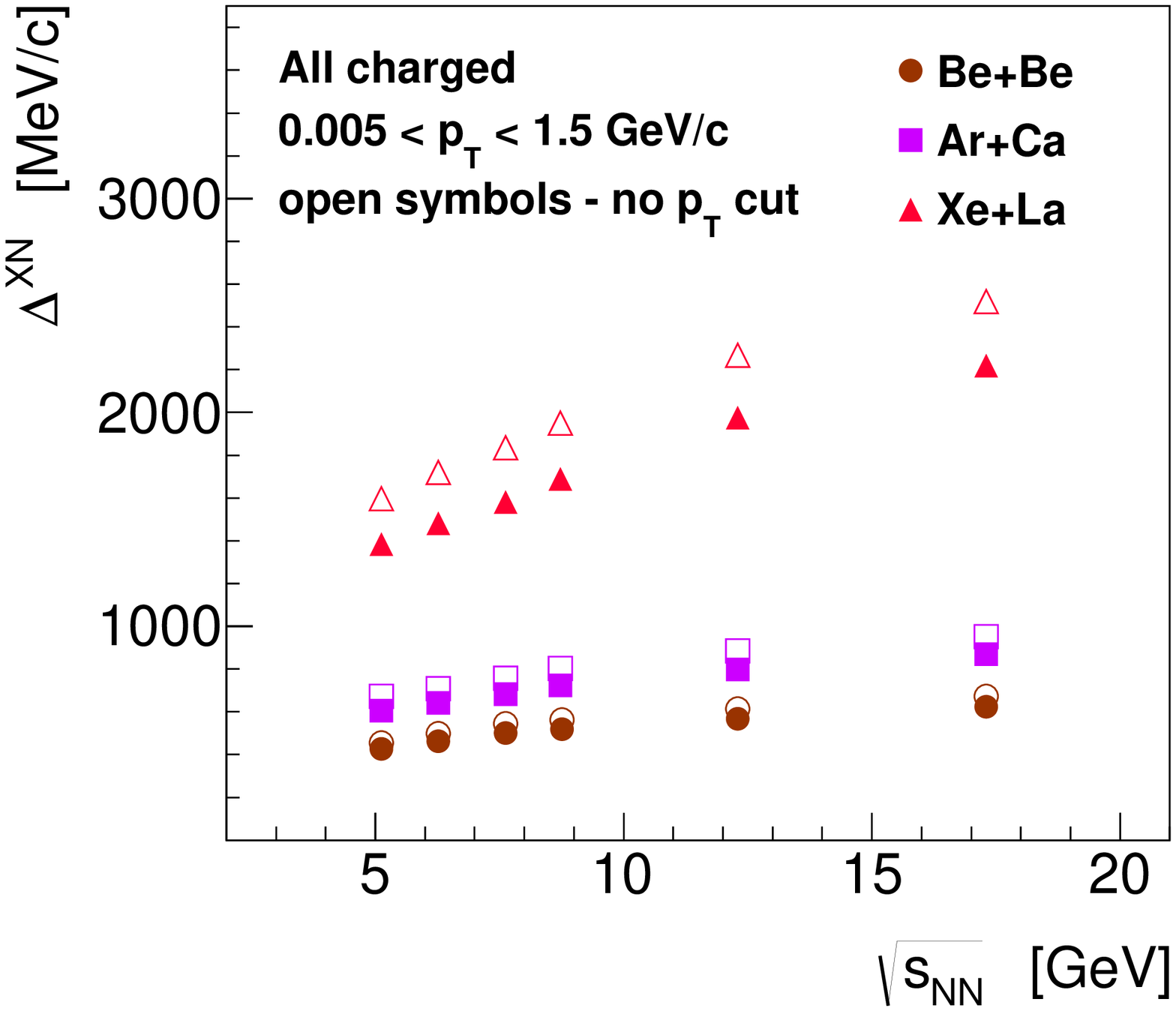}
\includegraphics[width=0.325\textwidth]{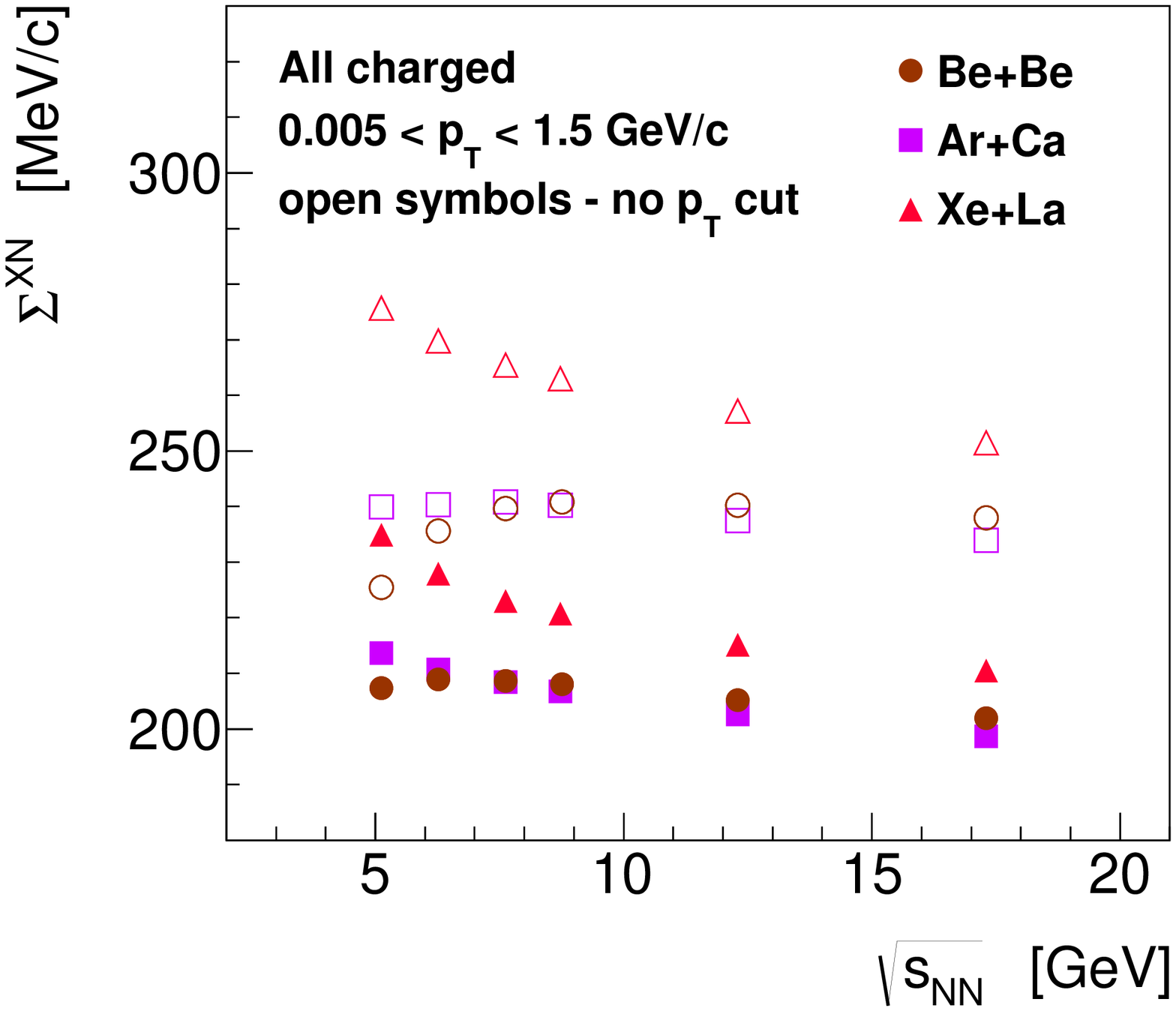}
\vspace{-0.6cm}
\caption[]{$\Phi_{p_{T}}$, $\Delta^{XN}$ and $\Sigma^{XN}$ measured for all charged particles produced in 20\% most central $A+A$ collisions in UrQMD.}
\label{urq_pt}
\end{figure}

\begin{figure}[ht]
\includegraphics[width=0.325\textwidth]{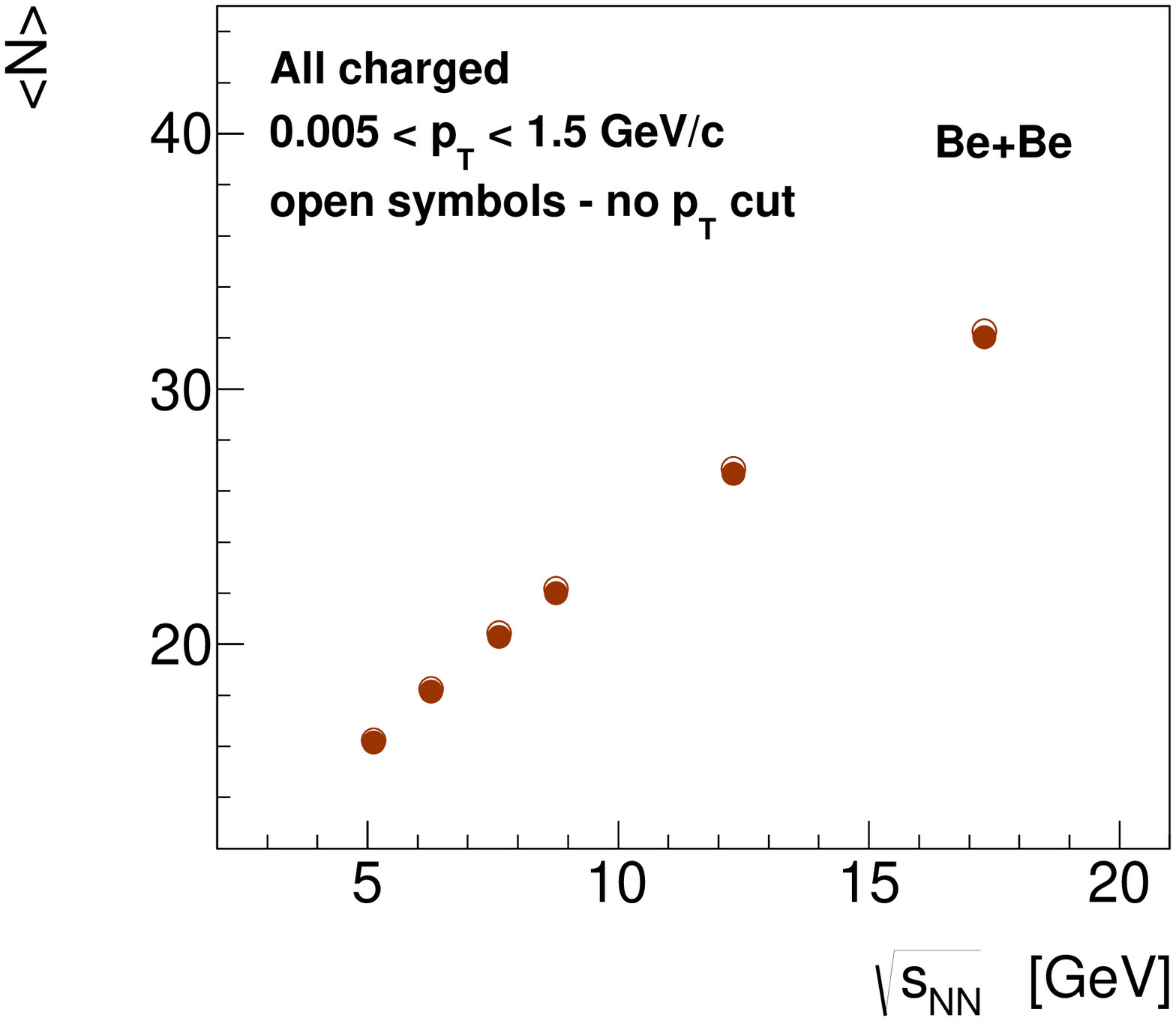}
\includegraphics[width=0.325\textwidth]{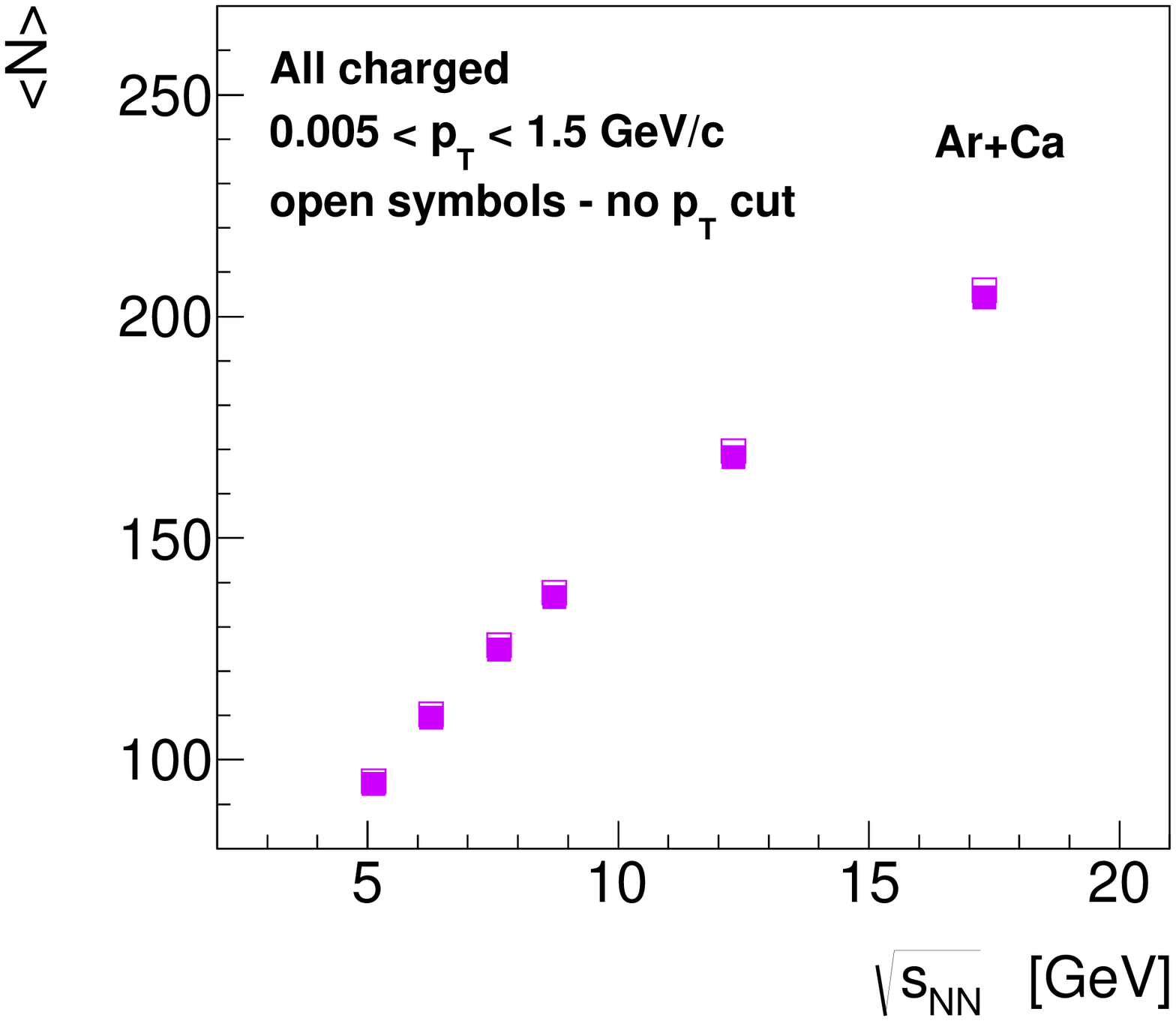}
\includegraphics[width=0.325\textwidth]{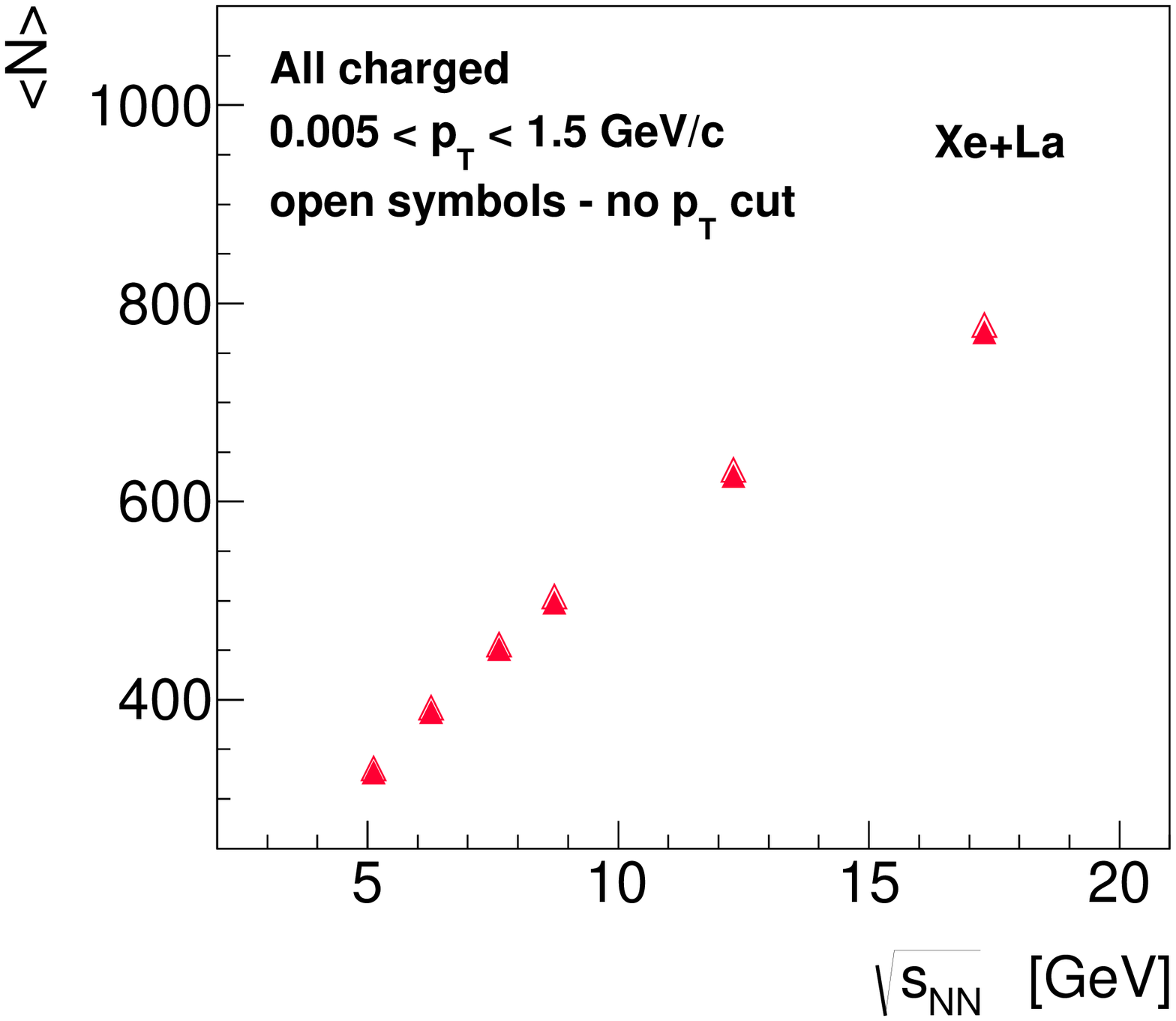}
\vspace{-0.6cm}
\caption[]{Mean multiplicities of charged particles produced in 20\% most central $A+A$ collisions in UrQMD.}
\label{urq_mult_pt}
\end{figure}

Figure \ref{urq_pt} presents $\Phi_{p_{T}}$, $\Delta^{XN}$ and $\Sigma^{XN}$ 
measures obtained for all charged particles produced in UrQMD. Open 
points are those without any kinematic restrictions. Full points 
correspond to results with $0.005 < p_T < 1.5$ GeV/c cut. Such cut was 
applied by the NA49 experiment \cite{fluct_size, fluct_energy, omega_size} in order 
to eliminate the possible effect originating from hard interactions (the 
lower cut was due to momentum resolution of the detector). Figure \ref{urq_mult_pt} shows mean multiplicities of charged particles with and 
without $0.005 < p_T < 1.5$ GeV/c cut. Although at SPS energies there 
is only a small fraction of particles coming from hard interactions (compare open and full points in Fig. \ref{urq_mult_pt}) such high-$p_T$ hadrons can increase 
the values of all considered fluctuations measures (Fig. \ref{urq_pt}). For $Xe+La$ collisions at the SPS this increase is on the level of 15-19\% for all three fluctuation measures.

In Fig. \ref{urq_pt} one also observes a slight decrease of $\Phi_{p_{T}}$ and $\Sigma^{XN}$ with increasing energy, while, in contrary, $\Delta^{XN}$ increases. 
It is also worth to notice that the values of all three measures are lower for lighter systems (increase when going from $Be+Be$ to $Xe+La$). In a simple superposition model no system size dependence is expected. However, the increase of $\Phi_{p_{T}}$, $\Delta^{XN}$ and $\Sigma^{XN}$ for heavier systems may be due to event-by-event impact parameter fluctuations. The sample of 0-20\% central collisions may be not narrow enough (especially for the heaviest $Xe+La$) and the characteristics of more and less central collisions may be significantly different leading to increased event-by-event fluctuations. In the next sections it will be shown that indeed the values of fluctuation measures are smaller when the centrality is restricted to 7.2\% most central data.

\subsection{Influence of hadrons different than $\pi$, $K$, $anti(p)$ }

In the event-by-event analysis of experimental data \cite{fluct_size, fluct_energy, omega_size} typically only charged particles originating from the main vertex are used. It practically means that only main vertex pions, protons, kaons and their antiparticles are used in the analysis, because particles coming from the decays of for example $\Lambda$, $\phi$, $\Sigma$, $\Xi$ and $\Omega$ are believed to be rejected by a set of track selection criteria. The time scale of the simulations performed within the UrQMD model did not allow for weak decays, therefore the UrQMD analysis of charged pions, protons, kaons and their antiparticles can be directly compared to the analysis of experimental data. 

\begin{figure}[ht]
\includegraphics[width=0.325\textwidth]{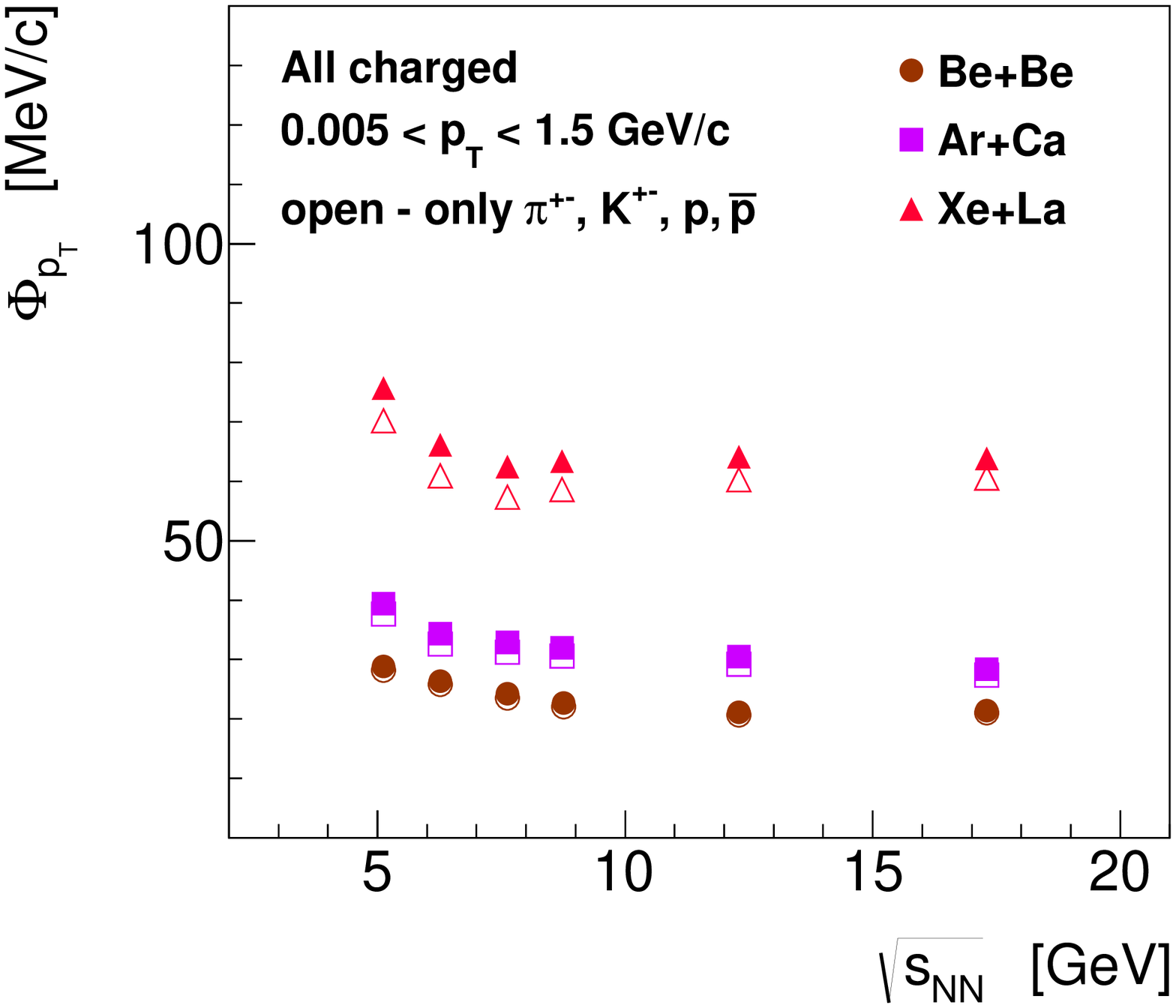}
\includegraphics[width=0.325\textwidth]{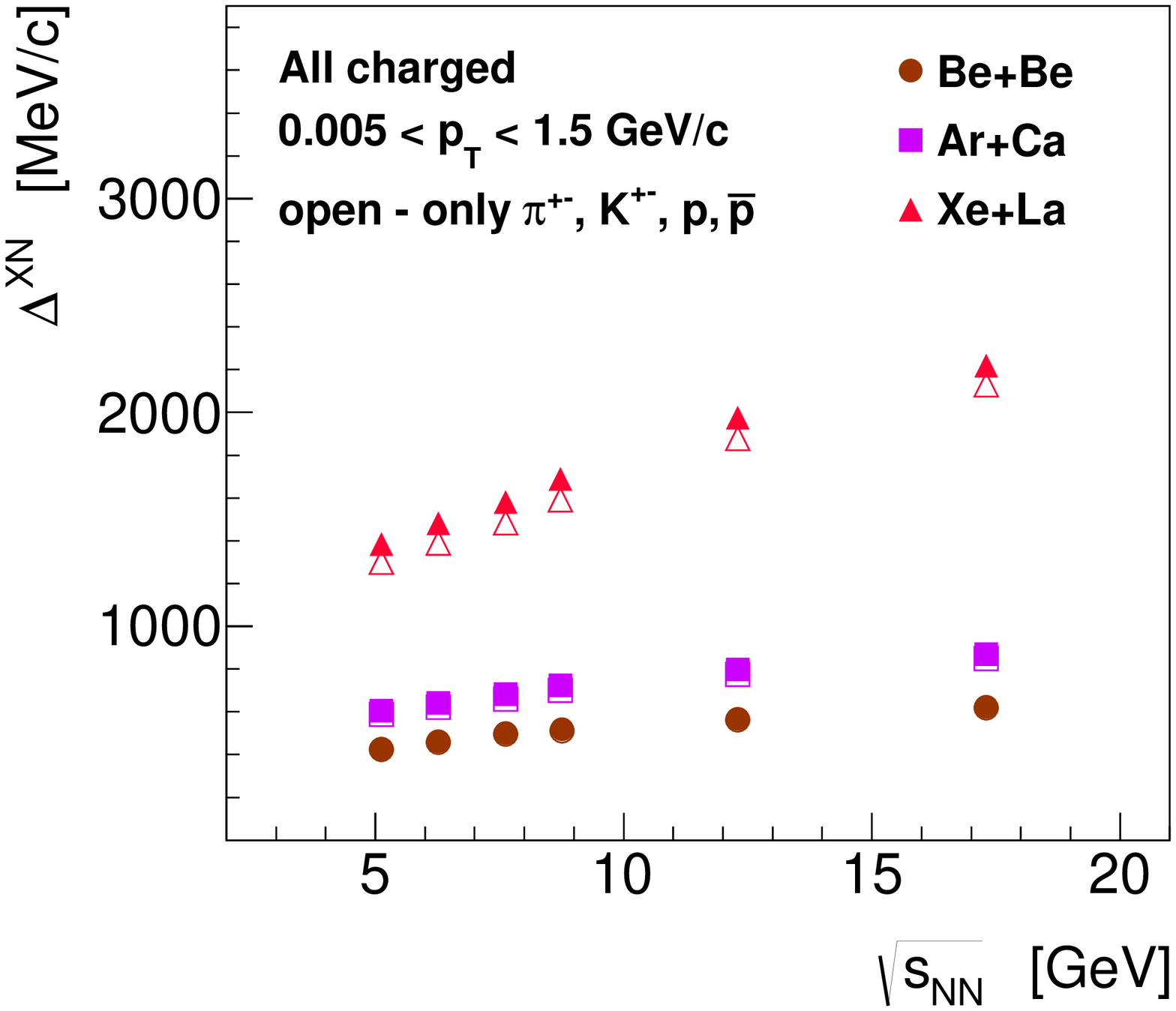}
\includegraphics[width=0.325\textwidth]{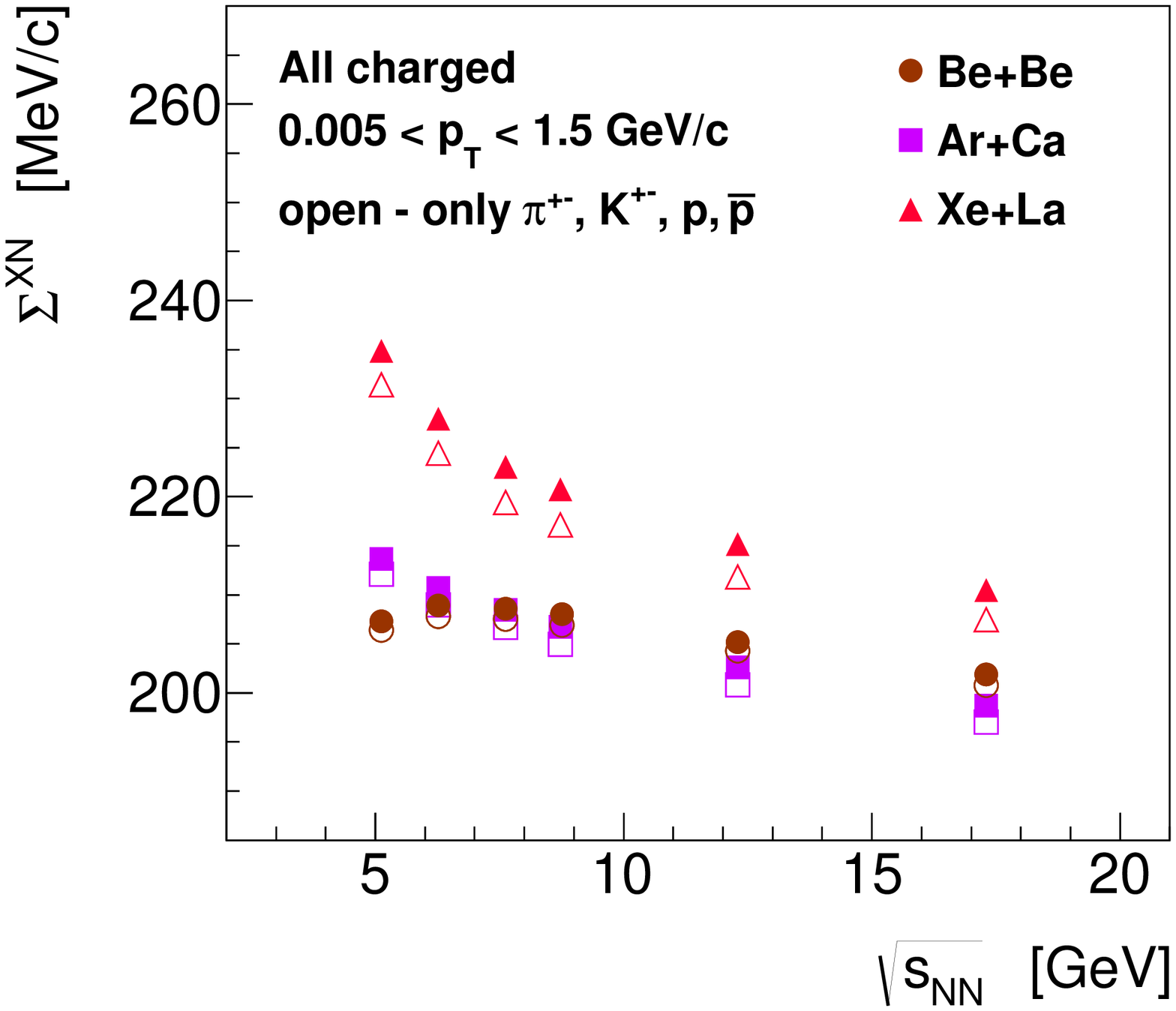}
\vspace{-0.6cm}
\caption[]{$\Phi_{p_{T}}$, $\Delta^{XN}$ and $\Sigma^{XN}$ measured for all charged particles and for charged pions, kaons and (anti)protons produced in 20\% most central $A+A$ collisions in UrQMD.}
\label{urq_kpip}
\end{figure}

Figure \ref{urq_kpip} compares $\Phi_{p_{T}}$, $\Delta^{XN}$ and $\Sigma^{XN}$ calculated for all charged particles produced in UrQMD (full symbols) and for only charged pions, protons, kaons and their antiparticles (open symbols). The small difference between these two cases is a reflection of the fact that the majority of final state particles produced in relativistic heavy ion collisions are pions, (anti)protons and kaons.

\subsection{$\Phi_{p_T}$ - dependence on energy and charge combination}

\begin{figure}[ht]
\includegraphics[width=0.325\textwidth]{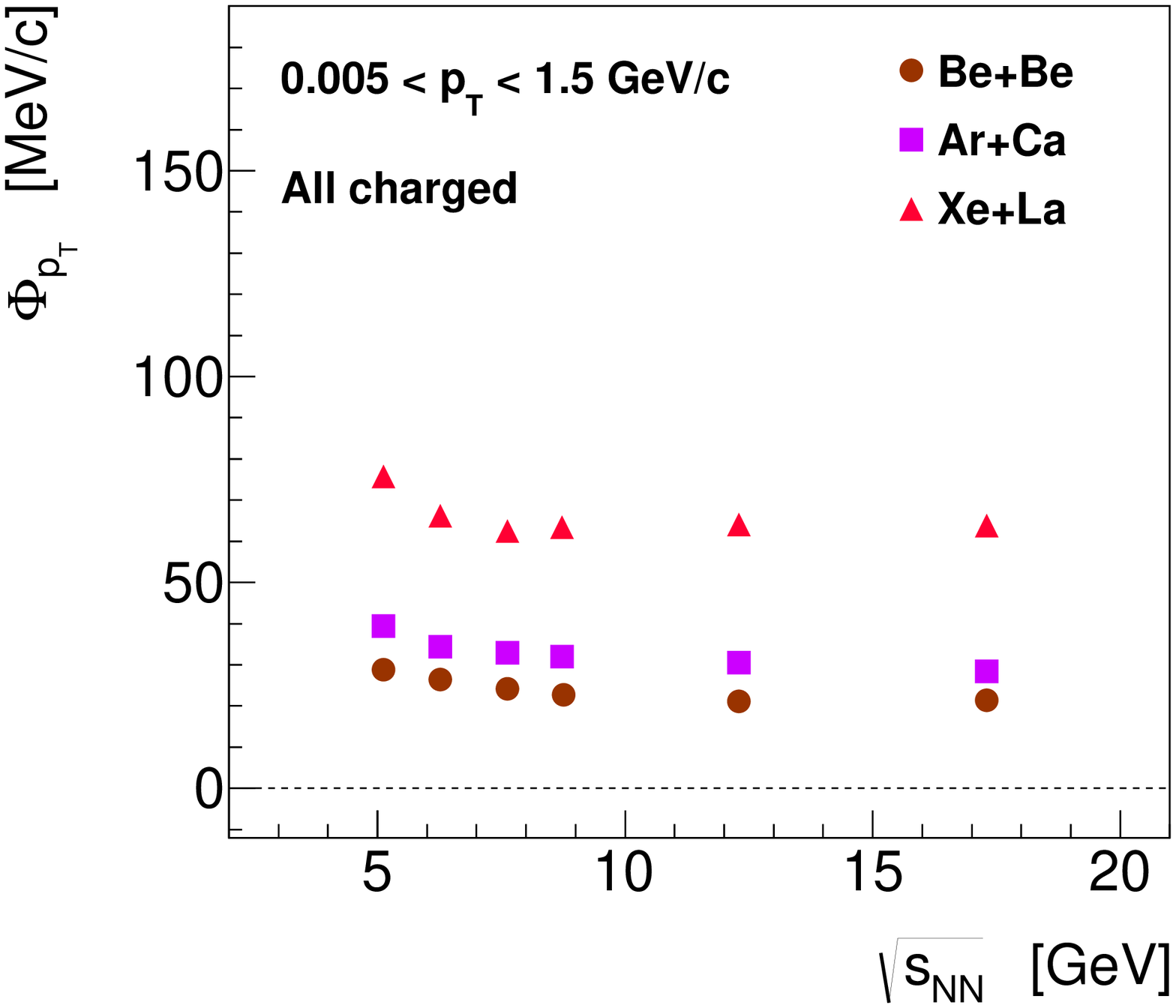}
\includegraphics[width=0.325\textwidth]{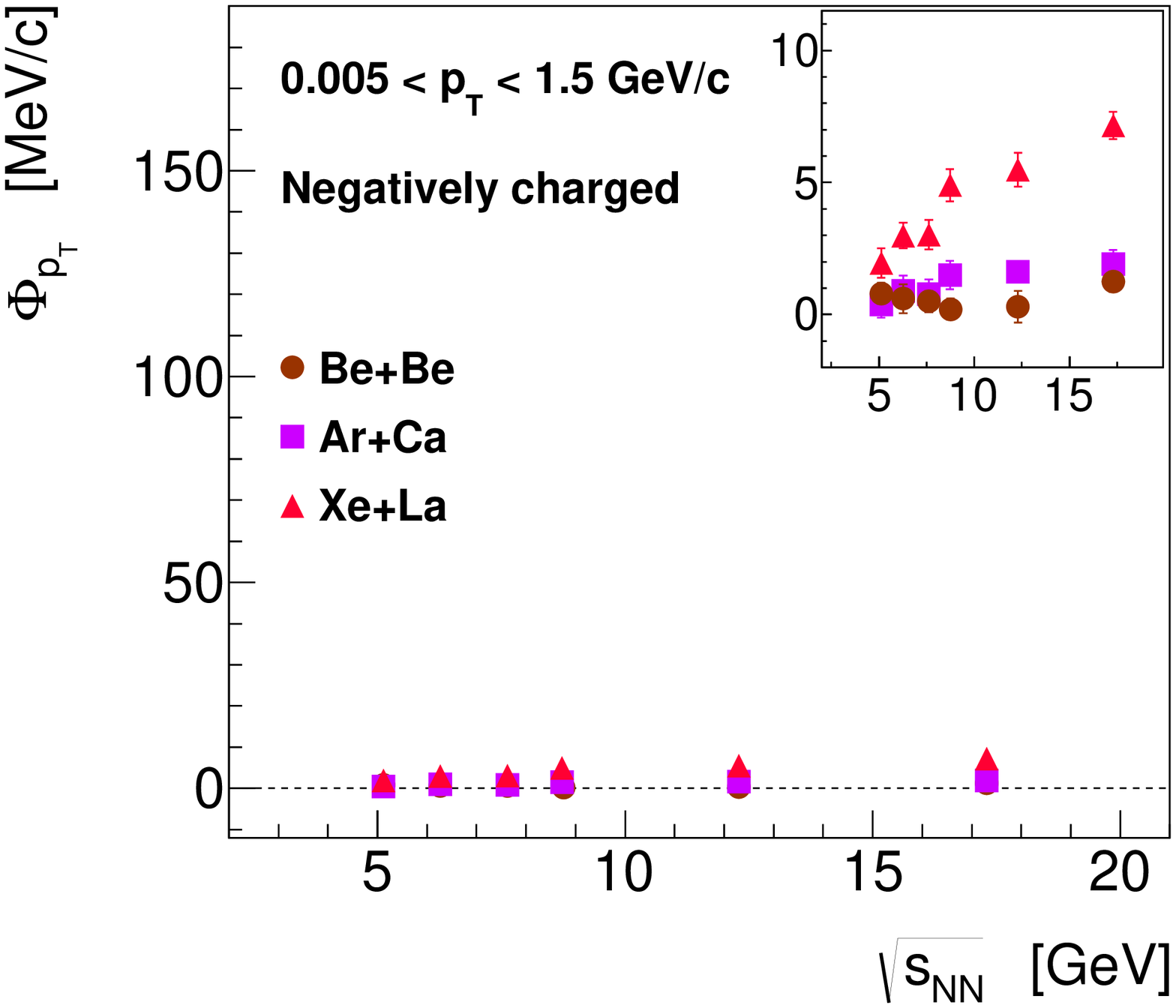}
\includegraphics[width=0.325\textwidth]{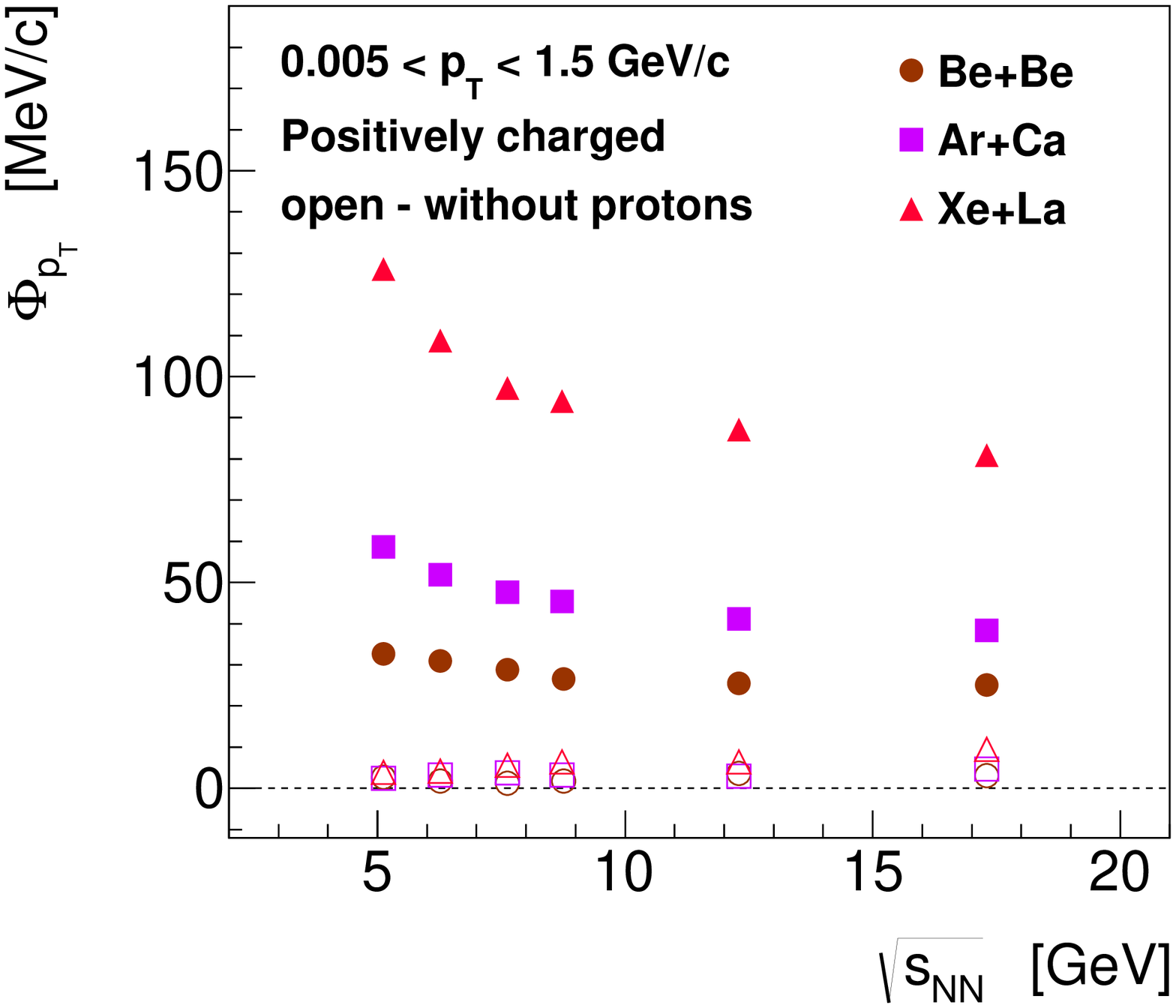}
\vspace{-0.6cm}
\caption[]{Energy dependence of $\Phi_{p_{T}}$ for different charge combinations of particles produced in 20\% most central $A+A$ collisions in UrQMD.}
\label{urq_phi_ptcut}
\end{figure}

Figure \ref{urq_phi_ptcut} shows the energy dependence of $\Phi_{p_{T}}$ for different charge combinations: all charged particles, negatively charged and  positively charged. The open symbols in the right panel represent positively charged particles where protons were removed from the sample. As already discussed in Fig. \ref{urq_pt} a decrease of $\Phi_{p_{T}}$ with energy can be observed for all charged particles. This decrease is even more pronounced when looking at positively charged particles (Fig. \ref{urq_phi_ptcut} right). In a contrary, negatively charged particles show values close to zero and only a very small increase with energy (see inset of Fig. \ref{urq_phi_ptcut} middle) can be observed for heavier systems. The most important difference between positively charged and negatively charged particles is the presence of protons in the sample of positively charged. Therefore the $\Phi_{p_{T}}$ values were calculated also for positively charged particles without protons (open points in the right panel). Their values are consistent with those for negatively charged particles thus confirming that the effect indeed originates from protons. It is also worth to remind that the highest effect from protons is observed for the lowest energies where the fraction of protons is the highest. For $Xe+La$ UrQMD data the fraction of protons is about 35\% of all charged for 13$A$ GeV and 15\% of all charged for 158$A$ GeV. For $Ar+Ca$ data these numbers are 36\% and 18\% for 13$A$ GeV and 158$A$ GeV, respectively, and for $Be+Be$ collisions 48\% and 24\%.

\begin{figure}[ht]
\includegraphics[width=0.325\textwidth]{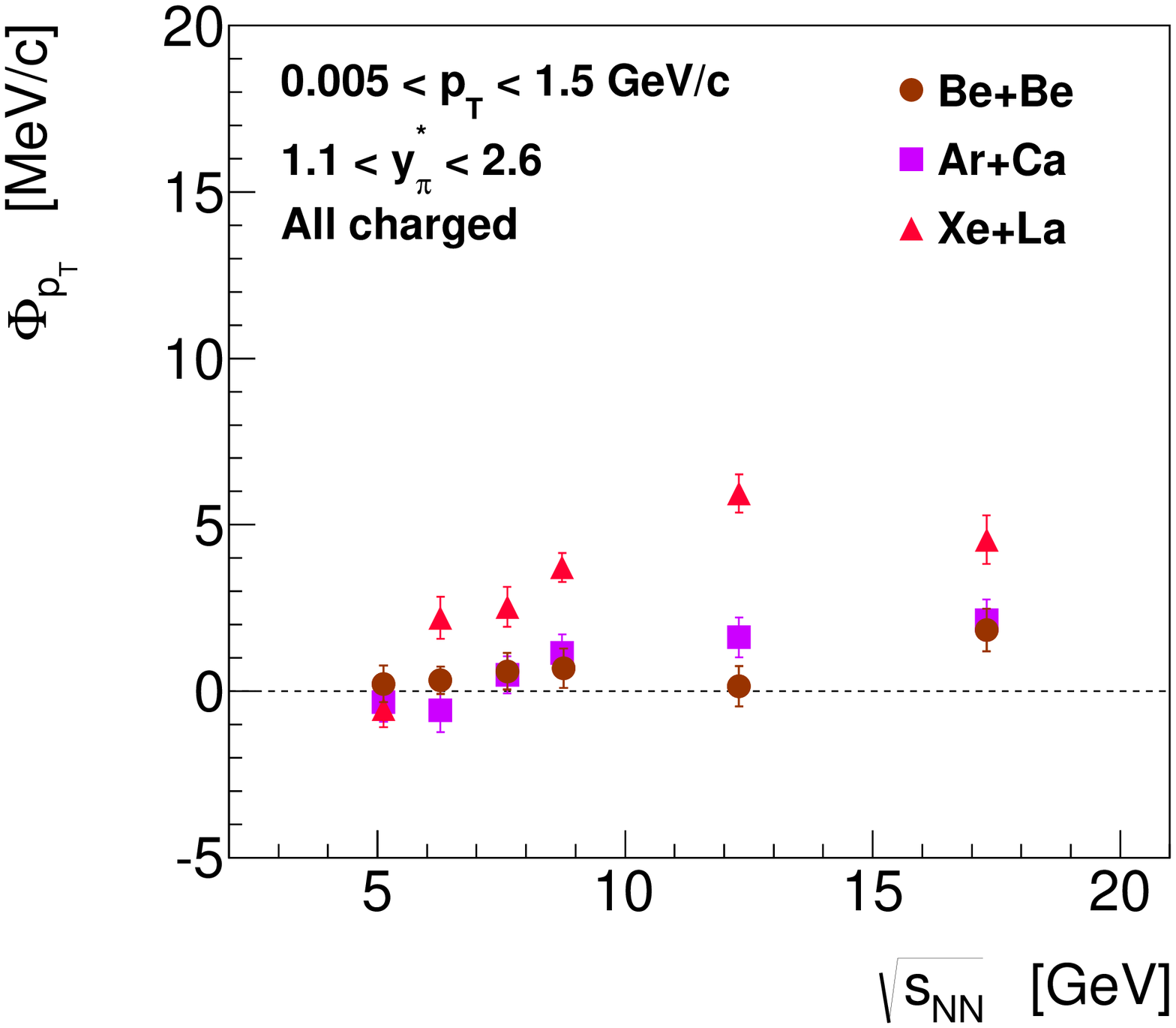}
\includegraphics[width=0.325\textwidth]{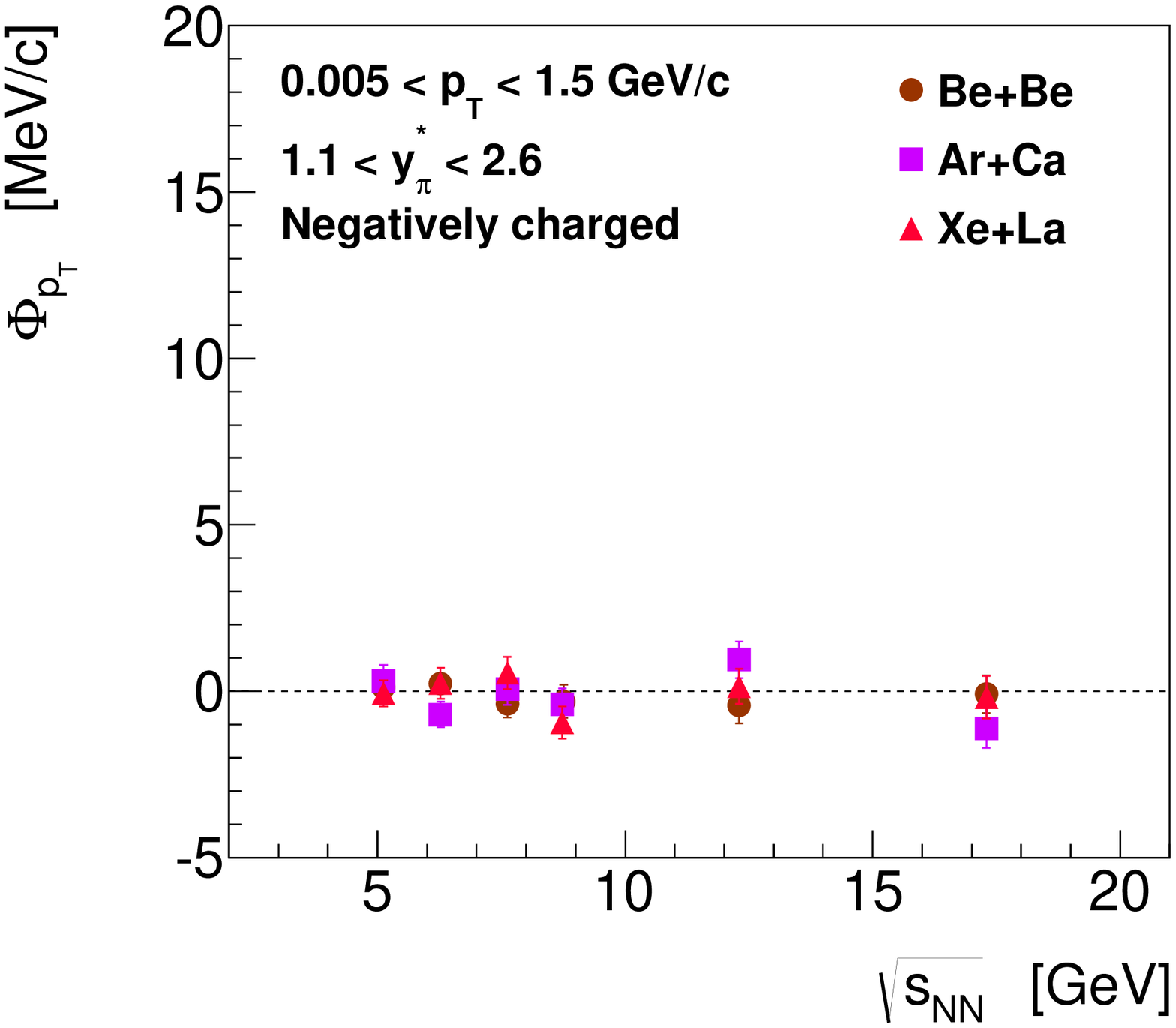}
\includegraphics[width=0.325\textwidth]{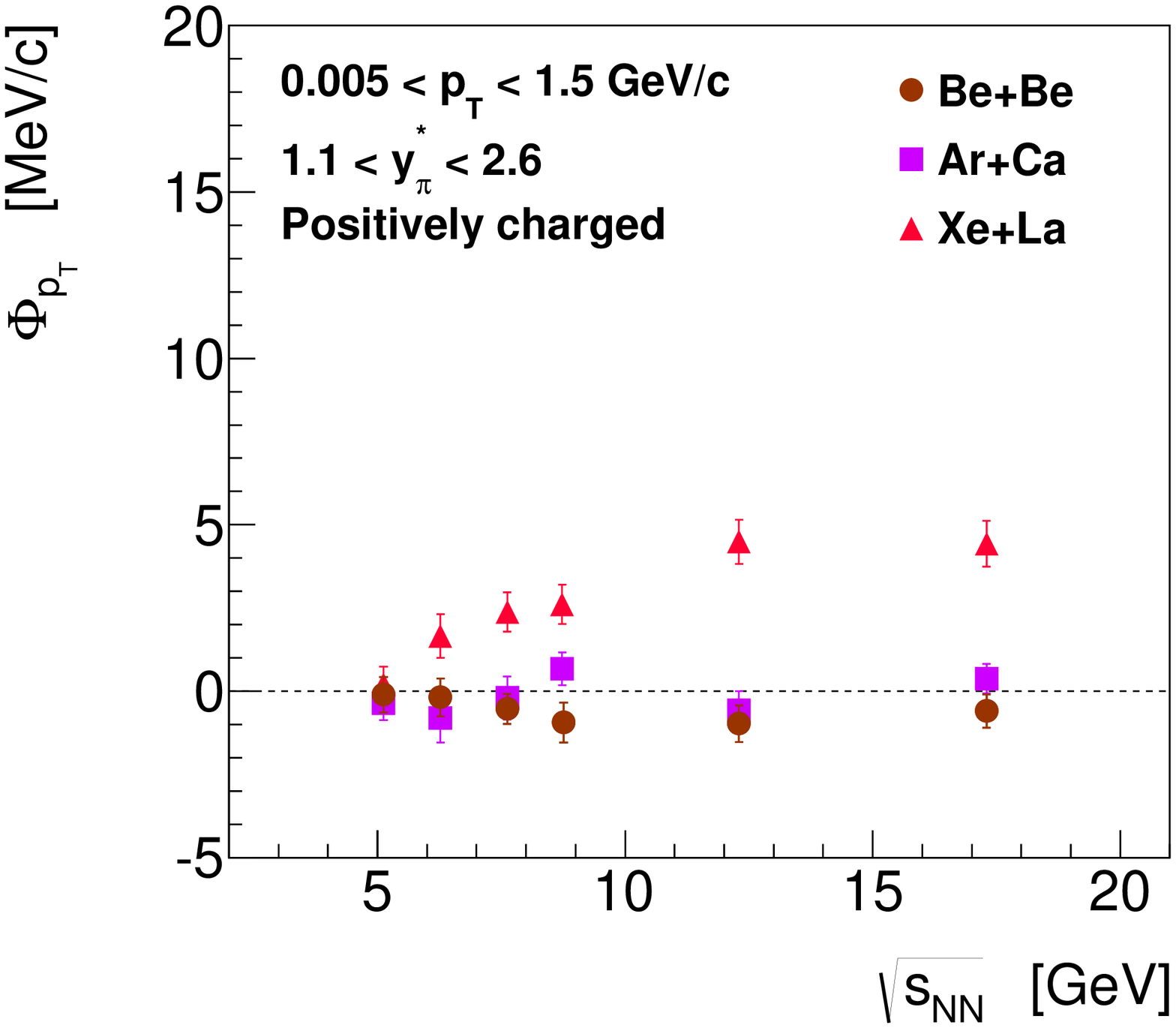}
\vspace{-0.6cm}
\caption[]{Energy dependence of $\Phi_{p_{T}}$ at forward rapidity for different charge combinations of particles produced in 20\% most central $A+A$ collisions in UrQMD. Additional cut $y^{*}_{p} < y^{*}_{beam} - 0.5$ was applied.}
\label{urq_phi_rapcut}
\end{figure}

In the next set of plots (Fig. \ref{urq_phi_rapcut}) the $\Phi_{p_{T}}$ values are shown for forward rapidity only ($1.1 < y^{*}_{\pi} < 2.6$, where $y^{*}_{\pi}$ is the particle rapidity calculated in the center-of-mass reference system assuming pion mass). The same rapidity cut was used by the NA49 experiment in the analysis of transverse momentum \cite{fluct_size, fluct_energy} and multiplicity \cite{omega_size} fluctuations. Additionally, only particles with $y^{*}_{p} < y^{*}_{beam} - 0.5$ were accepted ($y^{*}_{beam}$ is the rapidity of the beam calculated in the center-of-mass reference system). This cut allows to get rid of the effect of event-by-event impact parameter fluctuations when restricting the analysis to  forward rapidity only. The details of this cut were discussed in \cite{kg_imp} and in particular it was shown that event-by-event fluctuations of the number of protons in the forward hemisphere and the number of protons that are closer to the production region can lead to increased $\Phi_{p_{T}}$ values.

\begin{figure}[ht]
\includegraphics[width=0.325\textwidth]{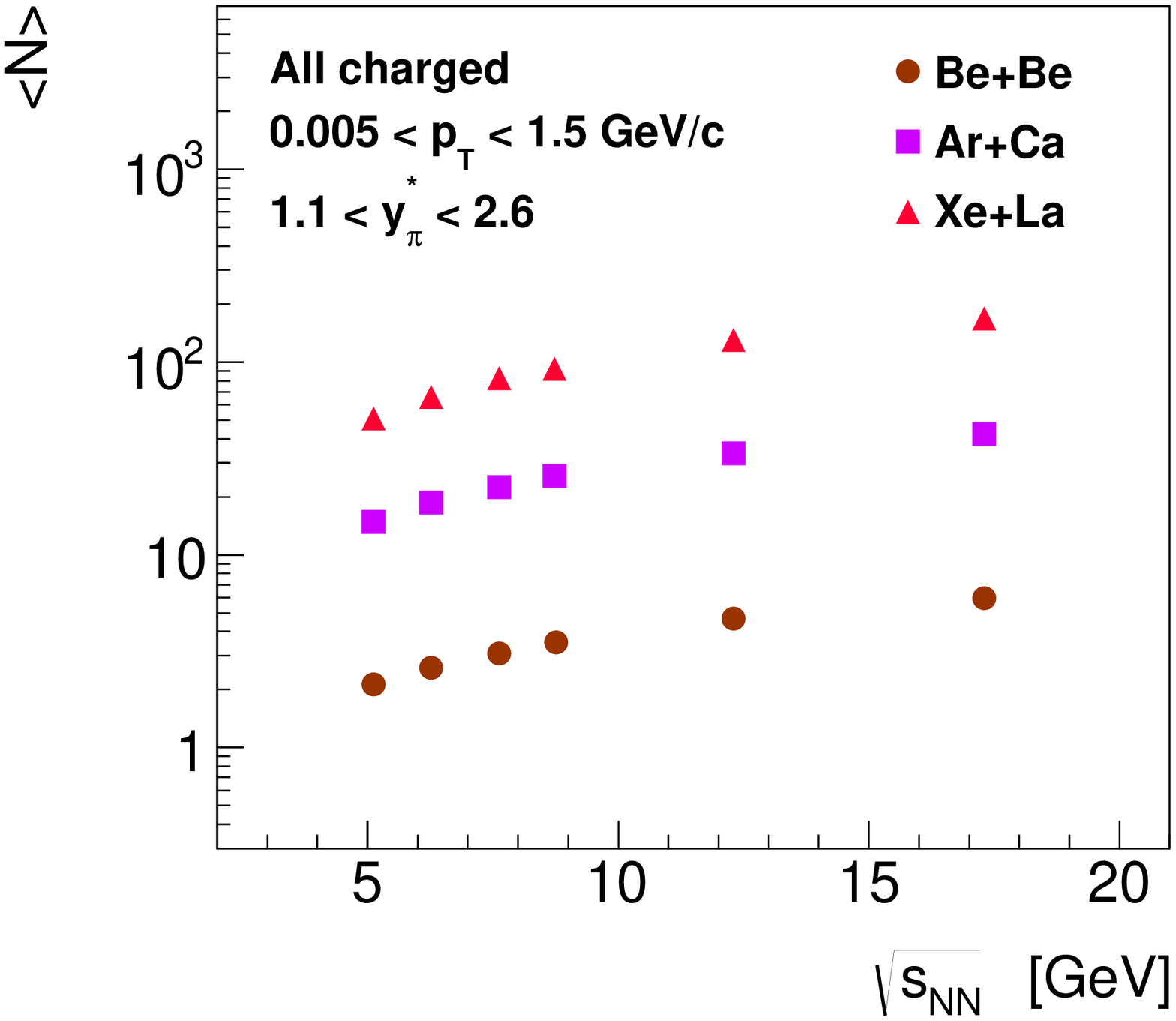}
\includegraphics[width=0.325\textwidth]{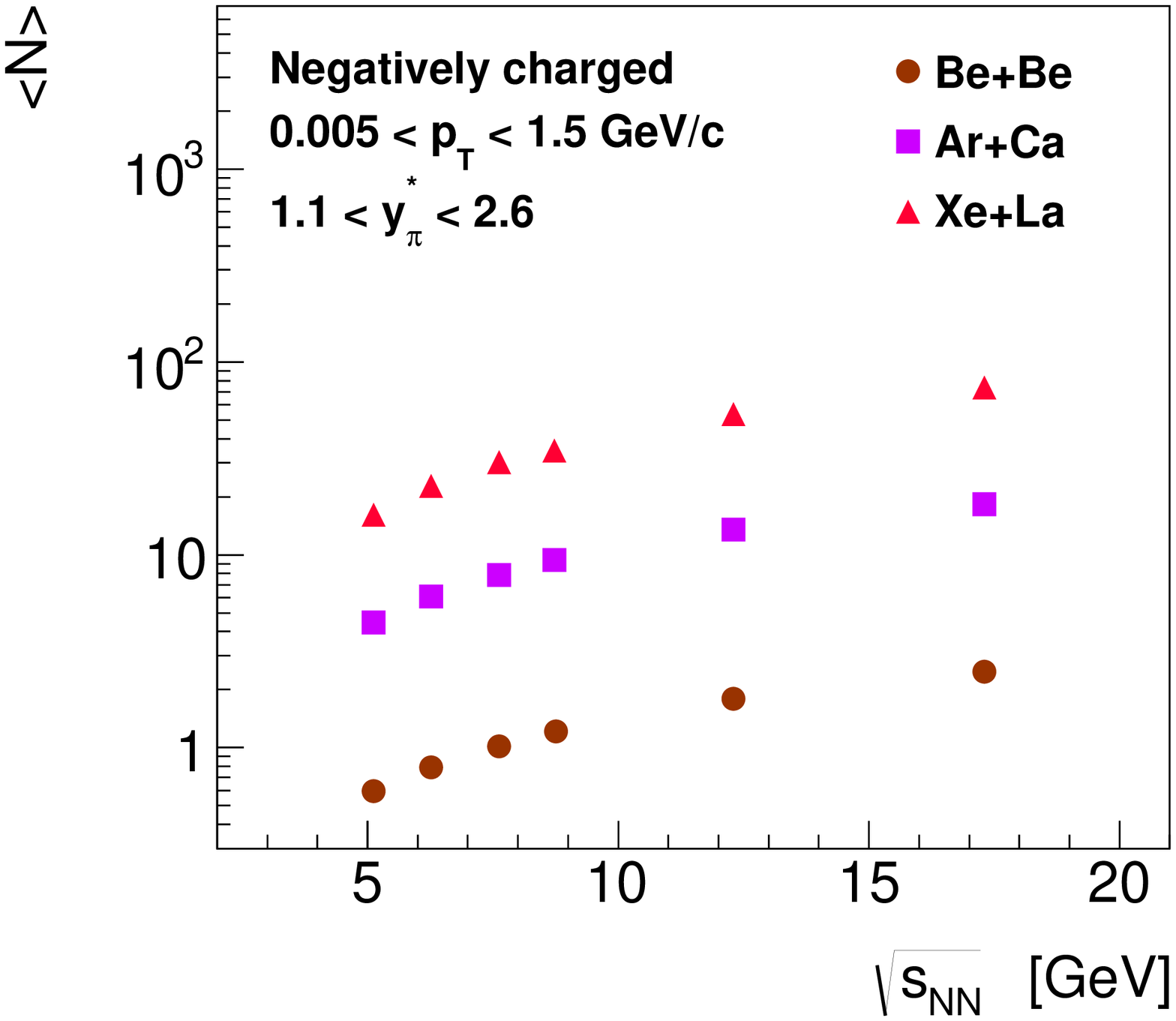}
\includegraphics[width=0.325\textwidth]{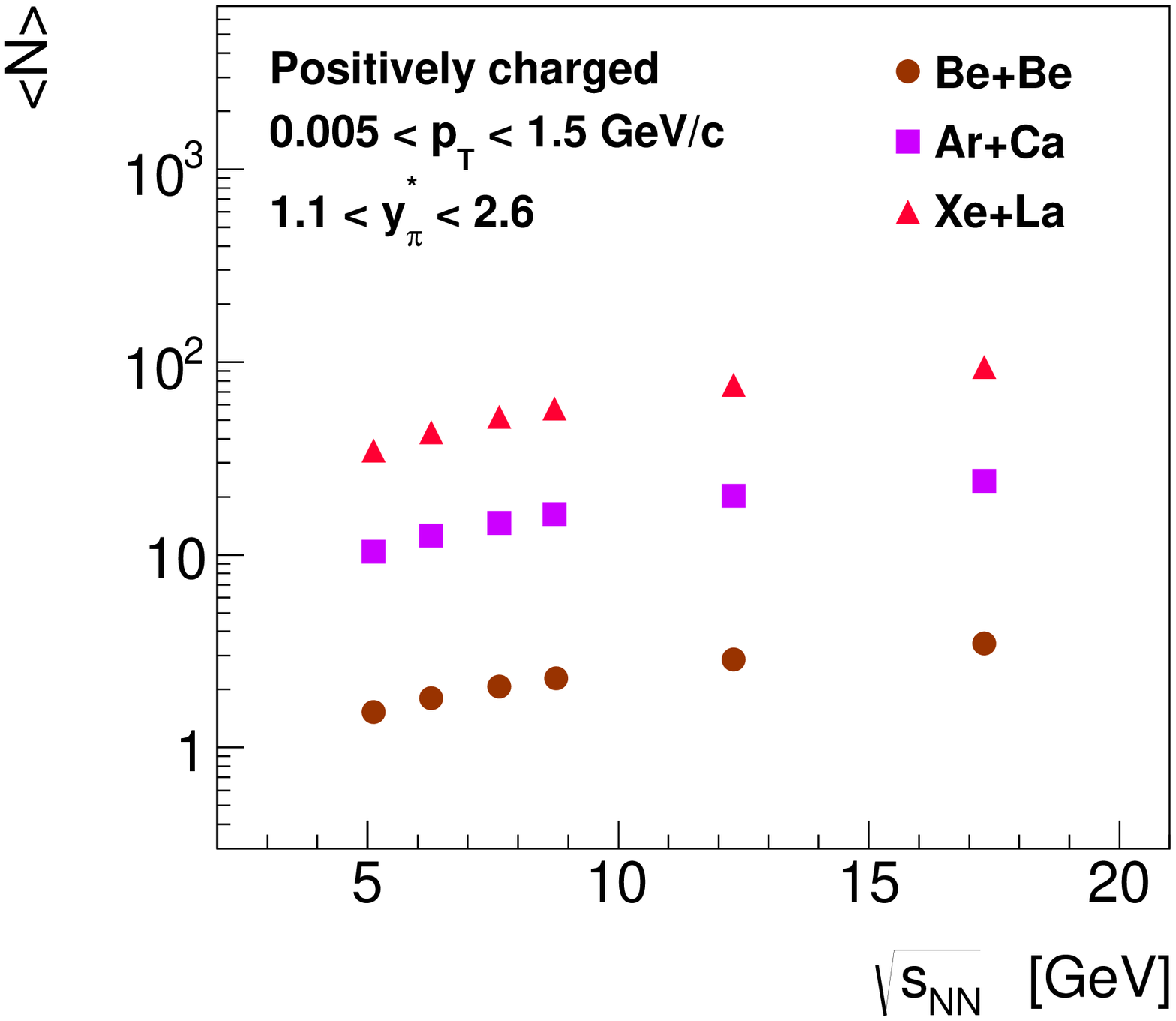}
\vspace{-0.6cm}
\caption[]{Mean multiplicities of all charged, negatively charged and positively charged particles produced at forward rapidity in 20\% most central $A+A$ collisions in UrQMD.. Additional cut $y^{*}_{p} < y^{*}_{beam} - 0.5$ was applied.}
\label{urq_multip_rapcut}
\end{figure}

Figure \ref{urq_phi_rapcut} shows that $\Phi_{p_{T}}$ values measured at forward rapidity are much smaller than those for complete rapidity region (see Fig. \ref{urq_phi_ptcut} for a comparison). Also the mean multiplicities are significantly smaller (Fig. \ref{urq_multip_rapcut}). The negatively charged particles show $\Phi_{p_{T}}$ consistent with zero for all three studied systems (the slight increase with energy observed for complete rapidity region is not seen here any more). The positively charged particles also result in $\Phi_{p_{T}}$ values close to zero and a weak increase with increasing energy can be observed for $Xe+La$ system only.

\subsubsection{Centrality restrictions}

\begin{figure}[ht]
\centering
\vspace{-0.3cm}
\includegraphics[width=0.325\textwidth]{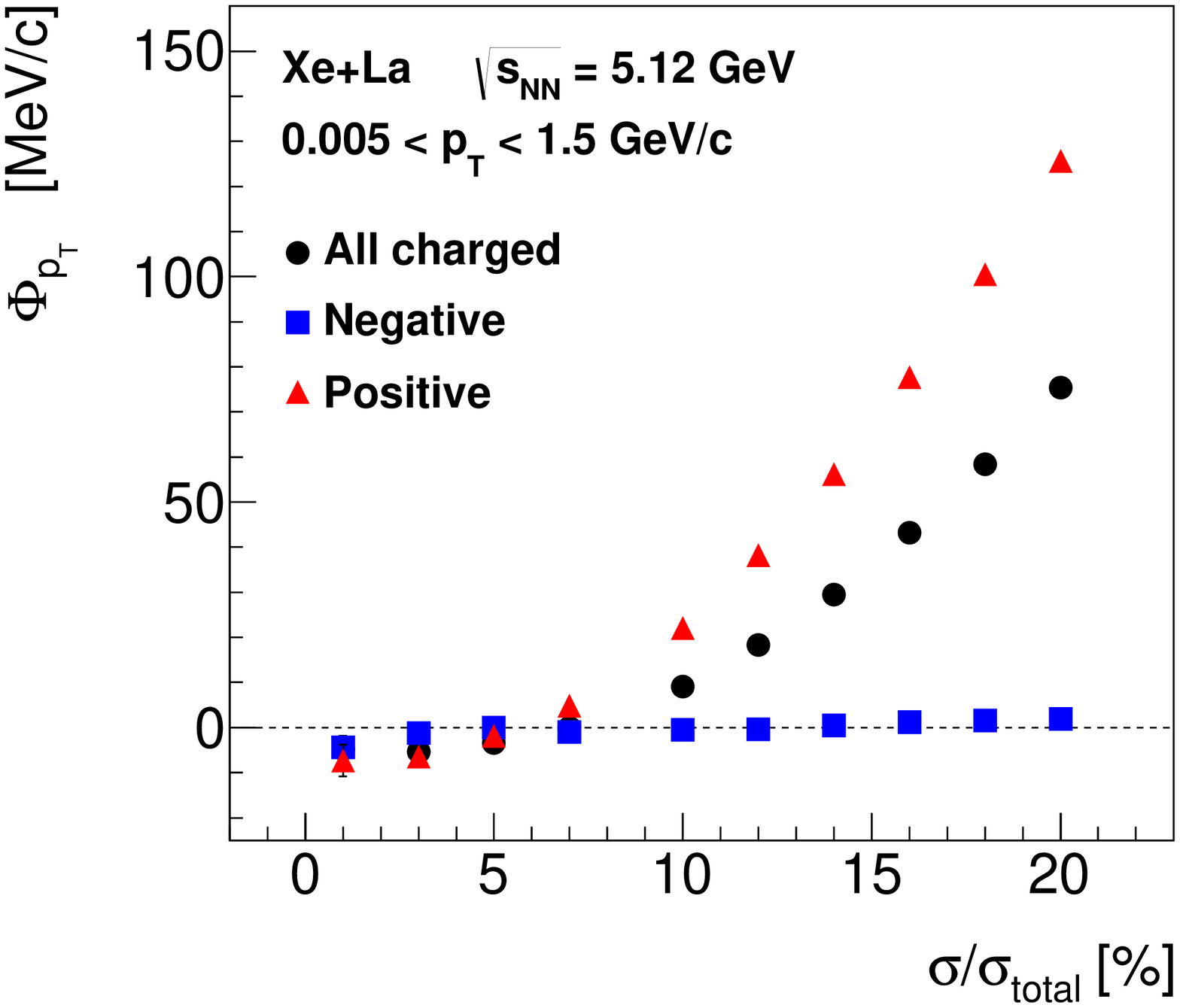}
\includegraphics[width=0.325\textwidth]{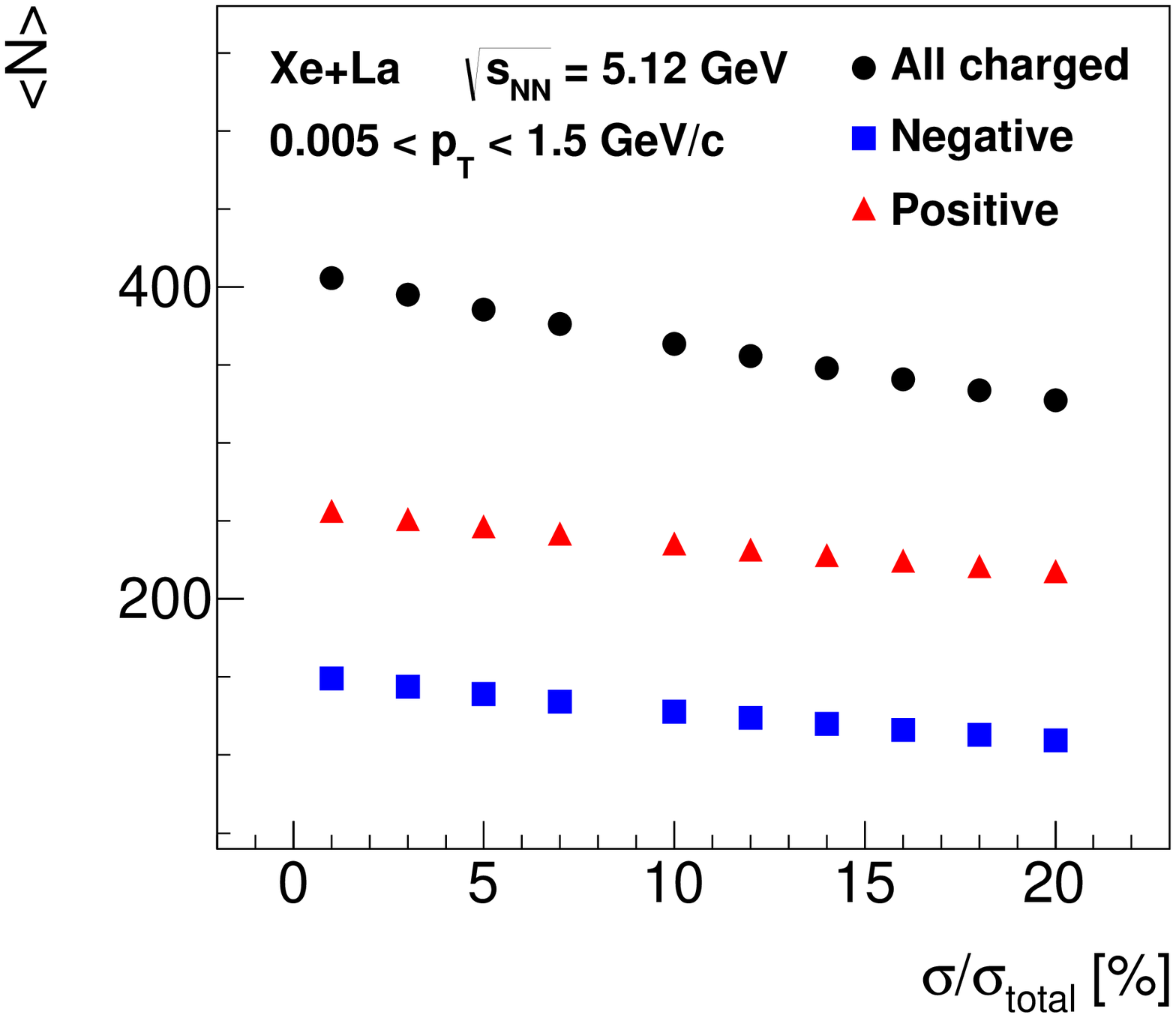}
\vspace{-0.4cm}
\caption[]{$\Phi_{p_{T}}$ and mean multiplicities as function of percent of total inelastic cross section for $Xe+La$ collisions at the lowest SPS energy. Note: the values and their errors are correlated.}
\label{urq_phi_centrality}
\end{figure}

Figure \ref{urq_phi_ptcut} suggests that for 20\% most central interactions the increased $\Phi_{p_{T}}$ values at lower SPS energies are mainly due to the significant fraction of protons present in the samples. The centrality bin 0-20\% is a relatively wide one 
and therefore the observed increase of $\Phi_{p_{T}}$ may be due to event-by-event impact parameter fluctuations (corresponding to event-by-event fluctuations in the number of participating protons). One can suppress this effect by selection of narrower centrality bins. In Fig. \ref{urq_phi_centrality} the centrality of $Xe+La$ at 13$A$ GeV beam energy is restricted from 0-20\% (rightmost points) down to 0-1\% most central (leftmost points). As the same (0-20\%) event sample was used points may be correlated. The negatively charged particles do not show any dependence on $\sigma / \sigma_{total}$. As expected, the $\Phi_{p_{T}}$ values for all charged paricles and for positively charged ones decrease when going to more central collisions, reaching the values similar to that for neagatively charged at aproximatelly 7\% most central interactions.

\begin{figure}[ht]
\includegraphics[width=0.325\textwidth]{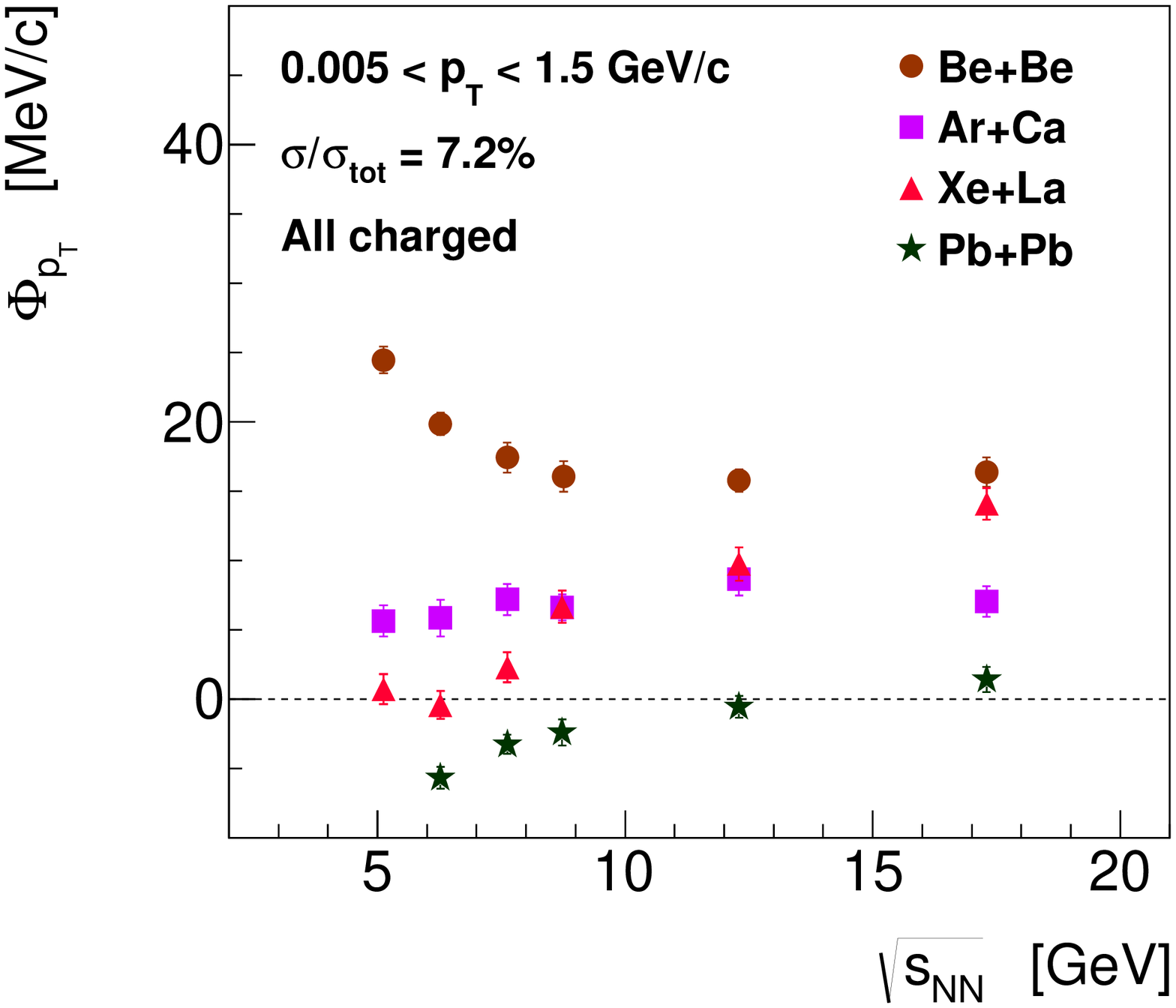}
\includegraphics[width=0.325\textwidth]{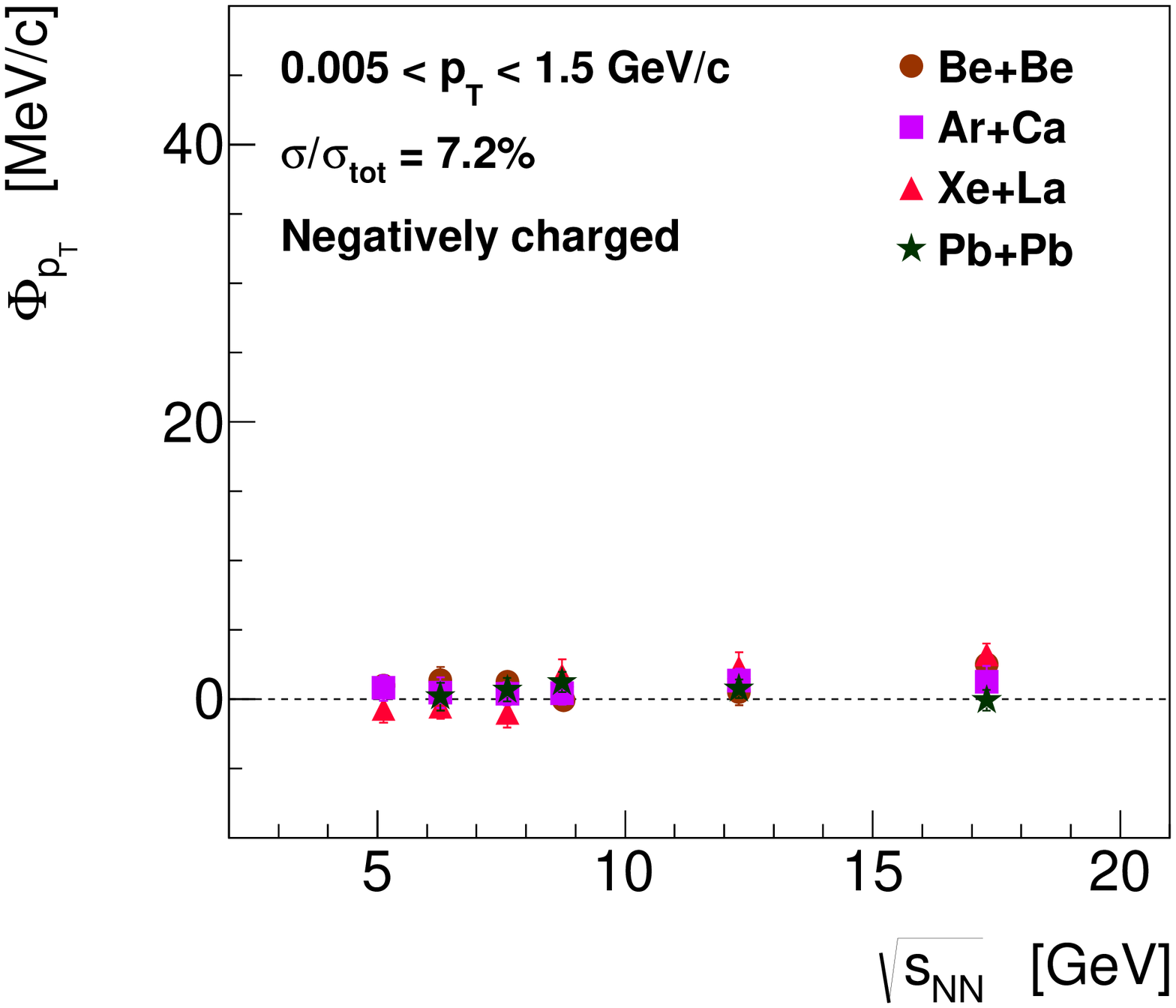}
\includegraphics[width=0.325\textwidth]{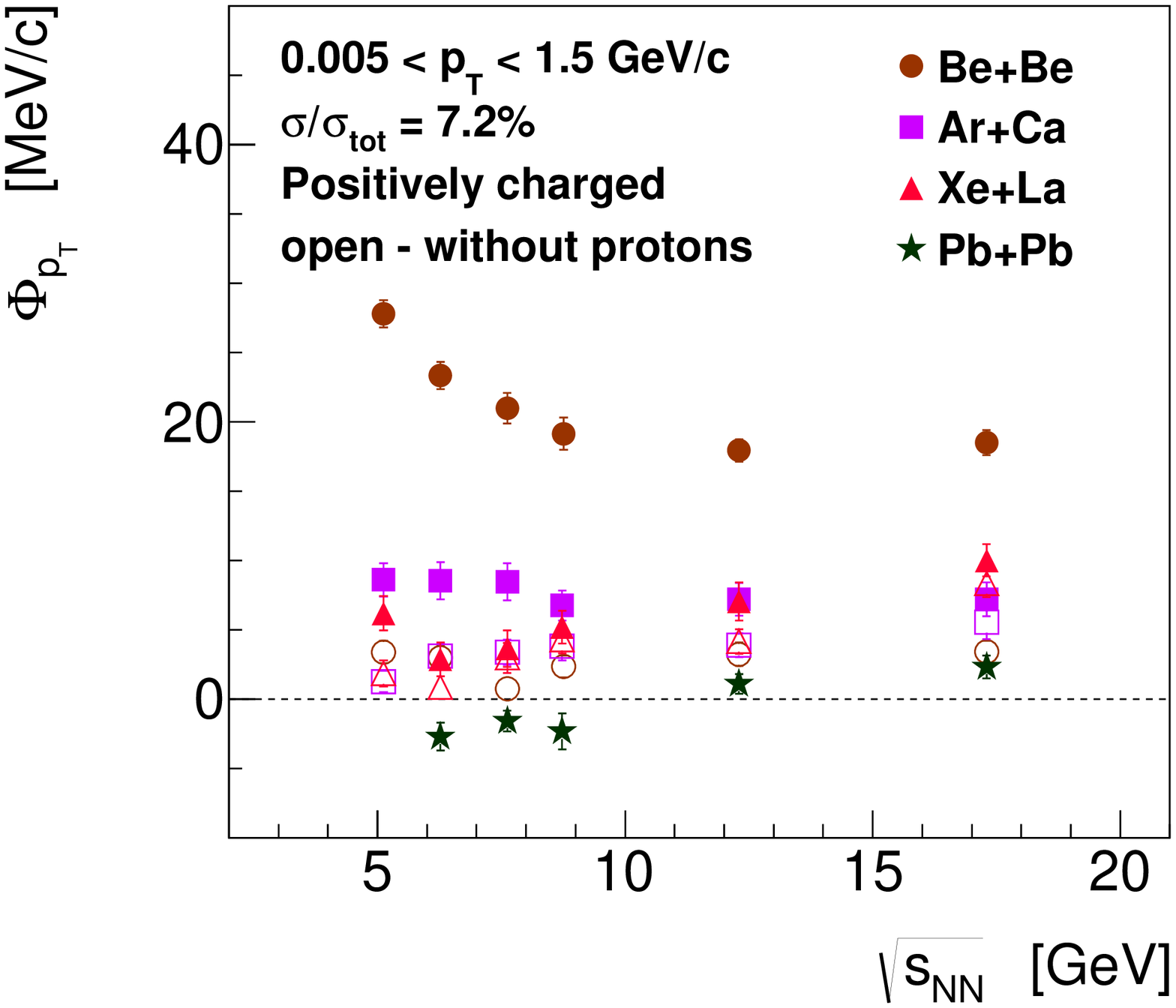}
\vspace{-0.6cm}
\caption[]{Energy dependence of $\Phi_{p_{T}}$ for different charge combinations of particles produced in UrQMD. 7.2\% most central interactions are selected.}
\label{urq_phi_72central}
\end{figure}

Figure \ref{urq_phi_centrality} confirms that the centrality range (0-20\%) of the events planned to be recorded by the NA61 experiment may need futher restrictions in order to calculate transverse momentum fluctuations. Therefore in Fig. \ref{urq_phi_72central} the same results as in Fig. \ref{urq_phi_ptcut} are presented but additional cut on selection of 7.2\% most central collisions was applied.
\footnote{In UrQMD simulations centrality 0-7.2\% corresponds to impact parametr cut $b<1.17$ fm for $Be+Be$, $b<2.29$ fm for $Ar+Ca$, and $b<3.42$ fm for $Xe+La$ interactions.}
This cut corresponds to centrality selection done by NA49 in the study of the energy dependence of $\Phi_{p_{T}}$ for Pb+Pb interactions \cite{fluct_energy}. In Fig. \ref{urq_phi_72central} additional points for UrQMD1.3 Pb+Pb collisions are included. For 7.2\% most central interactions $\Phi_{p_{T}}$ measure is close to zero but only for negatively charged particles. For positively charged it is close to zero \footnote{Very weak increase of $\Phi_{p_{T}}$ with energy may be observed for $Pb+Pb$ and $Xe+La$.} but only for heavier systems ($Pb+Pb$ and $Xe+La$), however for light systems (especially $Be+Be$) we still observe a significant increase at lower SPS energies. This increase is not visible any more when protons are removed from the sample. As the centrality is already restricted here we can guess that there may be another yet source of correlations (conservation laws?) that produces positive $\Phi_{p_{T}}$ values for central interactions at lower energies. This hypotesis needs further investigations however it is supported but the recent observation \cite{tobczo} that also in UrQMD $p+p$ interactions (no centrality restrictions required) $\Phi_{p_{T}}$ values for all charged and for positively charged particles are increased at lower SPS energies, wherease the negatively charged ones are consistent with zero in the whole SPS energy range.

\subsection{$\Delta^{XN}$ - dependence on energy and charge combination}

\begin{figure}[ht]
\includegraphics[width=0.325\textwidth]{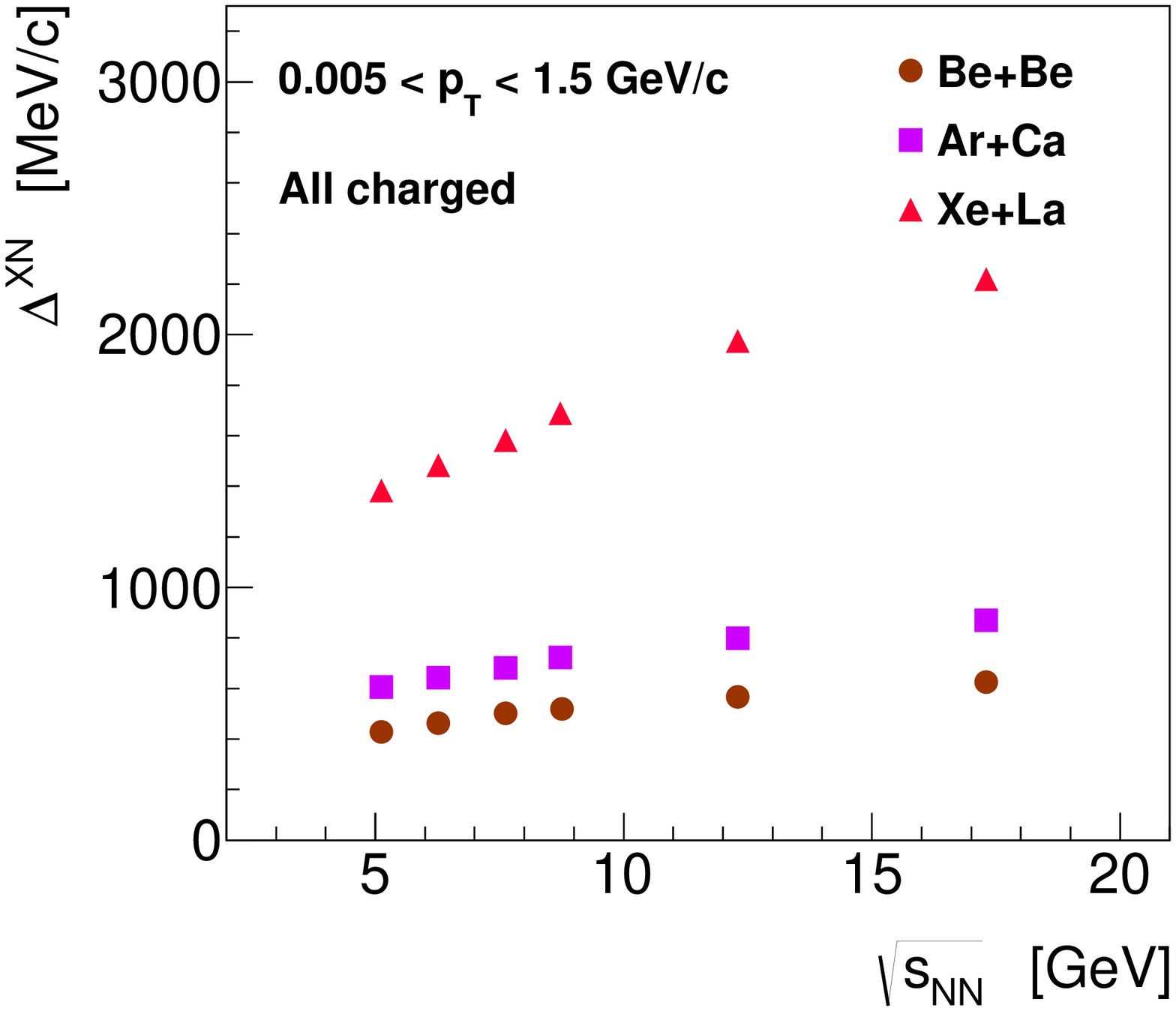}
\includegraphics[width=0.325\textwidth]{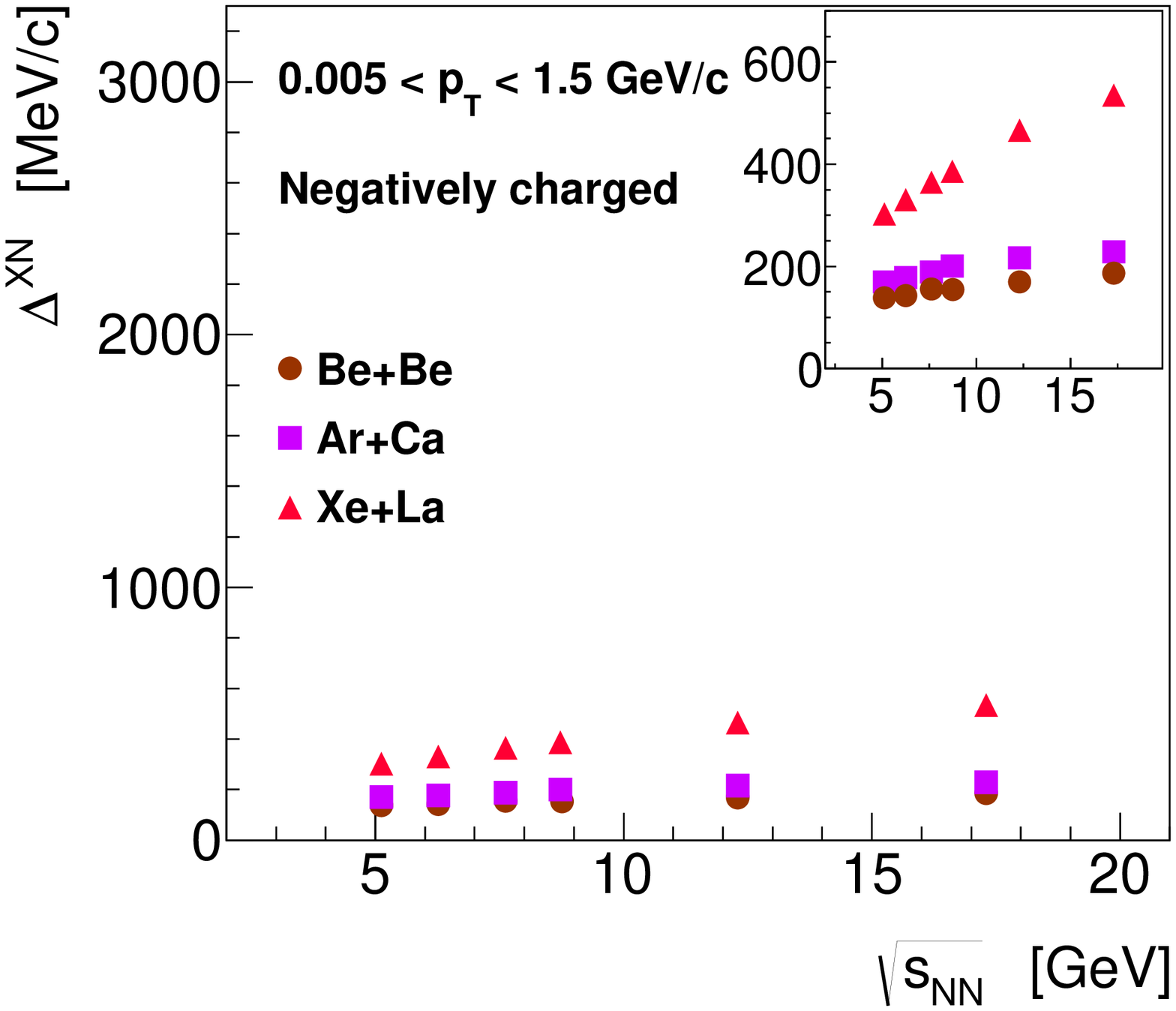}
\includegraphics[width=0.325\textwidth]{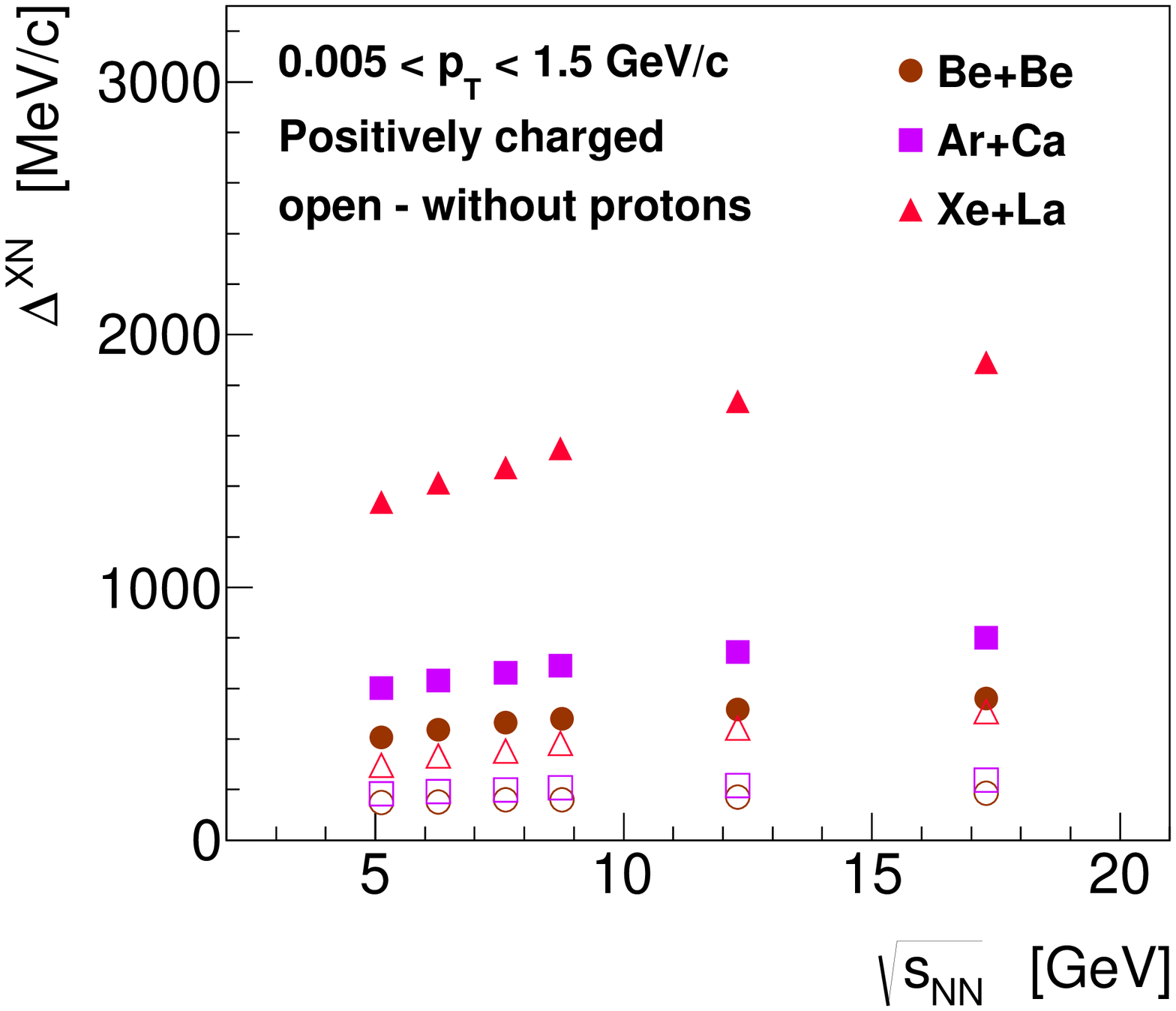}
\vspace{-0.6cm}
\caption[]{Energy dependence of $\Delta^{XN}$ for different charge combinations of particles produced in 20\% most central $A+A$ collisions in UrQMD.}
\label{urq_delta_ptcut}
\end{figure}

In Fig. \ref{urq_delta_ptcut} the energy dependence of $\Delta^{XN}$ is shown for 20\% most central $A+A$ interactions. Three different charge combinations are included: all charged particles, negatively charged and  positively charged. The open symbols in the right panel represent positively charged particles where protons were removed from the sample. In contrary to $\Phi_{p_{T}}$ measure (see Fig. \ref{urq_phi_ptcut}) the $\Delta^{XN}$ measure shows monotonic increase with increasing energy for all three charge combinations and for all studied systems. Similarly to $\Phi_{p_{T}}$ the values of $\Delta^{XN}$ for positively charged particles are comparable with those for negatively charged ones provided that protons are removed from the sample.

\begin{figure}[ht]
\includegraphics[width=0.325\textwidth]{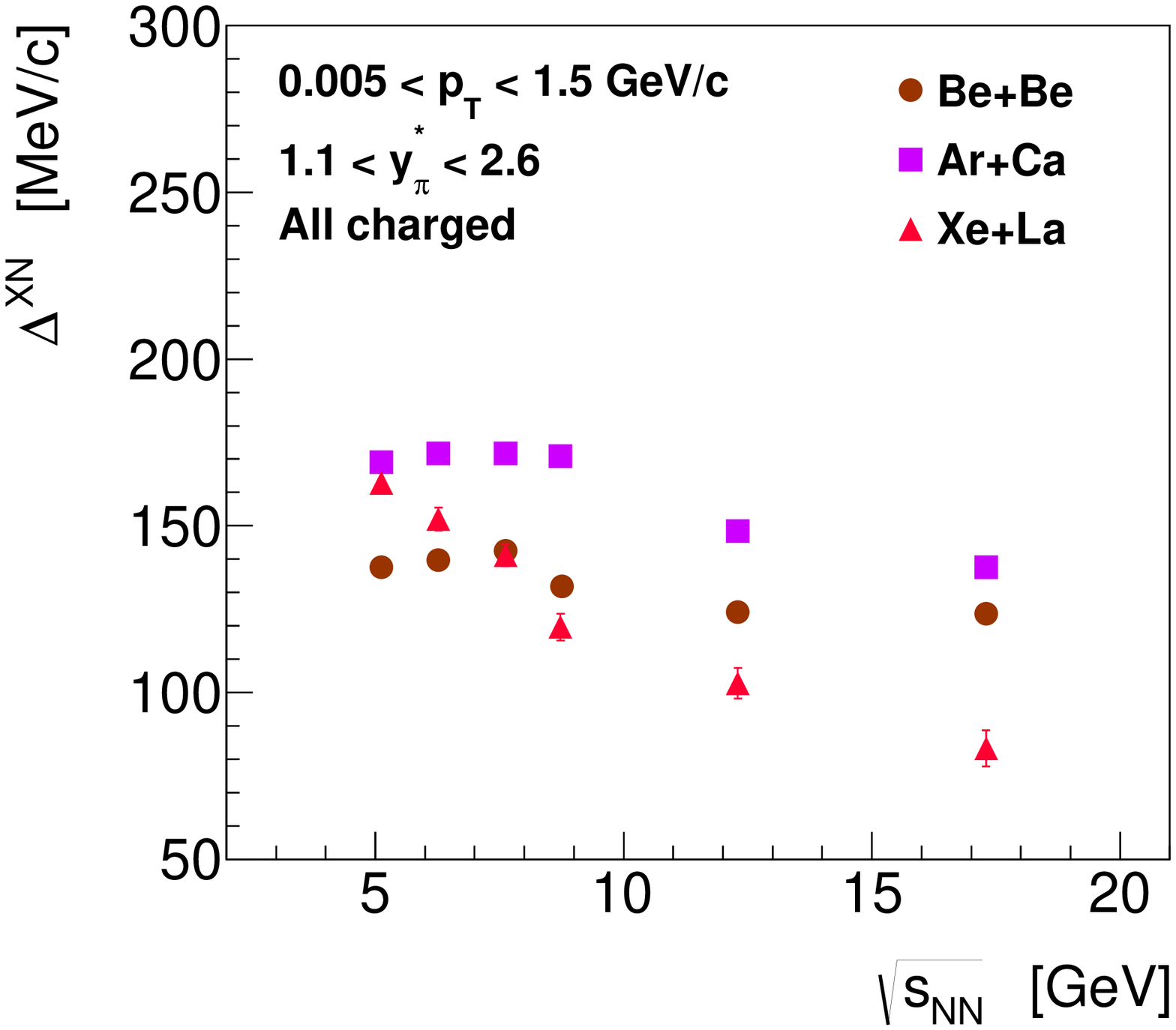}
\includegraphics[width=0.325\textwidth]{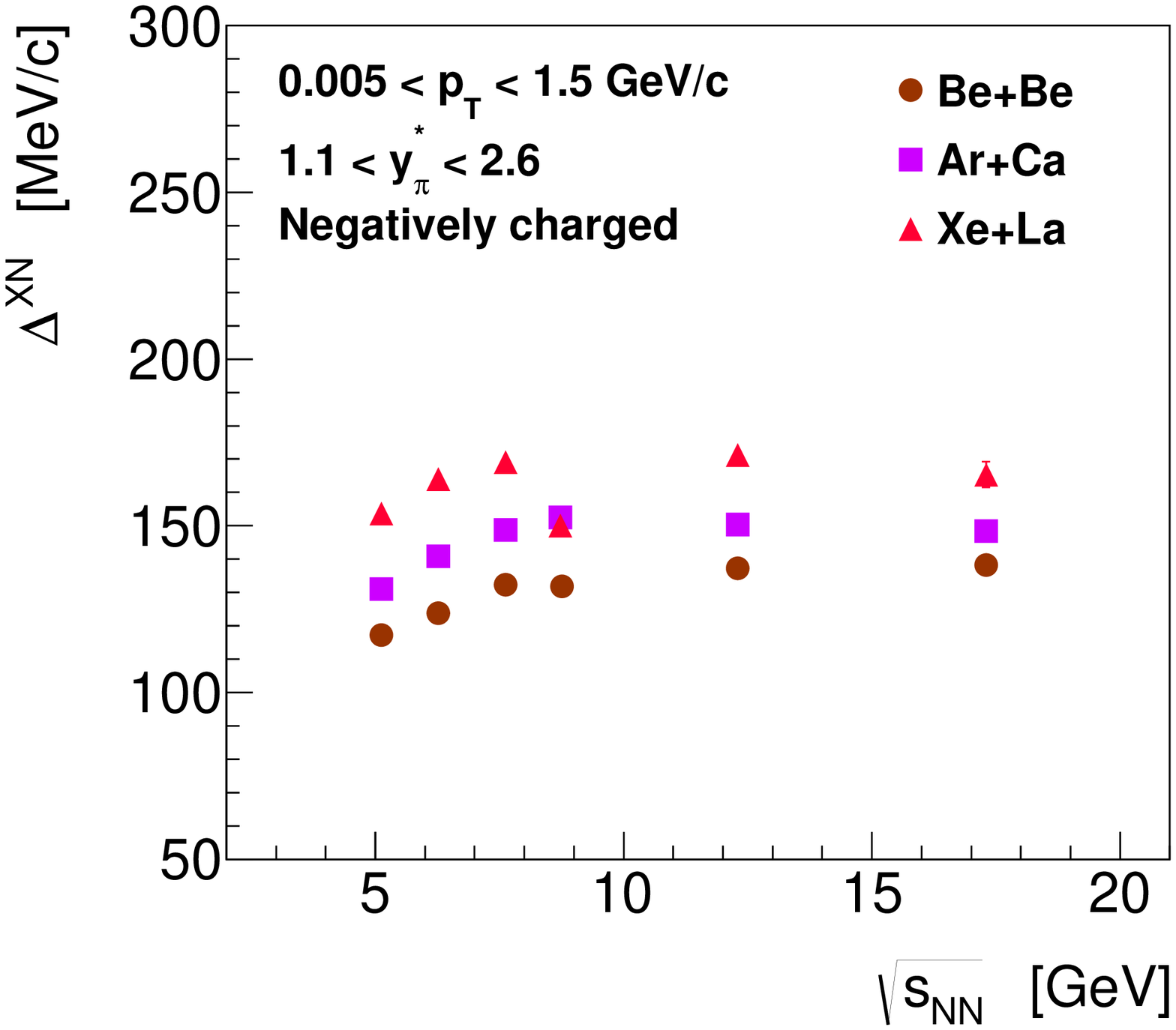}
\includegraphics[width=0.325\textwidth]{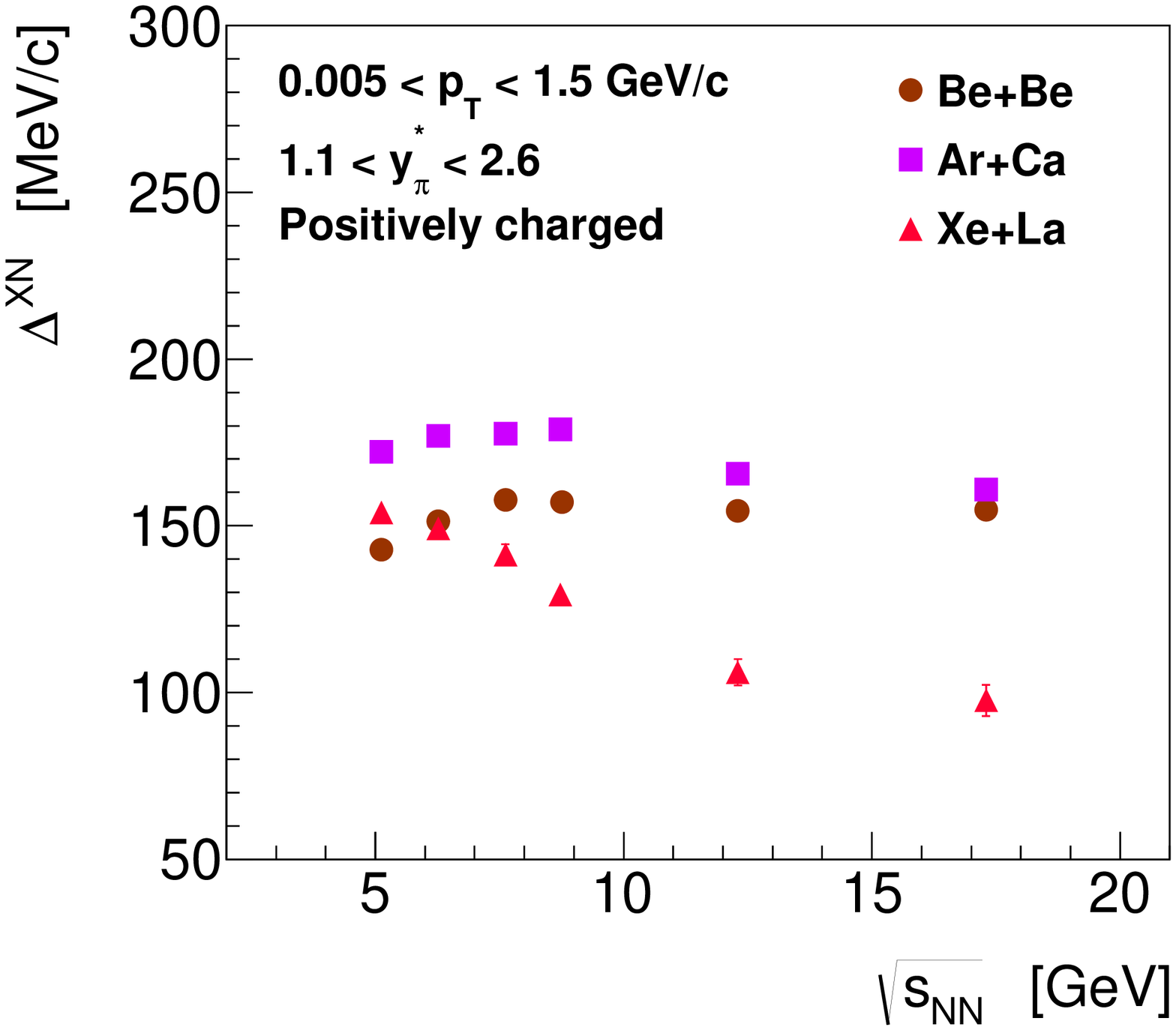}
\vspace{-0.6cm}
\caption[]{Energy dependence of $\Delta^{XN}$ at forward rapidity for different charge combinations of particles produced in 20\% most central $A+A$ collisions in UrQMD. Additional cut $y^{*}_{p} < y^{*}_{beam} - 0.5$ was applied.}
\label{urq_delta_rapcut}
\end{figure}

In the next set of plots (Fig. \ref{urq_delta_rapcut}) the $\Delta^{XN}$ values are show for forward rapidity only ($1.1 < y^{*}_{\pi} < 2.6$). Additionally, only particles with $y^{*}_{p} < y^{*}_{beam} - 0.5$ were accepted. As seen 
$\Delta^{XN}$ values measured at forward rapidity are much smaller than those for complete rapidity region (see Fig. \ref{urq_delta_ptcut} for a comparison). The negatively charged particles show only a very weak increase and then saturation of $\Delta^{XN}$ with increasing energy. Positively charged particles at forward rapidity show decrease of $\Delta^{XN}$ values with energy for heavier systems ($Ar+Ca$, $Xe+La$). This is in contrary to $\Phi_{p_{T}}$ (see Fig. \ref{urq_phi_rapcut}), which showed an increase with energy for positively charged particles in $Xe+La$ interactions.

\begin{wrapfigure}{r}{4.cm}
\vspace{-0.3cm}
\includegraphics[width=0.325\textwidth]{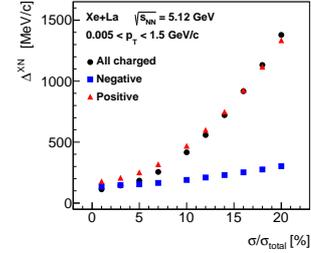}
\vspace{-0.6cm}
\caption[]{$\Delta^{XN}$ as function of percent of total inelastic cross section for $Xe+La$ collisions at the lowest SPS energy. Note: the values and their errors are correlated.}
\label{urq_delta_centrality}
\end{wrapfigure}

Similarly to what was done for $\Phi_{p_{T}}$ the centrality was also restricted when calculating $\Delta^{XN}$. In Fig. \ref{urq_delta_centrality} the centrality of $Xe+La$ at 13$A$ GeV beam energy is restricted from 0-20\% (rightmost points) down to 0-1\% most central (leftmost points). The negatively charged partciles show a small increase of $\Delta^{XN}$ with increasing $\sigma / \sigma_{total}$ ($\Phi_{p_{T}}$ values were constant - see Fig. \ref{urq_phi_centrality}). The $\Delta^{XN}$ values for all charged paricles and for positively charged ones decrease when going to more central collisions, thus confirming that a significant contribution to $\Delta^{XN}$ may be indeed from event-by-event fluctuations of the number of participating protons.

\begin{figure}[ht]
\includegraphics[width=0.325\textwidth]{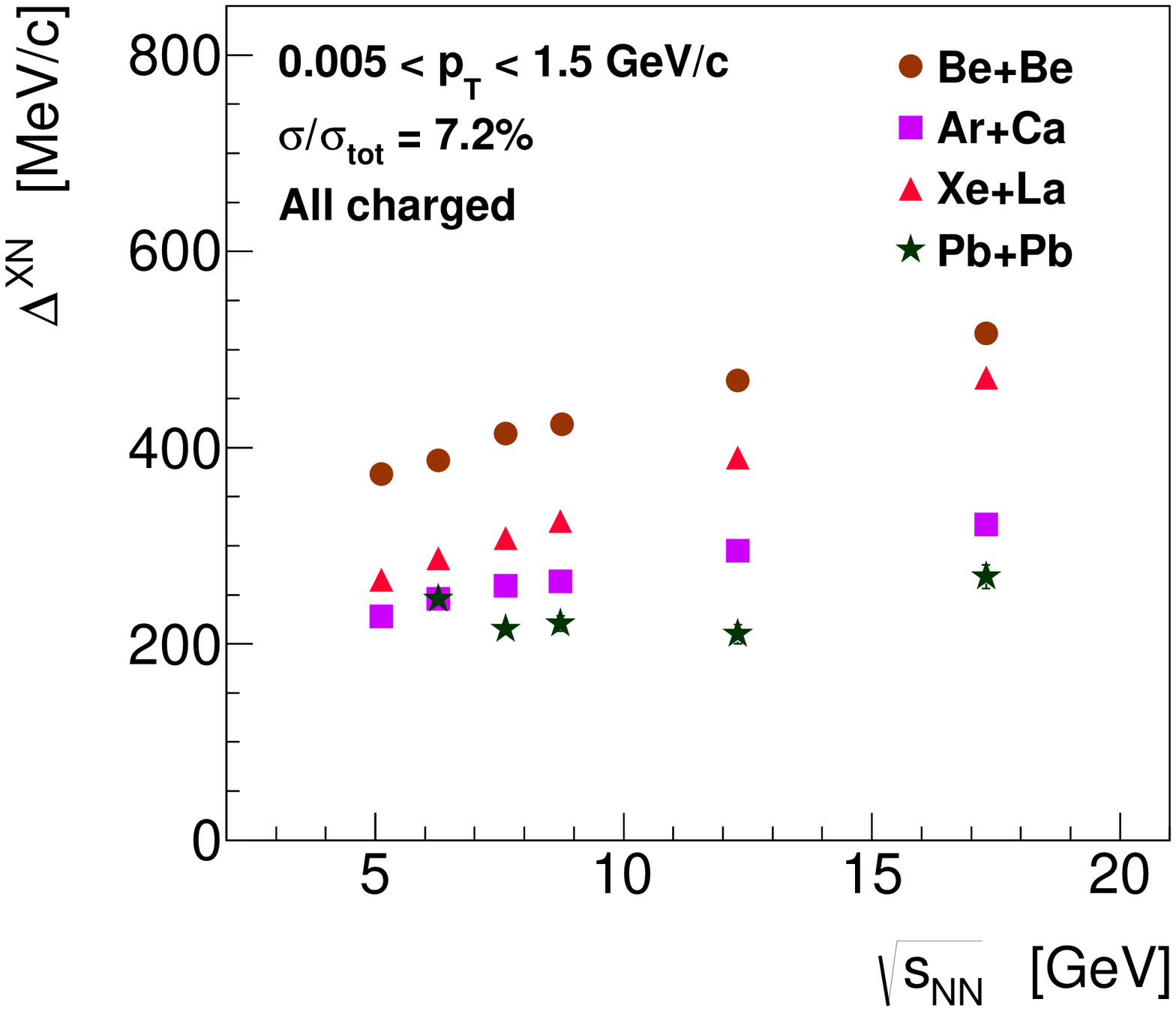}
\includegraphics[width=0.325\textwidth]{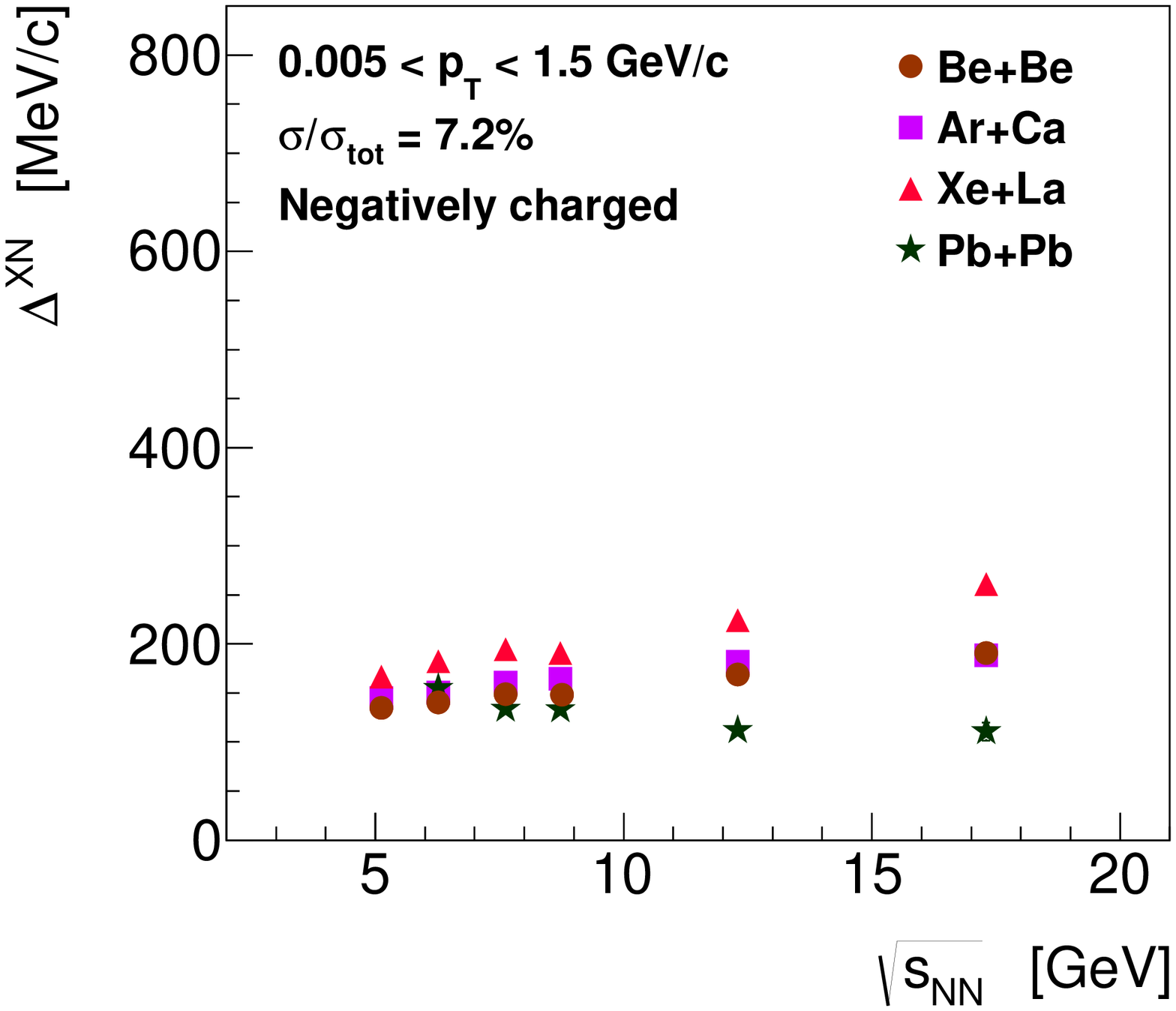}
\includegraphics[width=0.325\textwidth]{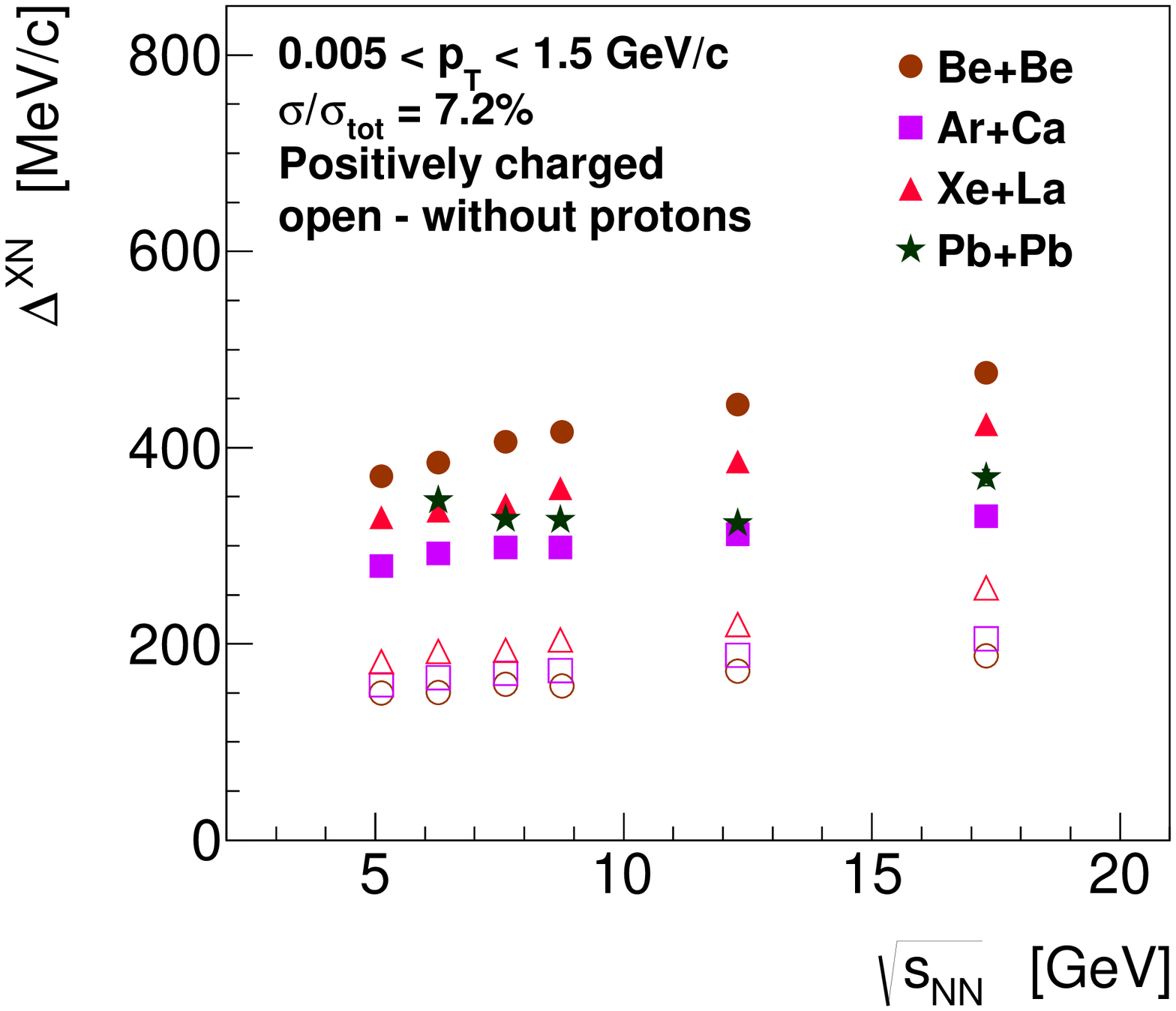}
\vspace{-0.6cm}
\caption[]{Energy dependence of $\Delta^{XN}$ for different charge combinations of particles produced in UrQMD. 7.2\% most central interactions are selected.}
\label{urq_delta_72central}
\end{figure}

Figure \ref{urq_delta_72central} presents the energy dependence of $\Delta^{XN}$ for 7.2\% most central $A+A$ collisions (additional points for UrQMD1.3 Pb+Pb collsions are included). In the case of $Xe+La$ and $Ar+Ca$ data the 7.2\% most central interactions result in $\Delta^{XN}$ values much smaller than those for 20\% most central (see Fig. \ref{urq_delta_ptcut} for a comparison). For $Be+Be$ interactions such drop is much less pronounces and in fact the magitudes for 7.2\% and 20\% most central interactions are comparable. For all systems the values of $\Delta^{XN}$ for 7.2\% most central collisions are higher for positivey charged partciles than for negatively charged ones and the rejection of protons can bring the magintudes of positively charged ones close to these for negatively charged. In contrary to what was observed for $\Phi_{p_{T}}$, for very central collisions $\Delta^{XN}$ for negatively charged particles show a rudimentary energy dependence. All systems and charge combinations show slight increase of $\Delta^{XN}$ with energy and only central Pb+Pb simulations seem to exhibit a very weak decrease/plateau of $\Delta^{XN}$ with energy. The different behaviour of $Pb+Pb$ collisions may be connected with the fact that the earlier version UrQMD1.3 was used for $Pb+Pb$ simulations, wherease UrQMD3.3 was employed to simulate $Be+Be$, $Ar+Ca$ and $Xe+La$ interactions.

In a simple superposition model all three fluctuation measures $\Delta^{XN}$, $\Sigma^{XN}$ and $\Phi_{p_T}$ should have the same value, independent of the system size (the number of sources). Therefore when canceling trivial effect of event-by-event impact parameter fluctuations (restricting to 7.2\% most central interactions) we should expect the same values of fluctuation measures for all studied systems. When considering "most save" sample of negatively charged particles one can see that it is indeed true for $\Phi_{p_T}$ measure (see middle panel of Fig. \ref{urq_phi_72central}), whereas for $\Delta^{XN}$ (middle panel of Fig. \ref{urq_delta_72central}) is is true only for lower SPS energies. As the results from the fast generators in the previous sections showed that $\Delta^{XN}$, $\Sigma^{XN}$ and $\Phi_{p_T}$ are indeed strongly intensive measures, it may suggest that the UrQMD model introduces somehow a small but systematic deviation from a simple superposition model beginning from middle SPS energies. $\Phi_{p_T}$ measure can be less sensitive than $\Delta^{XN}$ to such deviation and therefore its values are the same for all systems in the whole SPS energy range.

In the models the deviation from a simple superposition model (Wounded Nucleon Model) may be due to changes of the shape of the single particle spectrum and/or due to changes of fluctuations. 
It was suggested \cite{MGpriv} that an important difference between $\Phi$ and $\Delta$ or $\Sigma$ is that in case of $\Phi$ measure the single particle term is subtracted (the second term of $\Phi$ definition in Eq. \ref{Phi}), thus making $\Phi$ sensitive only to changes of fluctuations and insensitive to changes of the shape of the single particle spectrum. Such subtraction is not done for $\Delta$ and $\Sigma$, thus in this respect they are equivalent to the first (fluctuation) term of $\Phi$. As a consequence for an uncorrelated particle production $\Phi$ is zero and independent of a shape of the single particle spectrum. In a contrary, the $\Delta$ and $\Sigma$ are non-zero and their values depend on the shape of the single particle spectrum. If the UrQMD model introduces deviations from a simple superposition model by modifying the shape of the single particle spectrum only, then the large deviation from a superposition behavior should be visible for $\Delta$ and $\Sigma$ only but not for $\Phi$. Therefore, in the analysis of experimental data a simultaneous measurement of all three quantities $\Phi$, $\Delta$, and $\Sigma$ may help to understand the origin of superposition model violation.


\subsection{$\Sigma^{XN}$ - dependence on energy and charge combination}

\begin{figure}[ht]
\includegraphics[width=0.325\textwidth]{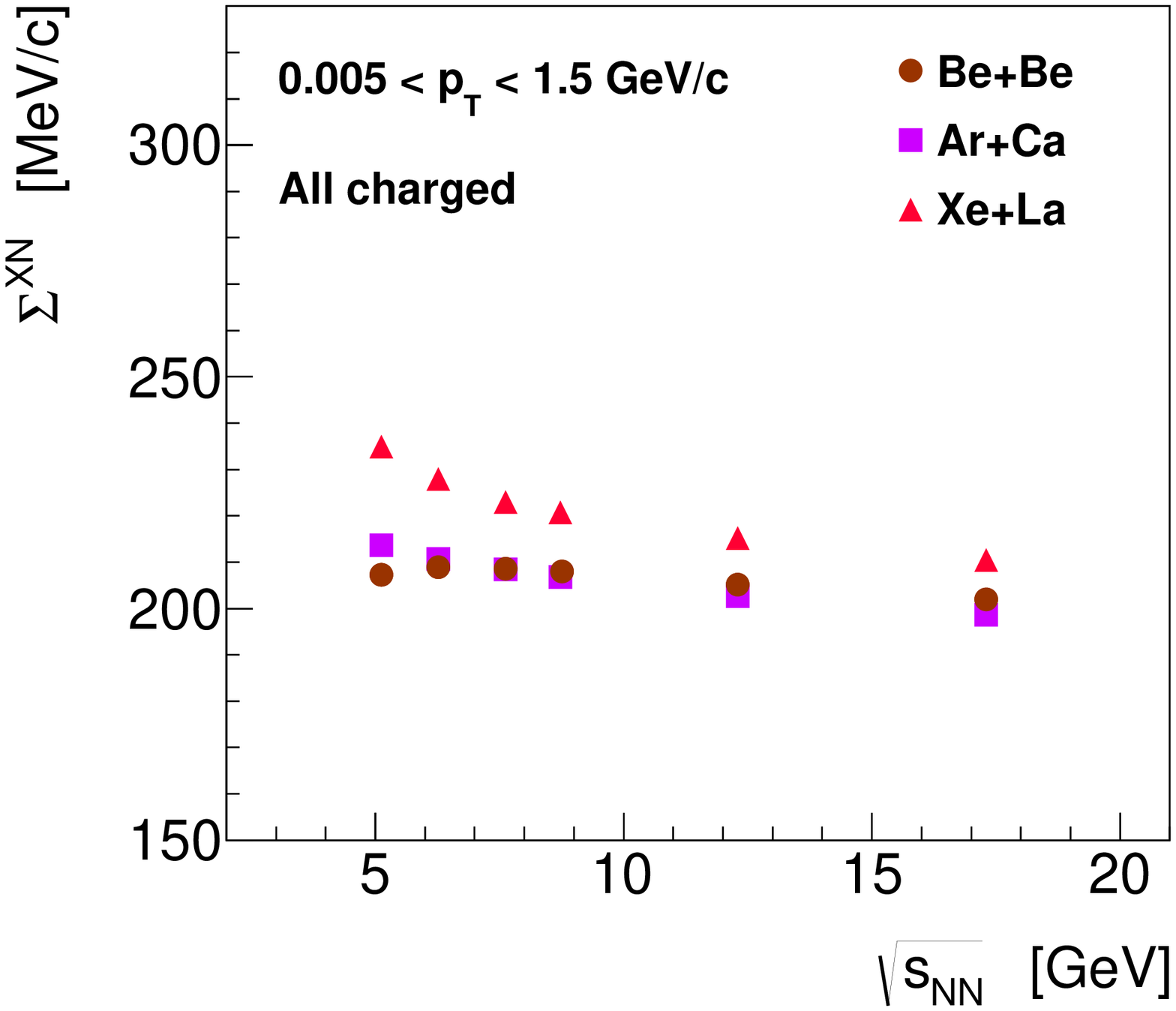}
\includegraphics[width=0.325\textwidth]{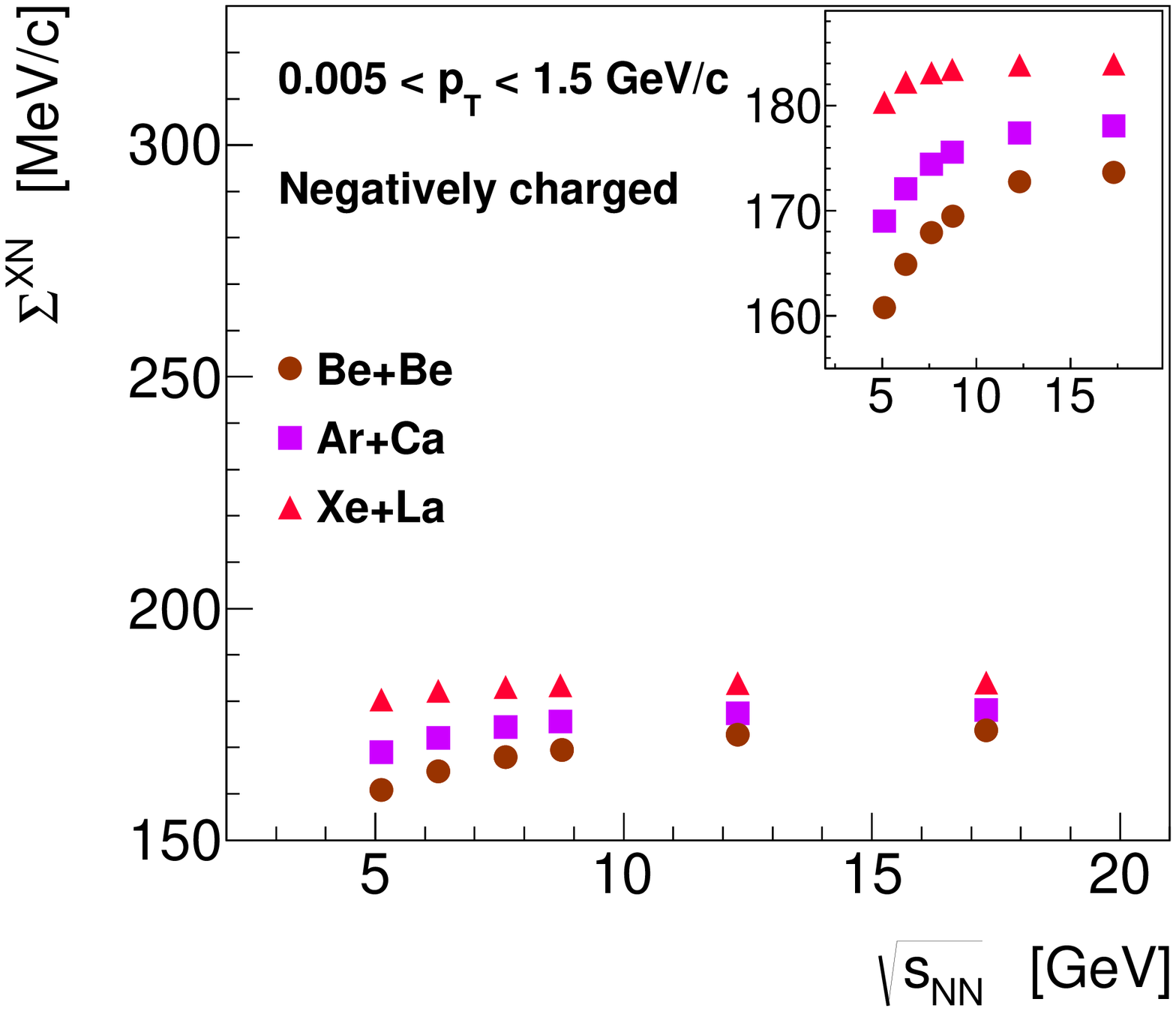}
\includegraphics[width=0.325\textwidth]{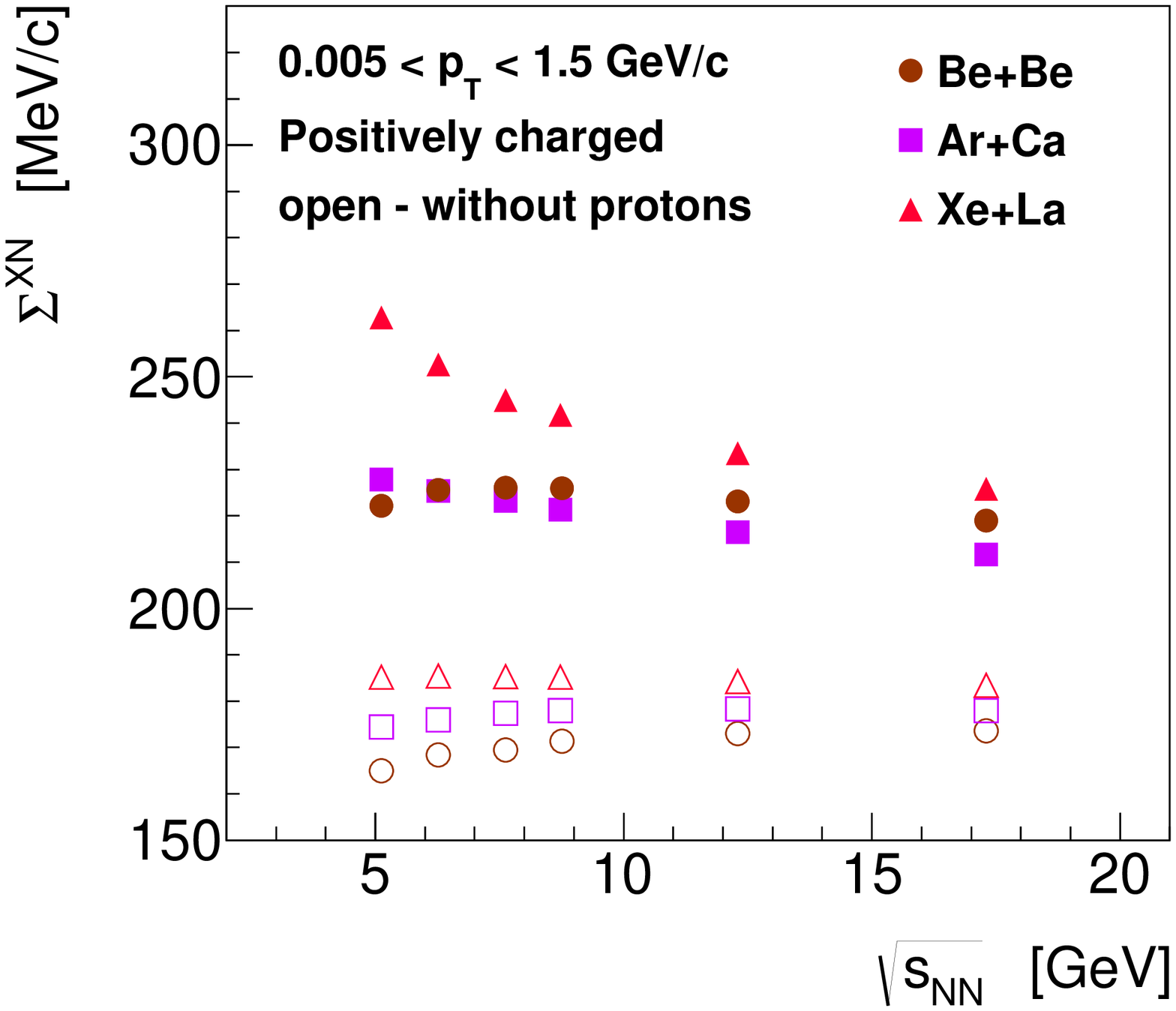}
\vspace{-0.6cm}
\caption[]{Energy dependence of $\Sigma^{XN}$ for different charge combinations of particles produced in 20\% most central $A+A$ collisions in UrQMD.}
\label{urq_sigma_ptcut}
\end{figure}

Figure \ref{urq_sigma_ptcut} shows the energy dependence of $\Sigma^{XN}$ for 20\% most central $A+A$ interactions. Three different charge combinations are presented: all charged particles, negatively charged and  positively charged. The open symbols in the right panel represent positively charged particles without protons. The samples of all charged and positively charged partciles show a monotonic decrease of $\Sigma^{XN}$ with increasing energy, whereas negatively charged rise and reach saturation at middle SPS energies. Similarly to $\Phi_{p_{T}}$ and $\Delta^{XN}$ the values of $\Sigma^{XN}$ for positively charged particles became comparable with those for negatively charged ones provided that protons are removed from the sample.

\begin{figure}[ht]
\includegraphics[width=0.325\textwidth]{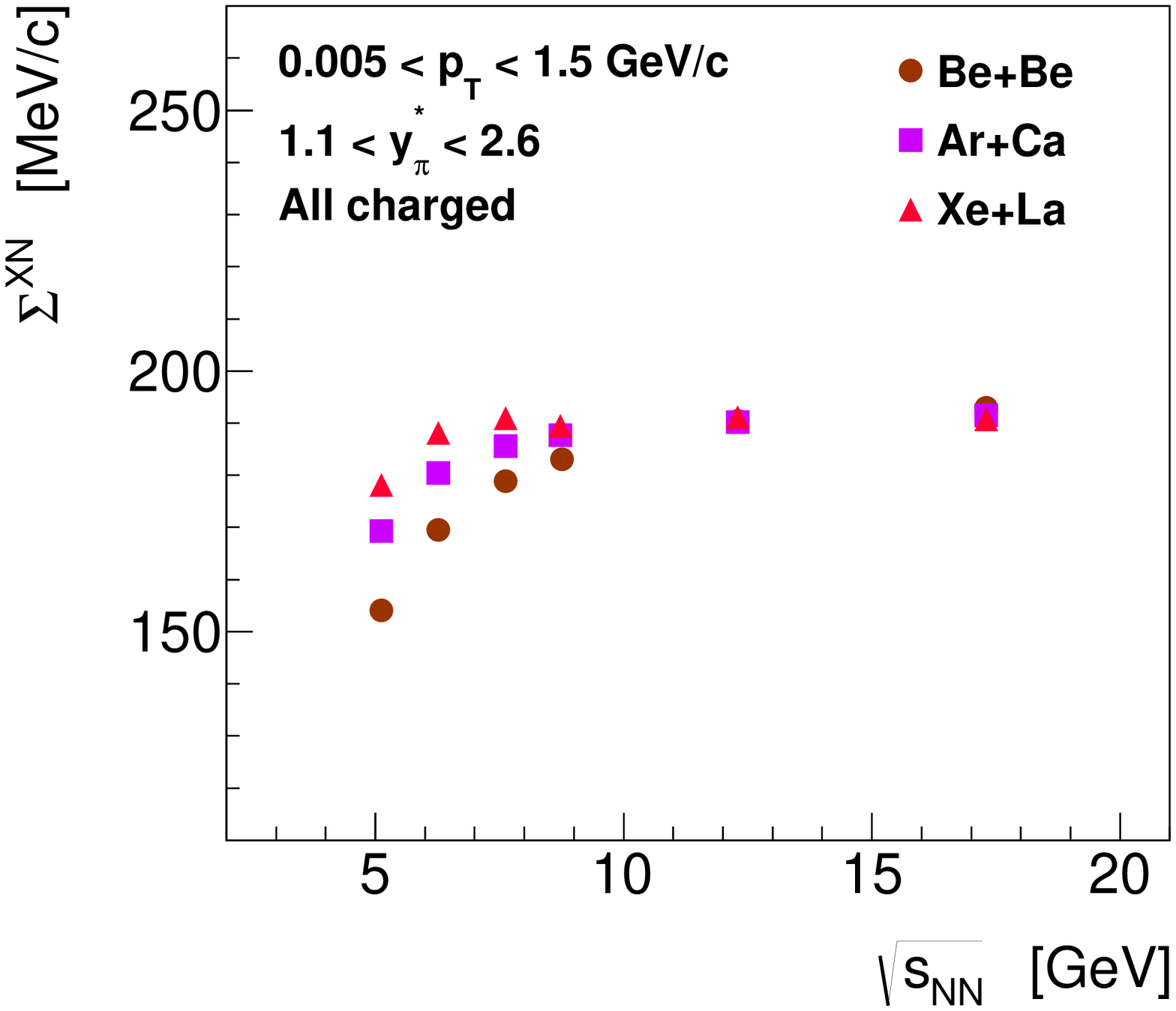}
\includegraphics[width=0.325\textwidth]{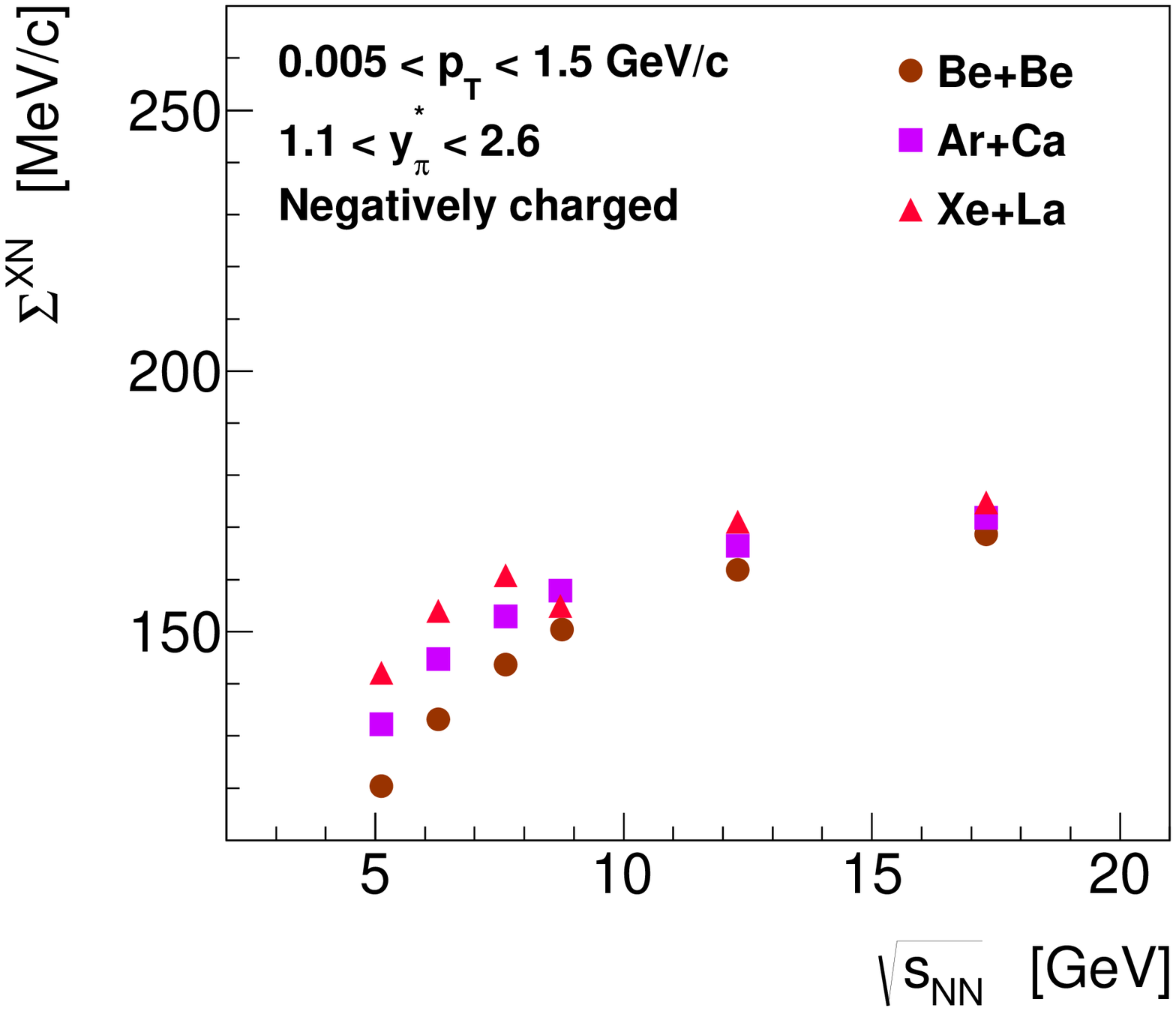}
\includegraphics[width=0.325\textwidth]{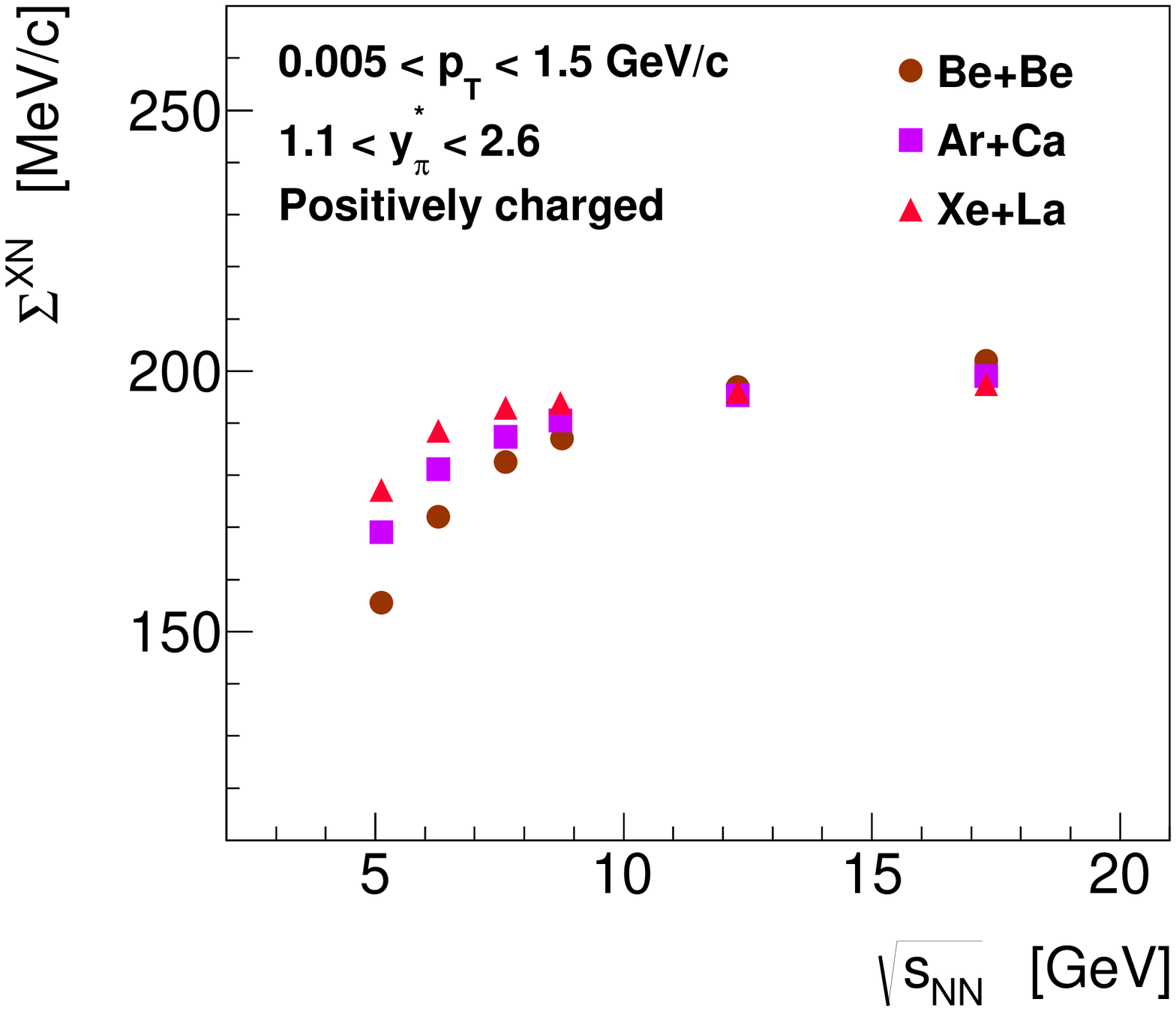}
\vspace{-0.6cm}
\caption[]{Energy dependence of $\Sigma^{XN}$ at forward rapidity for different charge combinations of particles produced in 20\% most central $A+A$ collisions in UrQMD. Additional cut $y^{*}_{p} < y^{*}_{beam} - 0.5$ was applied.}
\label{urq_sigma_rapcut}
\end{figure}

In Fig. \ref{urq_sigma_rapcut} the values of $\Sigma^{XN}$ are shown for forward rapidity only ($1.1 < y^{*}_{\pi} < 2.6$). Additionally, only particles with $y^{*}_{p} < y^{*}_{beam} - 0.5$ were accepted. $\Sigma^{XN}$ values measured at forward rapidity are slightly smaller than those for complete rapidity region. All three charge combinations and all systems show an increase and then saturation of $\Sigma^{XN}$ with increasing energy.

\begin{wrapfigure}{r}{4.cm}
\vspace{-0.6cm}
\includegraphics[width=0.325\textwidth]{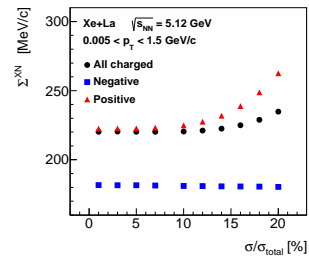}
\vspace{-0.6cm}
\caption[]{$\Sigma^{XN}$ as function of percent of total inelastic cross section for $Xe+La$ collisions at the lowest SPS energy. Note: the values and their errors are correlated.}
\label{urq_sigma_centrality}
\end{wrapfigure}

Figure \ref{urq_sigma_centrality} presents the values of $\Sigma^{XN}$ when the centrality of $Xe+La$ at 13$A$ GeV beam energy is restricted from 0-20\% (rightmost points) down to 0-1\% most central (leftmost points). The negatively charged particles exhibit nearly flat dependence of $\Sigma^{XN}$ on $\sigma / \sigma_{total}$ (similarly to $\Phi_{p_{T}}$ - see Fig. \ref{urq_phi_centrality}). The values of $\Sigma^{XN}$ for all charged paricles and for positively charged ones decrease when going to more central collisions and
reach a plateau for approximately 10\% most central collisions. There is however one significant difference between the centrality dependence of $\Phi_{p_{T}}$ or $\Delta^{XN}$ and $\Sigma^{XN}$ for $Xe+La$ at 13$A$ GeV. For extremely central collisions the values of both $\Phi_{p_{T}}$ and $\Delta^{XN}$ are similar for positively, negatively and all charged particles (see Figs. \ref{urq_phi_centrality} and \ref{urq_delta_centrality}). In contrary, the values of $\Sigma^{XN}$ for very central collisions are much different for negatively charged particles than for positively charged and all charged. It might indicate the presence of additional source of correlations in the sample of positively charged particles, which was not detected by use of $\Phi_{p_{T}}$ and $\Delta^{XN}$ measures. The origin of such correlation(s) was not investigated here.

\begin{figure}[ht]
\includegraphics[width=0.325\textwidth]{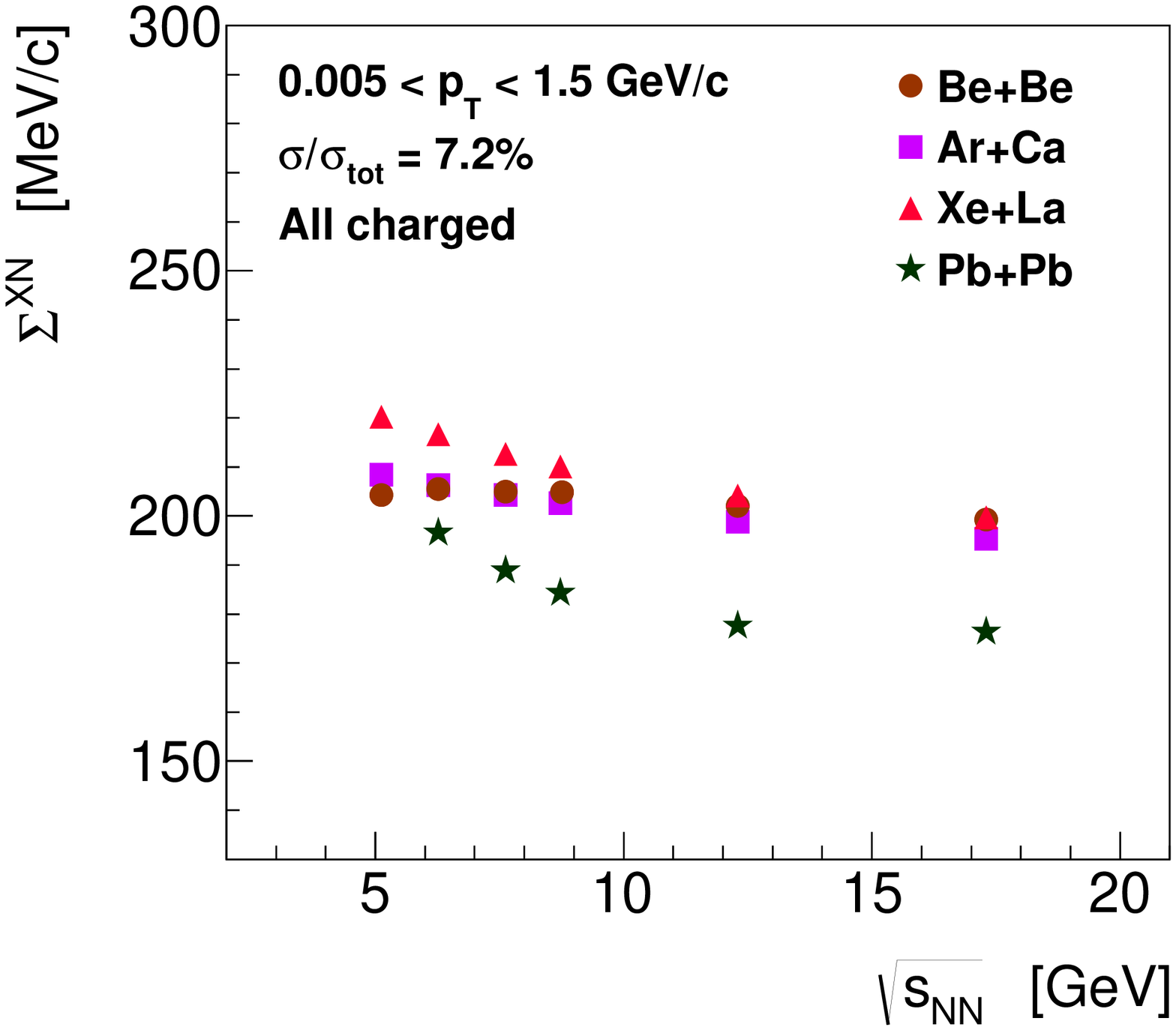}
\includegraphics[width=0.325\textwidth]{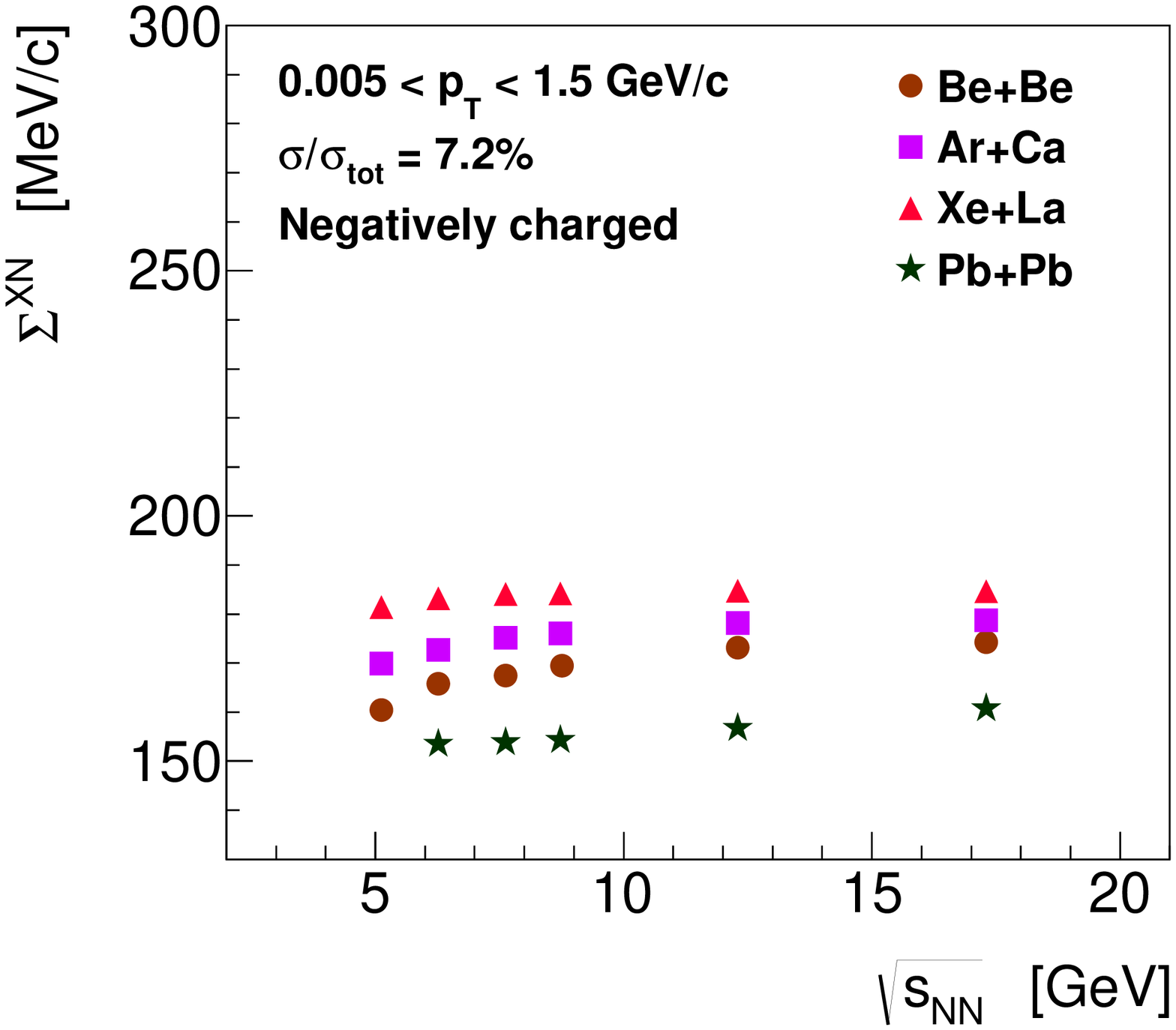}
\includegraphics[width=0.325\textwidth]{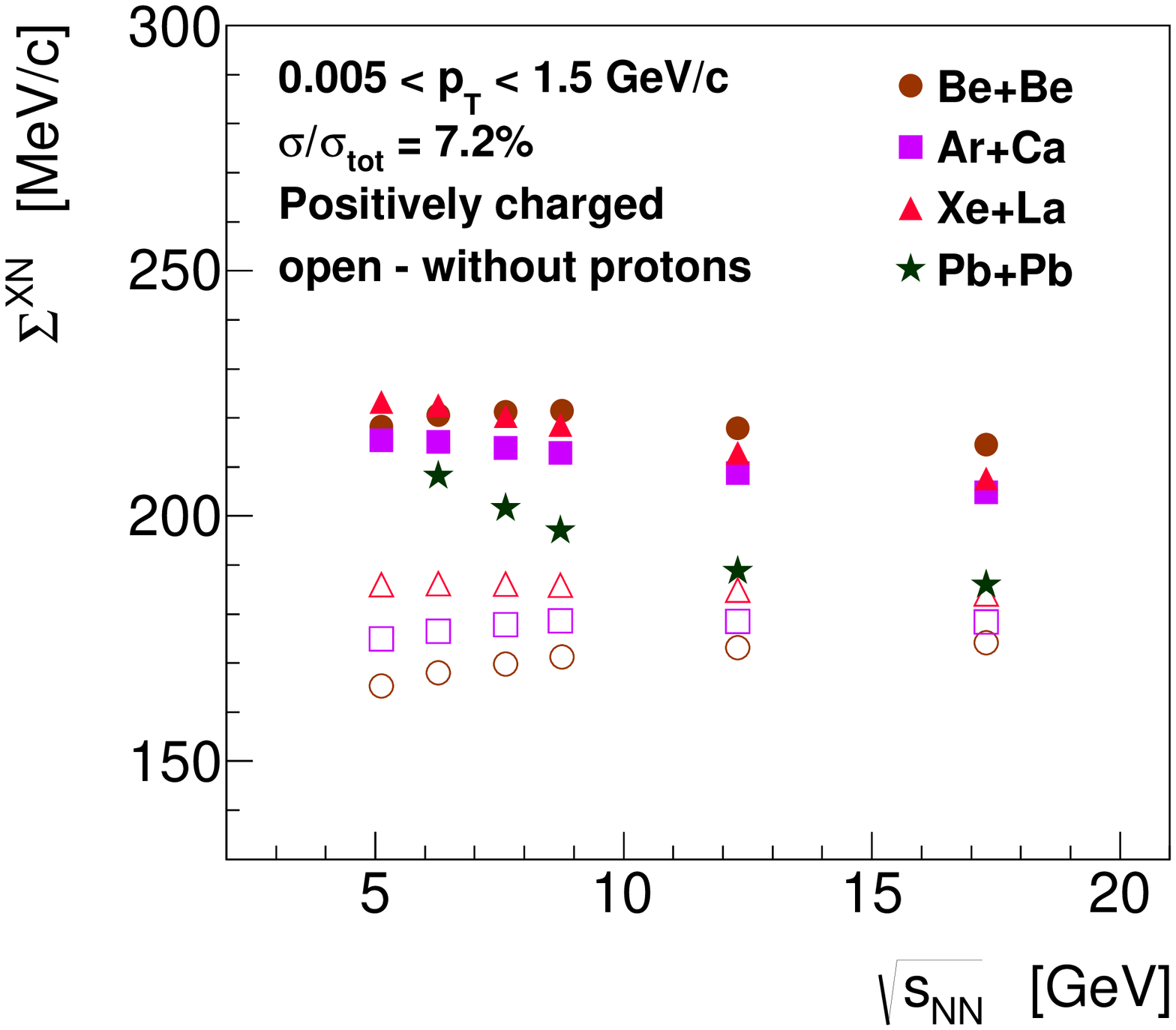}
\vspace{-0.6cm}
\caption[]{Energy dependence of $\Sigma^{XN}$ for different charge combinations of particles produced in UrQMD. 7.2\% most central interactions are selected.}
\label{urq_sigma_72central}
\end{figure}

The last set of plots (Fig. \ref{urq_sigma_72central}) presents the energy dependence of $\Sigma^{XN}$ for 7.2\% most central $A+A$ collisions. For all systems the values of $\Sigma^{XN}$ are higher for positivey charged partciles than for negatively charged ones and, as usually, the rejection of protons brings the magintudes of positively charged ones close to these for negatively charged. 

The 7.2\% most central collisions showed a difference between the values of fluctuation measures for positively charged particles and for positively charged without protons. In case of $\Phi_{p_{T}}$ (Fig. \ref{urq_phi_72central}) this difference is significant only for light systems ($Be+Be$, $Ar+Ca$) and in case of $\Delta^{XN}$ and $\Sigma^{XN}$ such deviation is observed for all systems. Therefore, the results suggest that either the centrality bin 0-7.2\% is still not narrow enough (it may be the case for $\Delta^{XN}$, see Fig. \ref{urq_delta_centrality}) and event-by-event impact parameter (number of protons) fluctuations can be still present or there is another yet source of correlations visible for positively charge particles only (then protons, of course, would have a significant contribution to such correlation). The $\Sigma^{XN}$ measure can be much more sensitive to such kind of correlation (conservation laws?) because a difference between $\Sigma^{XN}$ for positively charged particles and for positively charged without protons is visible for all systems. In case of $\Phi_{p_{T}}$, as already mentioned, only smaller systems show this deviation. The origin of such possible correlation is still under study and in particular the $\Phi_{p_{T}}$ and $\Sigma^{XN}$ measures will be carefully analysed for the energy scan of UrQMD $p+p$ collisions \cite{tobczo}.

Finally, in Fig. \ref{urq_sigma_72central}, one observes that for 7.2\% most central collisions the $\Sigma^{XN}$ values for negatively charged particles are nearly independent of energy. There is however rather significant system size ($A$) dependence suggesting the deviation of UrQMD from a simple superposition model. Such deviation was also observed for $\Delta^{XN}$ measure (middle panel of Fig. \ref{urq_delta_72central}), however only at higher SPS energies.


\section{Summary and conclusions}

In this publication the recently proposed $\Delta^{XN}$ and $\Sigma^{XN}$ measures of transverse momentum fluctuations are tested and compared to $\Phi_{p_{T}}$ measure. The fast generator simulations confirm that all three measures are both intensive and strongly intensive. The results suggest that all three measures are also very sensitive to the effect of event-by-event "temperature" fluctuations. Such effect should be better visible for heavier systems.  

Some preliminary tests were performed also within the UrQMD model, which showed many interesting effects, different for different measures. The complete understanding of all possible sources of fluctuations and correlations in UrQMD would require much more detailed investigations and it is out of the scope of this publication. However, a rather consistent picture can be drawn from these basic tests: if one wants to study the exotic effects such as critical point, onset of deconfinement, etc. one should find good reference values of the studied measures. The UrQMD model suggests that the best reference values are those for negatively charged particles produced in very central $A+A$ collisions, however even for these samples one should first try to better understand the origin of the rudimentary energy and/or system size dependence of $\Delta^{XN}$ and $\Sigma^{XN}$ observed in UrQMD. The simultaneous measurement of all three quantities suggests that the system size ($A$) dependence of $\Delta^{XN}$ and $\Sigma^{XN}$ can be due to change of the single particle spectrum mainly.     
The $\Phi_{p_{T}}$ values for negatively charged particles and for very central $A+A$ are consistent with zero and do not depend on the mass of the colliding nuclei. For a sample of positively charged particles the reference values of all three measures may be masked by some trivial sources of fluctuations and correlations caused by the presence of protons in the sample. The fluctuation measures for a sample of all charged particles is always a non trivial combination of measures for positively and negatively charged particles. Moreover, correlations between positively and negatively charged particles (i.e. due to particle decays) contribute here.       

Finally, one should mention that in most cases the restriction of the analysis to forward rapidity only reduces the values of fluctuation measures. Therefore the NA61/SHINE experiment should do its best in order to measure event-by-event fluctuations in a rapidity window as wide as possible.

\vspace{1cm}

{\bf Acknowledgments:} \newline
I am indebted to the authors of the UrQMD model for the permission to use their code in my analysis. I also would like to thank Bartosz Maksiak for a possibility of using his UrQMD3.3 $Be+Be$, $Ar+Ca$ and $Xe+La$ data sets. I am very grateful to Marek Ga\'zdzicki and Zbigniew W{\l}odarczyk for their useful comments concerning this analysis, and to Maja Ma\'ckowiak-Paw{\l}owska for careful reading of the manuscript.


\end{document}